\documentclass[12pt,english,oneside,openany]{book}
\usepackage{bookman}
\usepackage{mathrsfs}
\usepackage[T1]{fontenc}
\usepackage[latin1]{inputenc}
\usepackage{fancyhdr}
\usepackage[Sonny]{fncychap}
\usepackage{babel}
\usepackage{psfrag}
\usepackage{graphicx}
\usepackage{amsmath}
\usepackage{amssymb}
\usepackage{setspace}
\usepackage[nottoc]{tocbibind}
\usepackage[sectionbib]{chapterbib}
\usepackage[toc,titletoc]{appendix}

\makeatletter
\usepackage{verbatim}
\DeclareMathOperator{\csch}{csch}
\providecommand{\abs}[1]{\lvert#1\rvert}
\newcommand{\lra}{\ensuremath{\leftrightarrow}}
\newcommand{\etal}{\textit{et al}.}
\newcommand{\au}[1]{\textsc{#1}}
\newcommand{\yr}[1]{\textsf{#1}}
\newcommand{\jn}[1]{\textsl{#1}}
\newcommand{\vl}[1]{\textbf{#1}}
\newcommand{\pg}[1]{\texttt{#1}}
\newcommand{\vol}[1]{\textbf{#1}}

\newcommand{\ffill}[1]{\mbox{\textrm{\textit{#1}}}}
\makeatother
\begin{document}
\pagestyle{empty}
\begin{figure*}[p]
\begin{center}
\includegraphics{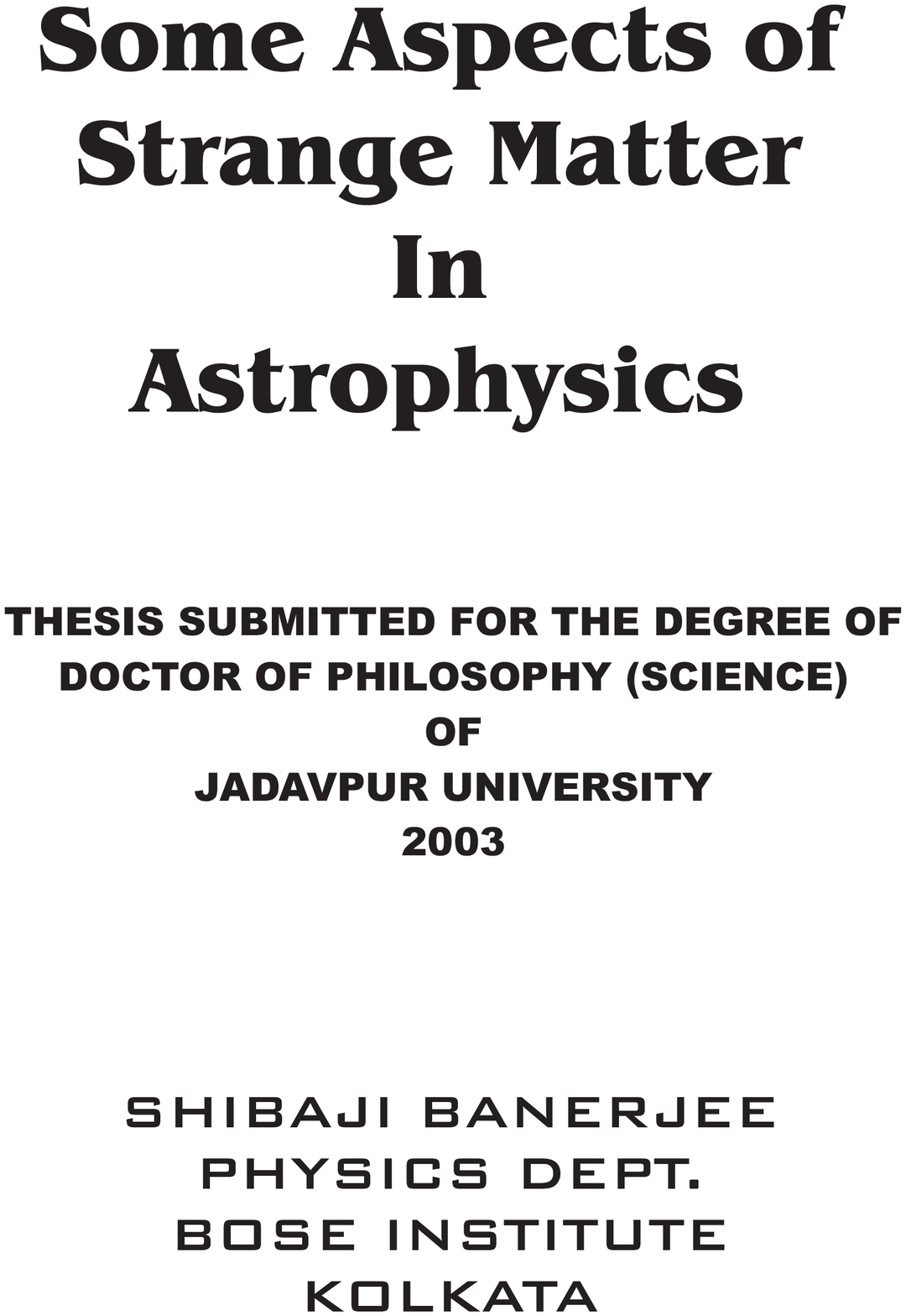}
\end{center}
\end{figure*}
\clearpage
\vspace{4in}
{\centering \Large{\underline{CERTIFICATE FROM THE SUPERVISOR}}\vspace{0.4in}\par}
\begin{flushleft}
\texttt{%
This is to certify that the thesis entitled\\ \ffill{"Some Aspects of Strange Matter in Astrophysics"}
submitted by \textrm{Sri} \ffill{Shibaji Banerjee}, who got his name registered on \ffill{16.11.2000} for the award of \textrm{Ph.D. (Science) degree} of \textrm{Jadavpur University}, is absolutely based upon his work under the supervision of
\ffill{Professor Sibaji Raha} and that neither this thesis nor any part of it has been submitted for any degree/diploma or any other academic award anywhere before.}
\end{flushleft}
\vspace{0.4in}
\begin{flushright}
\bfseries
\texttt{%
\underline{(Signature of the Supervisor} \\
\underline{\& date with official seal.})}
\end{flushright}
\clearpage
\begin{figure*}[p]
\begin{center}
\includegraphics{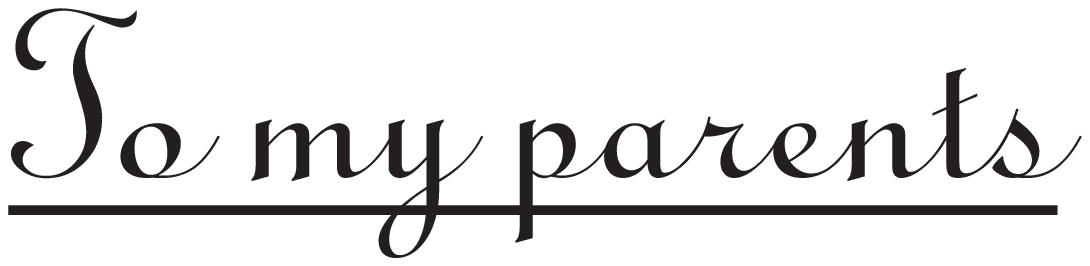}
\end{center}
\end{figure*}
\clearpage
{\centering \Large\textit{Acknowledgements}\vspace{0.2in}  \par}
{\it \small I would like to use this space to convey my sincere appreciation for
all the assistance and support that I have received during the course of
this work. 

The Council of Scientific and Industrial Research was the primary
source of support during the above period, backed up by the facilities
provided by Bose Institute.

I am indebted to Prof. Sibaji Raha, who, apart from being my supervisor
continues to act as a friend and a philosopher -- all in one unfragmented
self ! I remember being given complete freedom in choosing what, when
and how to go about a work, but rescued frequently from the heat and
frustration generated as a natural consequence of such enhanced degrees
of freedom. The most valuable thing that I consider having learnt
from him is the value of \emph{down to Earth physics} which can really
stand on it's own feet in the middle of technical hype and dust. 

I am equally grateful to Prof Debopriyo Syam whom I had the opportunity
to have as a teacher as well as a collaborator. Many of the viewpoints
expressed here are the results of long hours of discussion with him. 

No words of thanks are enough to express my gratitude towards the
local group consisting of my (senior) collaborators -- Dr. Abhijit
Bhattacharya and Dr. Sanjay Ghosh. It is from them that I learnt numerous
tricks of the trade, about the community and about making sense out
of a morass of numbers. They have indeed been my virtual mentors and
family throughout. 

I would also like to use this opportunity to thank Prof. Bikash Sinha
for valuable suggestions. I am also thankful to Prof G.Wilk, Prof.
Kamalesh Kar, Prof. Santosh Samaddar and Prof. Dipanjan Roy Chowdhury
for essential feedback. I want to thank Prof. Partha Mitra specially
for directing me to the right group and person at the beginning of
my pre-doctoral work. 

I would also like to thank all the faculty of the Bose Institute and
especially Prof. M.H Engineer, Prof. Indrani Bose and Prof. Dipankar
Home for inspiration. I earnestly thank the staff of the Physics Office
as well as the Institute Canteen, who have been very supportive during the
period. The thanks extend to all my friends and colleagues at the Bose
Institute -- Asim, Indra, Mala, Somsubhro, Emily, Sonali, Tamal, Rajesh
and Subhashis for tea, plain and technical chat, smoke, crosswords and
sharing this part of life together. Thanks also go to my colleagues at
the St.Xavier's College who have tried their best to give me a free hand
in the development of the work. No words of thanks will be sufficient to
express my sincere gratitude and appreciation to my family members for
their unfailing support, patience and endurance shown during the period.}

\vspace{0.2in}
\begin{center}
\includegraphics[scale=0.4]{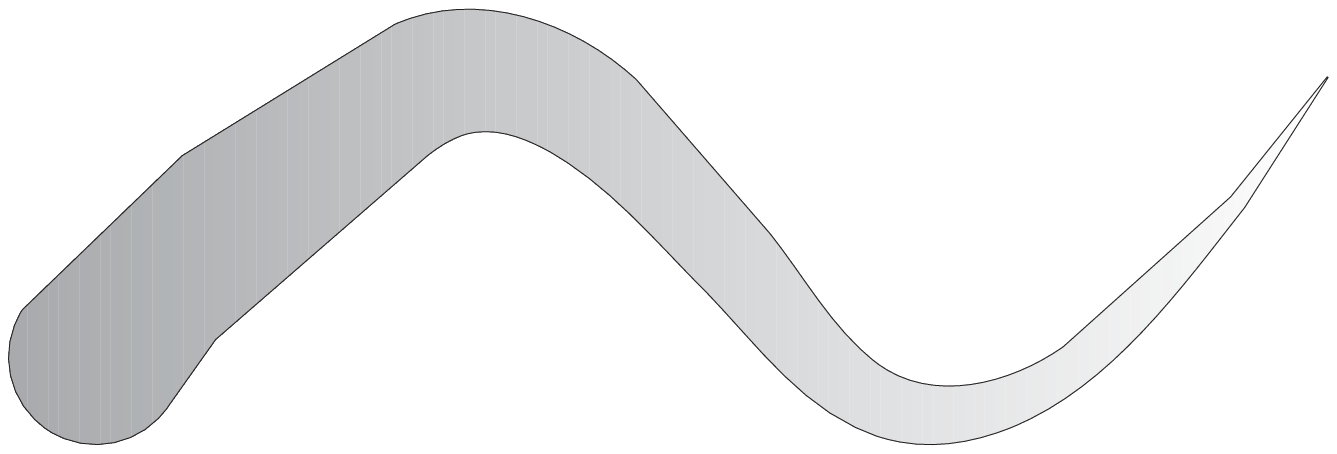}
\end{center}

\tableofcontents
\newpage
\pagestyle{fancy}
\chapter{\label{chap:intro}Introduction and Overview}
\section{\label{sec:outlook} Outlook}
The issues related to the origin, evolution, structure and constitution of
the universe have attracted the attention of specialists and non-specialists
for a very long time; however, the present epoch in the development of such
theories is marked by a capacity to nail down everything to an astonishing
precision. This has become possible largely due to the availability of
today's observational and computational resources and it appears that
at this time, the community has reached a consensus as to the 
queries related to the origin and composition of the universe. The 
major agreements (not yet really unanimous) belong mainly to three important
areas, summarized in the following statements.
\begin{enumerate}
      \item The Universe had its origin in a hot (and dense) big bang.
	\item In the large scale the Universe appears to be flat, isotropic, 
              homogeneous and expanding, possibly at an accelerated pace.
	\item Most of the material content of the universe is carried by dark
              matter (i.e non luminous matter which marks its presence only
              through gravitational interaction.)
\end{enumerate}
A large number of theories have been put forward to explain and interpret
the above facts. Most of them seem to exist independently of the others,
even when one takes into account constraints put on the individual
theories. It appears that further experimental data are required before one
is able to discern between them in an effective manner. In contrast to
the many proposals on the composition of the Universe, which often border
on exoticism, the present work lies entirely within the framework of the
Standard Model of particle interactions. 

The unifying theme of the current work is the fact that the STANDARD MODEL 
allows the existence of \mbox{QUASIBARYONIC} objects (hypothetical quark matter
forms to which one can assign a baryon number, and in which strange quarks 
are an essential ingredient ). This work focuses on the aspect of finding 
the relevance of such quasibaryonic objects in the issues related to the 
structure and composition of the universe in the light of modern astronomical 
observations. 

The remaining part of this chapter is essentially a compact summary of the 
current ideas regarding the structure and composition of the universe along 
with an overview of the ideas leading to and linked with the strange (quark) 
matter hypothesis -- how it can be correlated with the facts known  
(and being continually revealed) about the dark matter dominated universe. 
Specifically, I will try to outline the ways in which strange quark matter 
may form and manifest itself, current searches for such matter and other 
related issues which lend perspective to the present work.

\section{\label{sec:bricks}Bricks of the World}
\begin{figure}[p]
\begin{center}
\includegraphics{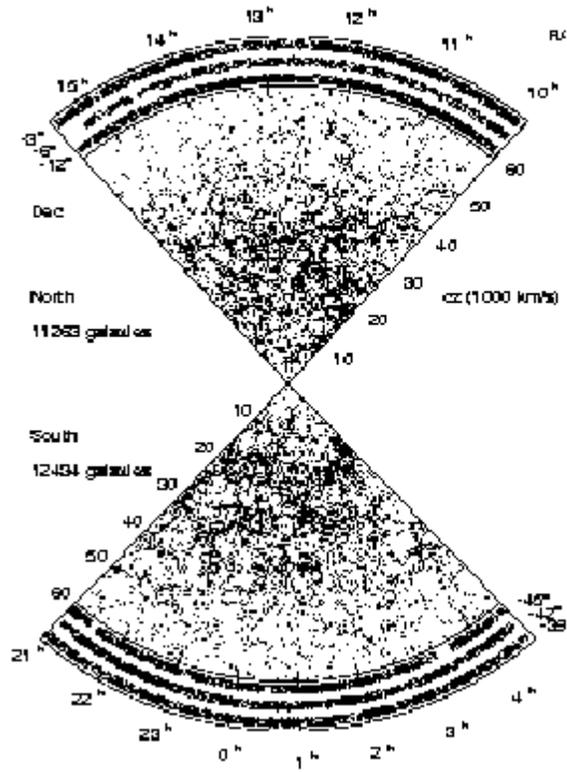}
\end{center}
\caption{\label{fig:ustruct} Pie-diagram showing the hierarchial Structure of the Universe
as revealed by the Las Campanas Redshift Survey \cite{shectman}}
\end{figure}

Not so long ago, the picture of the universe consisted of a single galaxy
(the Milky Way) and was believed to be few million of years old. Today, we 
believe that the origin of everything can be traced back to the hot Big-Bang 
event that took place about 14 billion (\( 13-15\times 10^{9} \)) years 
ago. The structure of the universe (\cite{structu}) depends on the scale at
which it is being observed : at scale sizes lower than \( \sim \) 100 Mpc
\footnote{Mpc = 3.26 \( \times 10^6 \) Light Years} 
there exists hierarchical structures consisting of great \textit{walls}
of galaxies(accommodating more than 90\% of the observed galaxies),
super-clusters accommodating clusters of groups of galaxies, inter-spaced by
\emph{voids} or spaces devoid of any bright galaxies(Fig.\ref{fig:ustruct}),
although at higher scales, all these structures appears to be replaced
by a uniform homogeneous mass distribution. The questions regarding the
structure and evolution of these objects appears to be intimately related. 

Much of what is currently accepted as standard cosmological model \cite{grtext}
can be traced back to the discovery of the cosmic microwave background radiation (CMBR) in 1964. 
This, together with the observed Hubble expansion of the universe, had established 
the hot big bang model as a viable model of the universe. The ultimate acceptance of the 
standard cosmological model was mostly due to the success of the theory of 
nucleosynthesis in reproducing the observed pattern of abundance of the 
light elements along with the proof of the black body character of the CMBR. 

For cosmic times larger than the Planck time
\footnote{%
The Planck Mass is the mass value obtained by appropriately combining 
the fundamental constants $G$, $c$ and and $\hbar$ to $\sqrt{ \frac{\hbar c}{G} }$
In Naturalized units,where $\hbar = c= 1$, $M_{\mathrm{Pl}}^{-1}=\sqrt{G}$.The
Planck time is the time it takes light to travel a length equal to the Compton
wavelength of a Planck mass particle, i.e $t_{\mathrm{Pl}}=\sqrt{G}$. 
}%
(\( t_{\mathrm{Pl}}=M_{\mathrm{Pl}}^{-1}=10^{-44} \) sec) gravitation can be 
described adequately by classical general relativity. 
The (Null) experimental evidence (\cite{smoot}-\cite{wilkinson}) regarding the 
anisotropy of the CBR (cosmic background radiation) allows one to assume the 
\emph{cosmological principle} i.e the universe is homogeneous and 
isotropic (in the large scale). 
The four dimensional spacetime in the universe is then simply described by the 
Robertson - Walker metric \cite{grtext}: 
\[ ds^{2}= -dt^{2}+a^{2}(t)\left( \frac{dr^{2}}{1-kr^{2}}+r^{2}\left( d\theta ^{2}+\sin ^{2}\phi d\phi ^{2}\right) \right) \]
where, \( r,\theta ,\phi  \) are the co-moving polar coordinates which are fixed
for objects that have no other motion other than the general expansion of the
universe, \( a(t) \) is the scale factor normalized to 
\( a_{0}=a(t_{0})=0 \), \( t_{0} \) is the present time and \( k \) is the 
scalar curvature. The instantaneous \emph{physical} radial distance is given by : 
\[ R(t)=a(t)\int _{0}^{r}\frac{dr}{\sqrt{1-kr^{2}}}\] 
and the physical velocity of an object (with no peculiar velocity with respect 
to the co-moving frame) then turns out to be 
\[\vec{V}=\frac{\dot{a}}{a}\vec{R}\equiv H(t)\vec{R}\]
where \( H(t) \) is the Hubble parameter. For a flat universe (\( k=0 \))
the relationship between the physical vectors \( \vec{R} \) and the
co-moving vectors \( \vec{r} \) is simple revealed as \( \vec{R}=a\vec{r} \)
. The radius of curvature is given by 
\[R^{2}_{\mathrm{curv}}=\frac{H_{0}^{-2}}{\Omega _{0}-1}\]
where \( H_{0} \) is the present value of the Hubble parameter 
(\( H_{0}=100h\, \mathrm{Km}\, \mathrm{S}^{-1}\, \mathrm{Mpc}^{-1} \)
with \( 0.4\leq h\leq 0.8 \)) and \( \Omega _{0} \) is the ratio
of the energy density \( \rho _{\mathrm{tot}} \) contributed by all
forms of matter and energy to present value of the critical density
\( \rho _{\mathrm{crit}} \). In other words,  
\[\Omega _{0}=\sum _{i}\Omega _{i}=\sum _{i}\frac{\rho _{i}}{\rho _{\mathrm{crit}}}=\frac{\rho _{\mathrm{tot}}}{\rho _{\mathrm{crit}}}\]
where \( \rho _{i} \) is the energy density contributed by the \( i^{th} \)
entity. The critical density, which appears in the above equation can be
expressed in terms of the the constants \( H_0 \) and \( G \) in the following
form --
\[\rho _{\mathrm{crit}}\equiv \frac{3H_{0}^{2}}{8\pi G}\simeq 1.88h^{2}\times 10^{-29}\mathrm{g}\, \mathrm{cm}^{-3}\]
and can as well be identified with the closure density.  It follows that one can set the scalar
curvature \( k \) to zero if it is found that \( \Omega _{0}=1 \).

\begin{figure}[p]
\begin{center}
\begin{minipage}[t]{5in}
\scalebox{0.4}{\includegraphics{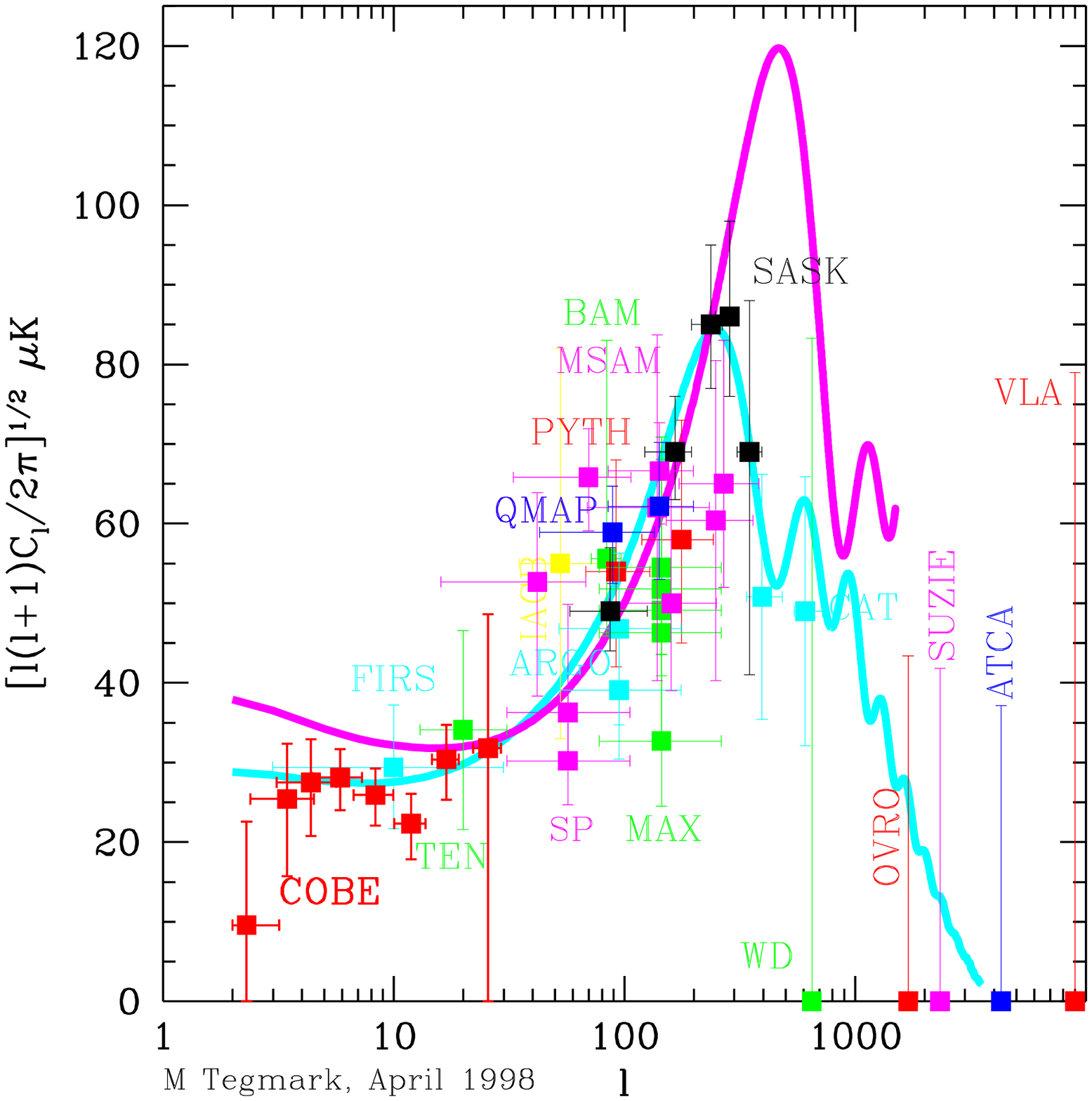}}
\caption{\label{fig:cmbpeak} The power spectrum of CMB, as obtained by recent experiments. 
The light curve is the one preffered by data and corresponds to $\Omega_0=1$ \cite{tegmark}.}
\end{minipage}\\[20pt]
\begin{minipage}[b]{5in}
\scalebox{0.45}{\rotatebox{90}{\includegraphics{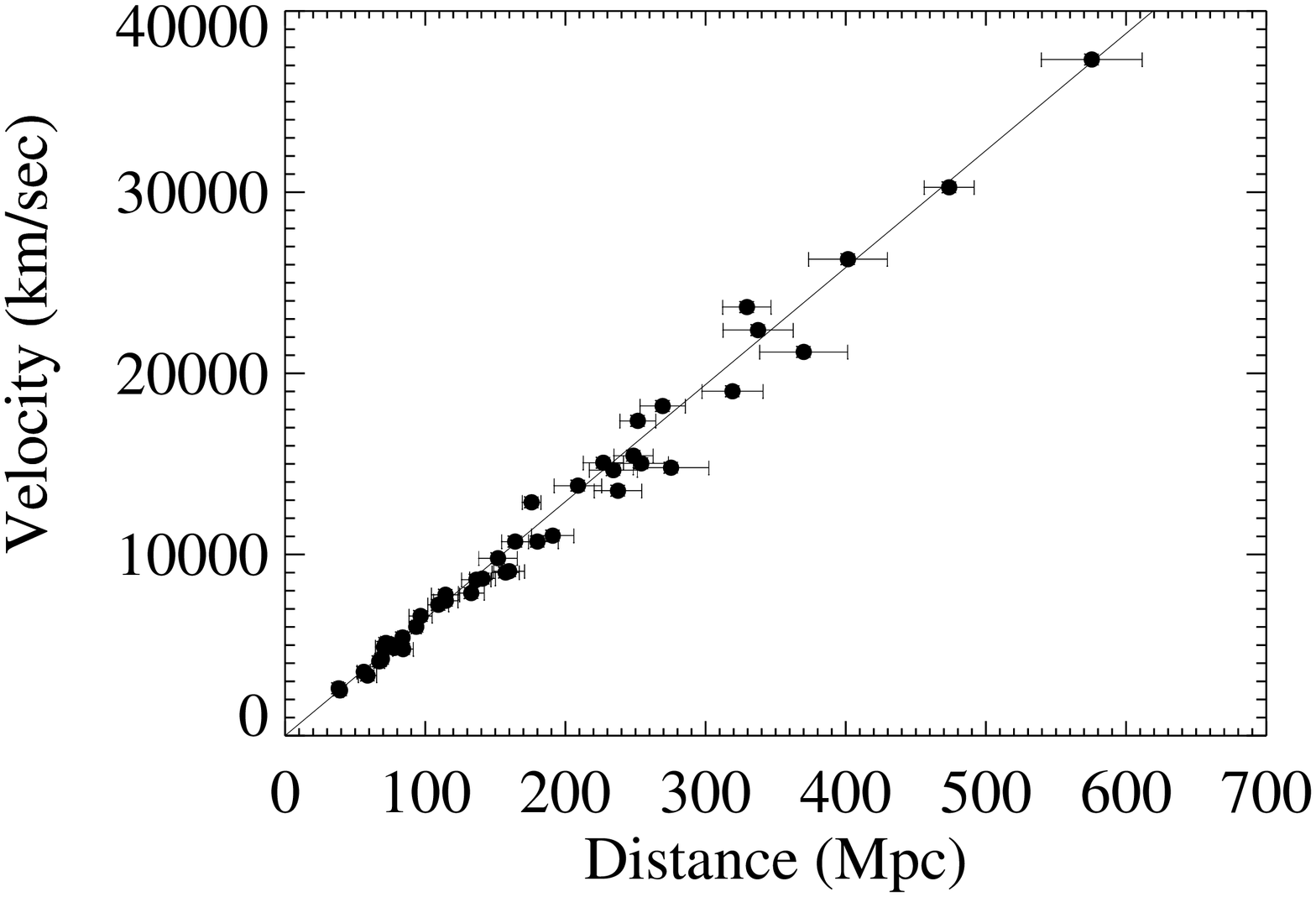}}}
\caption{\label{fig:sne} Hubble diagram based upon distances to supernovae of 
type 1a (SNe1a). The slope of the line is the presently 
accepted value of the Hubble constant $H_0 = 64 \mathrm{Km}\mathrm{s}^{-1}\mathrm{Mpc}$,
\cite{riess}}
\end{minipage}
\end{center}
\end{figure}

The value of \( \Omega _{0} \) is most reliably estimated from the
anisotropy of CBR (see \cite{jung,koso} and recent DASI observations \cite{cmbnew})
(Fig.\ref{fig:cmbpeak}) suggest that its magnitude can be quoted as 
\( \Omega _{0}=1\pm 0.2 \)\footnote{\( \Omega _{0}=1\)
is a \emph{requirement} for the theory of Inflation\cite{infl1}-\cite{infl4}}.
It is thus most likely the case that we really live in a flat 
universe.
The universe in its early stage of evolution is temperature dominated
and the temperature - time relationship in this era can be found from
the Friedman equations (with \( k=0 \)) (assuming adiabatic universe
evolution, i.e constant entropy in a co-moving volume ): 
\[T^{2}=\frac{M_{\mathrm{Pl}}}{2(8\pi c/3)^{1/2}t}\]
and classically the beginning of time \( t=0 \) coincides with \( T=\infty \).
However, due to quantum effects, one can only say that the
classical universe emerges at a cosmic time \( t\sim t_{\mathrm{Pl}} \) with a temperature 
\( T\sim M_{\mathrm{Pl}} \). The important events in the history of the universe 
are summarized in table ( \ref{tab:universeevents})
\begin{table}[h]
\begin{tabular}{|l|cl|}
\hline 
Time&
Temperature&
Event\\
\hline
\hline 
\( 10^{-37} \) sec&
\( 10^{16} \) GeV&
Gut group G breaks down to the \\
&
&
standard model gauge group\\
&
&
\( G\rightarrow G_{S}=SU(3)_{c}\times SU(2)_{L}\times U(1)_{Y} \)\\
\hline 
\( 10^{-10} \) sec&
\( 100 \) GeV&
The Electroweak phase transition\\
&
&
\( G_{S}\rightarrow SU(3)_{c}\times U(1)_{em} \)\\
\hline 
\( 10^{-6} \) sec&
\( 100 \) MeV&
QCD phase transition in which \\
&
&
quarks became bound into hadrons\\
\hline 
\( 180 \) sec&
\( 1 \) MeV&
Nucleosynthesis, protons and \\
&
&
neutrons begin to form nuclei\\
\hline 
\( 3000 \) year&
&
Equidensity Point, \\
&
&
matter begins to dominate over radiation.\\
\hline 
\( 200,000 \) year&
\( 3000 \) K&
Decoupling of matter and radiation\\
&
&
and subsequent evolution of radiation\\
&
&
as independent component.\\
\hline 
\( 2\times 10^{8} \) year&
&
Structure formation starts\\
\hline
\end{tabular}
\caption{\label{tab:universeevents}Events in the history of the Universe}
\end{table}

The scope of the present work necessarily excludes the events that
occurred before the cosmic Quark - Hadron phase transition but addresses
the cosmic history from the subsequent time (i.e after \( 10^{-6} \)
sec) to the current time (i.e \( 13\times 10^{9} \) years)

\subsection{\label{ssec:dm}Dark Matter}

\begin{figure}[p]
\begin{center}
\begin{minipage}[t]{5in}
\centering
\resizebox{!}{3in}{\includegraphics{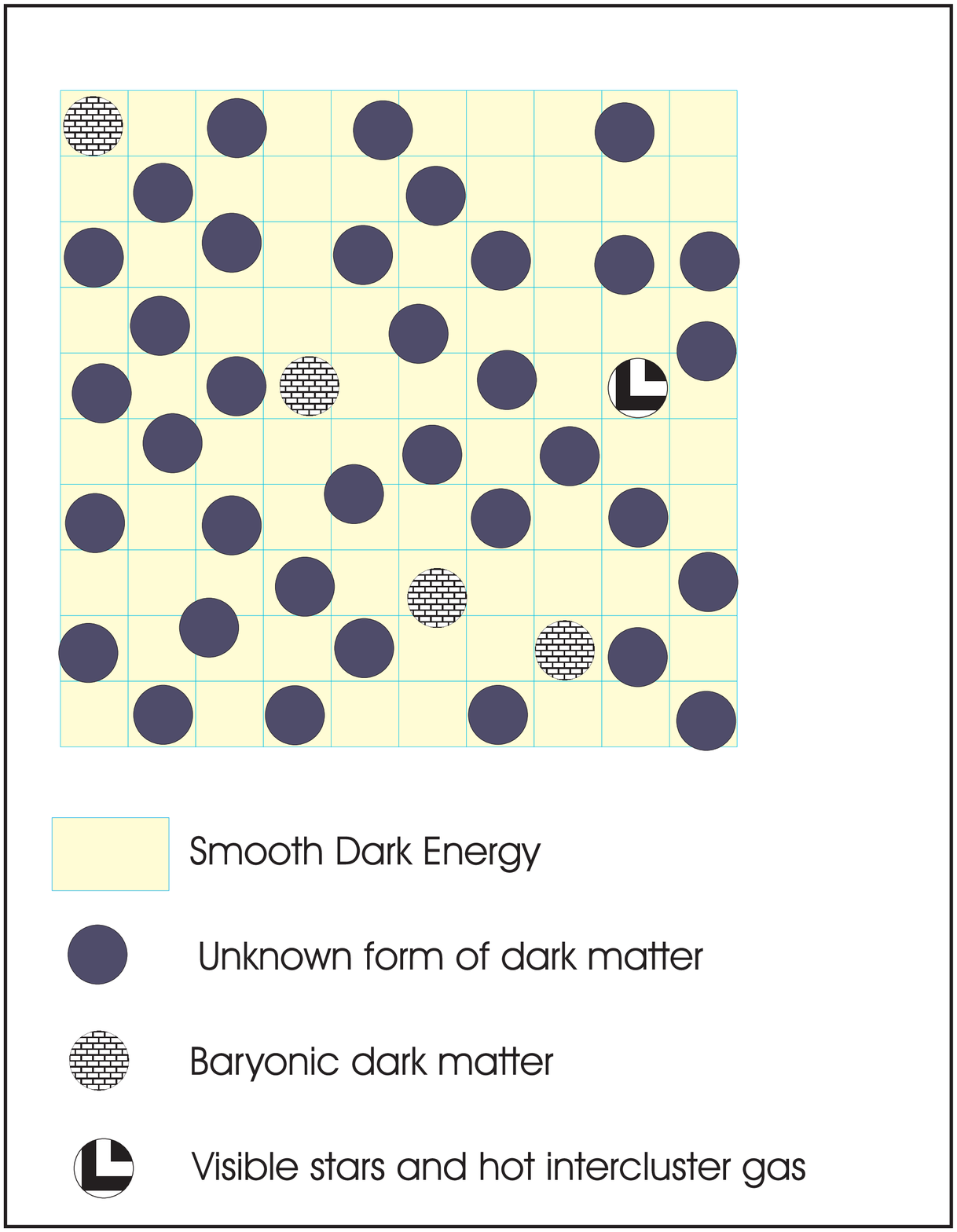}}
\caption{\label{fig:cakecdr} A basic summary of the matter/energy content of the Universe. For
a detailed account see fig.~\ref{fig:me}, below.}
\end{minipage}\\[20pt]
\begin{minipage}[b]{5in}
\scalebox{0.5}{\includegraphics{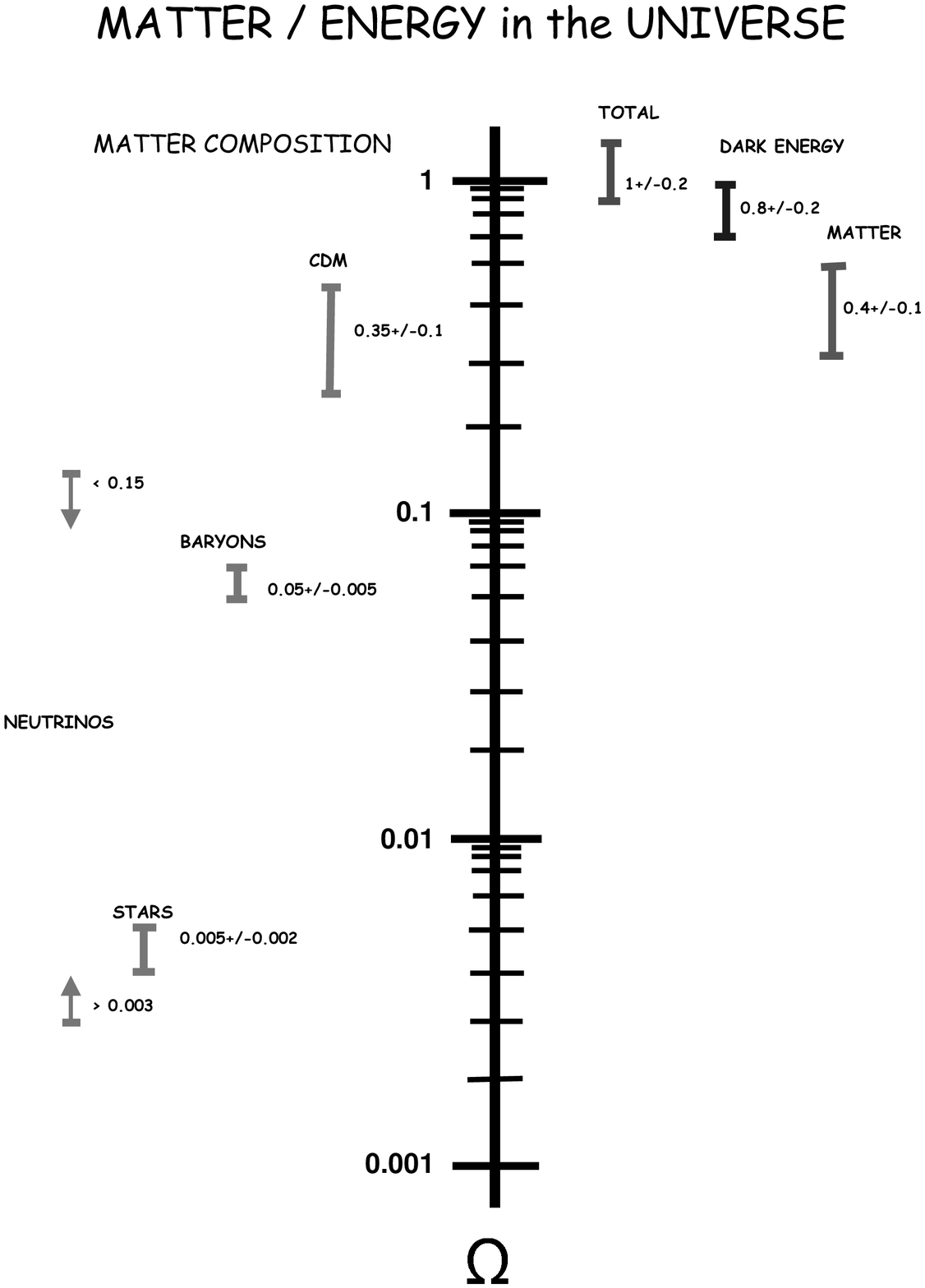}}
\caption{\label{fig:me} A standard summary of the composition of the universe, as given in \cite{structu}}
\end{minipage}
\end{center}
\end{figure}

The total contribution to \( \Omega _{0} \) can be split up into
the matter and energy part 
\[ \Omega _{0}=\Omega _{M}+\Omega _{E}\] (see Fig.\ref{fig:cakecdr} - \ref{fig:me})
The contribution of matter has been estimated in various ways : a
short summary is given in table (\ref{tab:matterdensity}), the generally
accepted value being \( \Omega _{M}\simeq 0.4 \). The apparent contradiction
between the results \( \Omega _{0}\simeq 1 \) and this value is attributed
to a form of a \emph{smooth dark energy} component; the evidence
for which is taken from the accelerated expansion of the universe,
as shown by the Hubble diagrams (Fig.\ref{fig:sne}) for several type Ia Supernovae (SNe
Ia). The energy density contributed by the CBR and the massless neutrinos
are too small to figure significantly in this part (energy sector)
of the energy budget. Turning one's attention to the matter sector one
finds that the big-bang nucleosynthesis can provide the most precise
determination of the baryon density. Comparison of the primeval abundances
of \( \mathrm{D},\, ^{3}\mathrm{He},\, ^{4}\mathrm{He} \) and \( ^{7}\mathrm{Li} \)
with their big-bang predictions defines a interval \( \Omega _{B}h^{2}=0.007-0.024 \)
or \( \Omega _{B}\approx 0.05-0.1 \) \cite{copi,tytler,omeara,burles}. For the most part then, the
baryons are unable to contribute significantly to the \( \Omega _{M} \)
and thus, most of the matter in the universe is dark.

There are actually
two kinds of dark matter problems. The baryons carry only up to 5\%
of the total budget - it is however difficult to fulfill even this
meager amount by collecting the baryons form all the visible stars
(0.3 - 0.6 \%) and hot intercluster gas (0.5 \%). The nature of the
remaining 90 \% dark baryons is unknown : this is the first dark matter
problem. In this work we are however concerned with the second problem
in which one needs to account for the remaining 35 \% contribution
to the matter density (cold dark matter). The baryons that took part
in nucleosynthesis are excluded from this sector and so it appears
that the only possibilities are relic (non-baryonic) elementary particles left over
from the big bang. One needs to \emph{postulate} the existence of
long lived or stable particles with very weak interactions so that
their annihilations cease before their numbers are too small. The
three most discussed particles are a neutrino ( of mass 30 eV) \cite{jung}, an
axion \cite{aximad} of mass \( 10^{-5\pm 1} \) eV and a neutralino \cite{neulmad} 
of mass between 50 - 500 GeV . These and the more exotic possibilities are invoked
primarily to cater to the viewpoint that CDM (Cold Dark Matter) must be nonbaryonic (where
the term nonbaryonic is usually taken as synonymous with non participants
in nucleosynthesis).

\begin{figure}[p]
\centering
\includegraphics[scale=0.75]{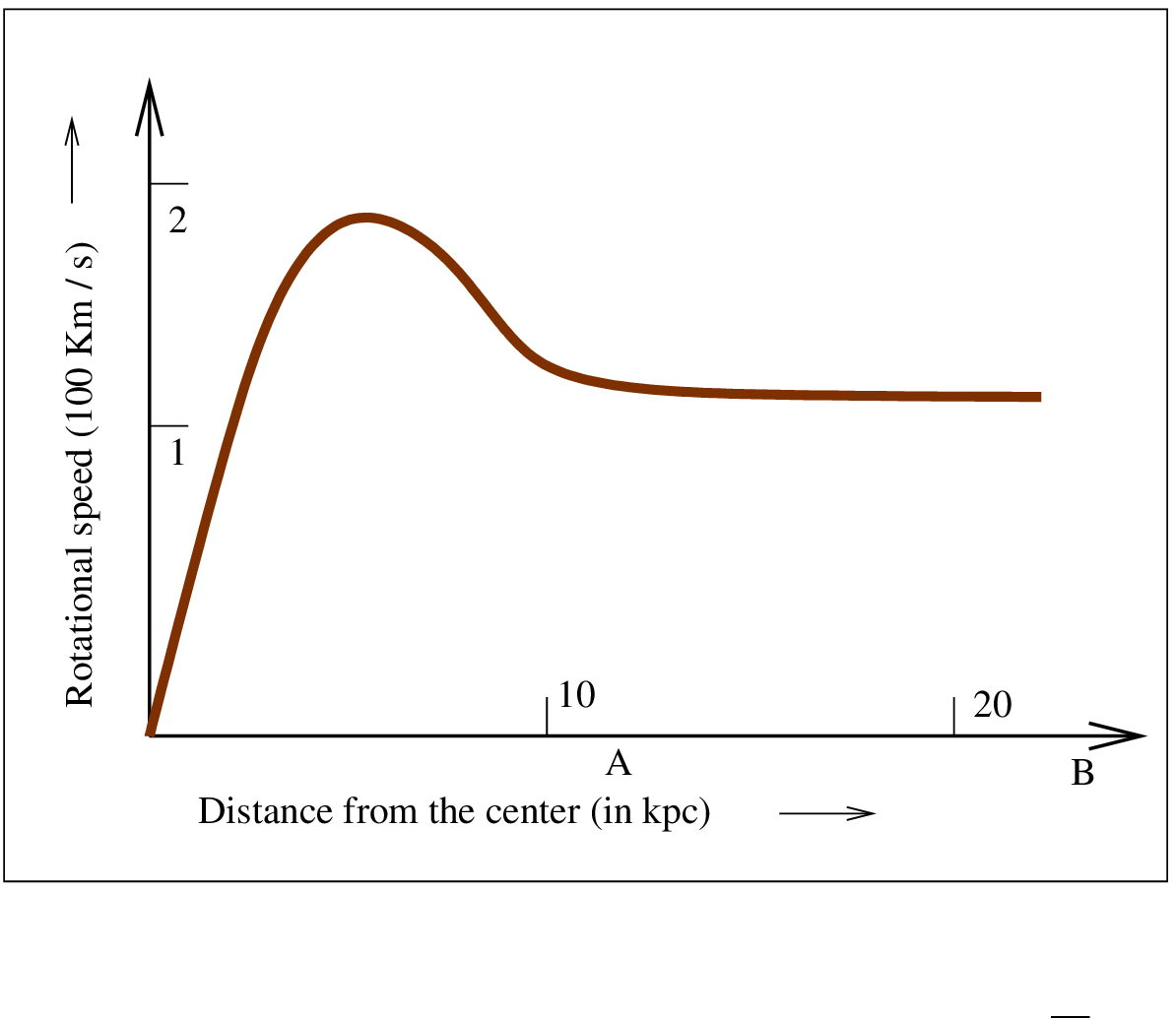}
\caption{\label{fig:rotcurve} Schematic diagram for the rotation curve of typical spiral galaxies.}
\end{figure}

In the present work we present an alternative scenario for the origin of dark matter. The most robust 
evidence for dark matter is related to the nature of the rotation curve of spiral galaxies, in which the 
velocities of some galactic component are plotted against the distance from the galactic center. In 
practice one can obtain the velocities of neutral hydrogen clouds using 21 cm emission. The 
typical form of the curve appears as in Fig.~\ref{fig:rotcurve}. The mass distribution of 
the galaxy can be inferred from the Newton's law of circular motion \( G M /r^2 = v^2 /r \).
The linear rise of \( v(r) \) with \( r \) near \( r = 0 \) show that the mass density is 
essentially constant there. After this brief spell, it is seen that the velocities
remain constant out as far as can be measured. This implies that the density drops like 
\( r^{-2} \) at large radius and that the mass \( M(r) \propto r \) at large radii. Once \( r \) 
becomes larger than the extent of mass contributed by luminous stars (The \textit{luminosity radii}) 
the velocities should drop like \( \propto r^{-1/2} \), but this behaviour is not seen. Thus the 
spiral galaxies seem to contain matter in its dark halo beyond its visible limit. Study of 
the Milky Way Galaxy show that it's rotation curves are consistent with a flat rotation curve 
with \( v = 220 \) km /sec all way out to 50 kpc, so that it is a typical spiral galaxy with 
a large dark halo \cite{kochanek}.

Recent experimental findings on the gravitational lenses in the halo of our galaxy have lend support to this fact. Using 
the cue that the Standard Model of particle interaction allows the existence of objects which can 
contribute baryon numbers but do not participate in the process of nucleosynthesis we put forward 
a theory in which the relationship of quasibaryonic objects to the dark matter content of the universe 
has been explored in some detail. This issue is specifically addressed in Chap.~\ref{chap:macho}.

\begin{table}[h]
\begin{tabular}{|l|c|l|}
\hline 
Type of Analysis/Experiment&
Specific&
Value of \( \Omega _{M} \)\\
\hline
\hline 
Gas to total mass ratio in rich clusters&
X Ray&
\multicolumn{1}{c|}{0.3\( \pm  \)0.05\( h^{-\frac{1}{2}} \)}\\
\cline{3-3} 
\cline{2-2} 
&
\multicolumn{1}{c|}{S-Z}&
0.25\( \pm  \)0.1\( h^{-1} \)\\
\hline 
Evolution of abundance of rich clusters with redshift&
&
0.45\( \pm  \)0.1\\
\hline 
Outflow of material from voids&
&
\( > \)0.3\\
\hline 
Evidence from structure formation&
&
0.4\\
\hline 
Power Spectrum&
&
0.4\\
\hline
Mass to Light Ratio of Clusters&
&
0.2\( \pm  \)0.04\\
\hline
\end{tabular}
\caption{\label{tab:matterdensity}Contribution of matter to \protect\( \Omega _{0}\protect \)}
\end{table}

\section{\label{sec:struniv}A Strange Universe ?}

\begin{figure}[p]
\begin{center}
\includegraphics{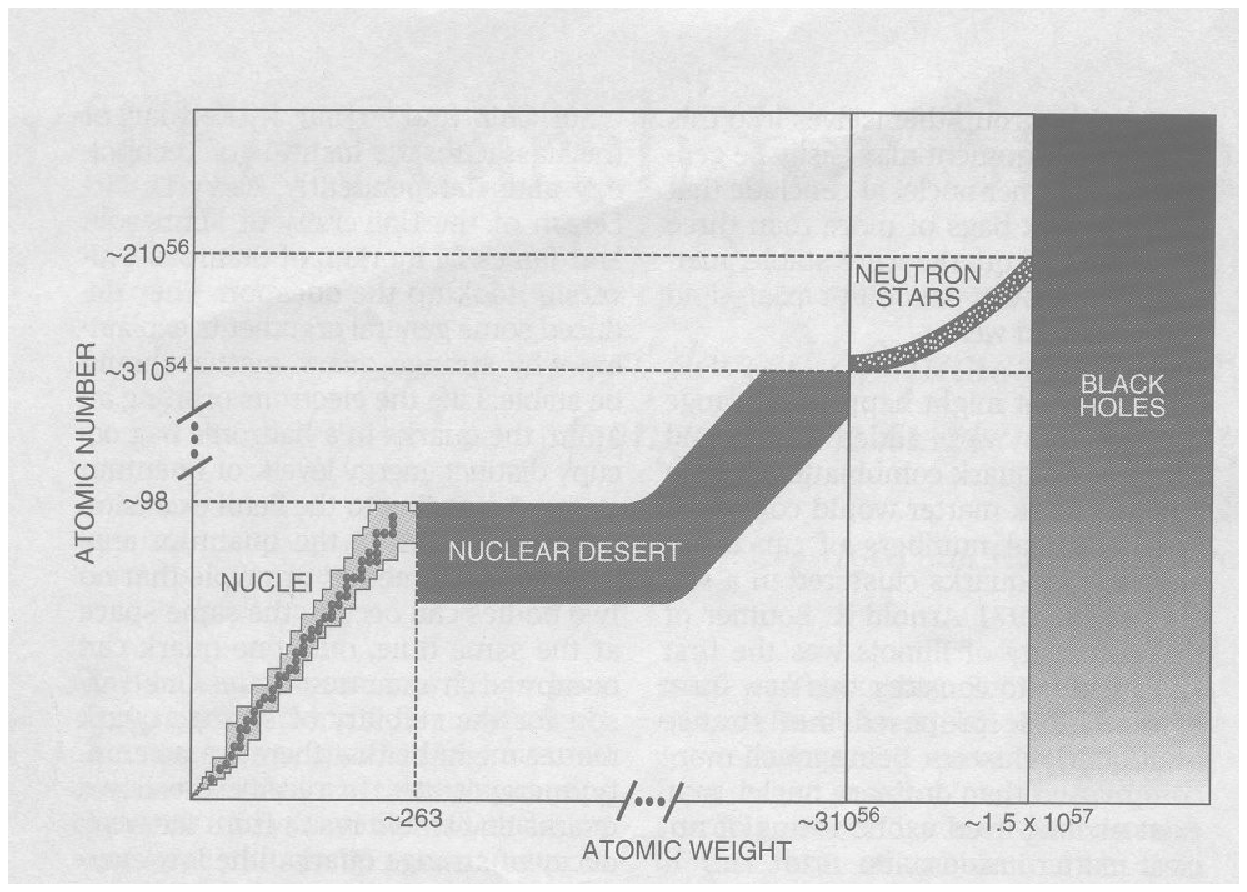}
\end{center}
\caption{\label{fig:sciam} In between the heaviest elements and neutron stars, 
there is a large range of atomic weights which does not contain any known forms of matter ; 
this stretch can easily be filled with SQM \cite{sciam}.}
\end{figure}

\begin{figure}[p]
\begin{center}
\scalebox{0.5}{\includegraphics{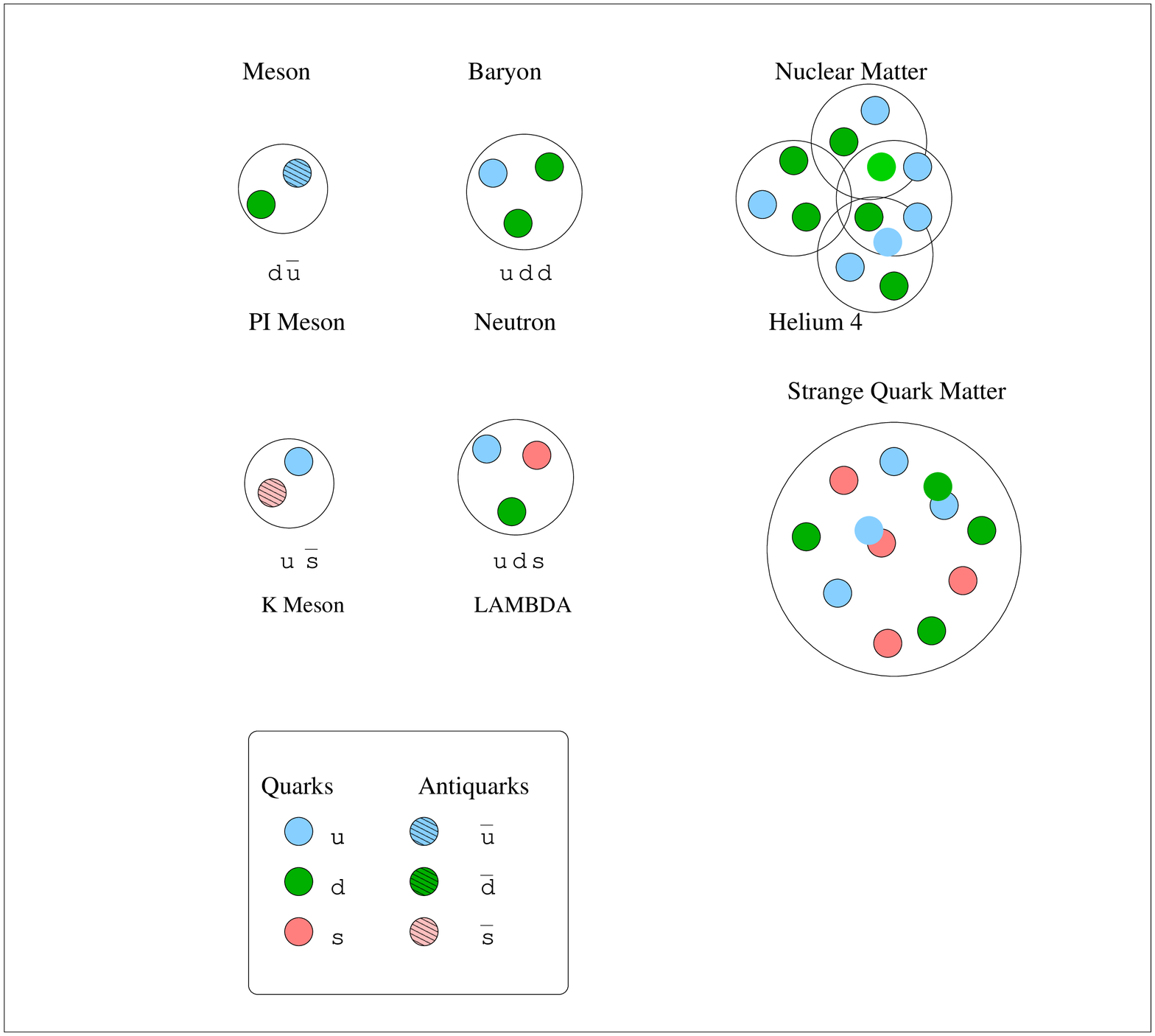}}
\end{center}
\caption{\label{fig:sqm} Various combinations of quarks in hadrons. The lightest two, up and down
are needed to make up ordinary matter. The strange quark has so far been found only in unstable particles.
In ordinary nuclear matter, the individual hadrons retain their identity. According to Witten's \cite{wittenci} 
conjecture, stable multiquark bags, in which the individual hadronic bounderies have dissolved, may exist 
as a more stable form of matter. }
\end{figure}

At the present moment, there is not enough experimental evidence 
to suggest that strangeness is a key ingredient for the matter
present in the universe. Indeed, the local \emph{visible} universe seems
to be made entirely out of nuclear matter. The protons and neutrons
(which form the majority of the baryons) readily form either tiny
clumps of matter in the form of atomic nuclei or very large and ultradense
conglomerates of neutron stars. There is a large {}``nuclear desert''
in the middle mass range of Fig.~\ref{fig:sciam}, where no form of
nuclear matter has been detected.
All the more, as outlined above in Sec.~\ref{sec:bricks}, the visible universe
consisting of visible stars and hot intercluster gas cannot even account
for the full baryonic mass budget, which itself lends a very nominal
quantity to the total matter density of the universe. The Standard
Model of particle interactions is consistent with the existence of
new forms of matter in which stable entities can form from the combination
of the third quark flavor (strange quark) with the two flavors (up and down) of
quark matter found in ordinary matter. There are however, important
structural differences between the two forms of matter. The difference
lies in the content of the hadronic bags which hold the quarks free
within the bag like enclosures but do not allow them to escape (Fig.\ref{fig:sqm}). 

Experience suggests that it is unlikely to find stable bags of more
than three up and down quarks. The deuteron, for example, exists in a stable
configuration, comprising two distinct quark bags representing the
proton and the neutron. If a quark bag capable of holding all these
six quarks had a lower energy than deuteron, then the deuteron's quarks
would have spontaneously regrouped themselves in this state and matter
as we know it would not exist. In 1971, A.R Bodmer \cite{bodmer} investigated the
possibility of what might happen if strange quarks were added to the
quark bag of up and down varieties and concluded that such forms of
matter might exist as long-lived exotic forms of matter within compact
stars, where they would be compressed much more than ordinary nuclei.
S.A Chin and A.K. Kerman \cite{chinker} and independently L.D McLerran 
and J.D. Bjorken \cite{larry} put forward
some general arguments why strange matter should be stable, regardless
of the possibility of the stabilization under pressure. Their the main argument
was that there would be no empty states to receive the down quarks
that would result from the weak decay of the strange quark : this
is the same principle that explains why the neutron is stable inside
nuclei but decays into a proton in about 11 minutes outside the nuclei.
In nuclei, the long range electrostatic repulsion between the protons
will ultimately break it up, if the size grows beyond the attractive range
of the short range internuclear forces. In contrast the different quark
flavors in a hadronic bag shares the energy equally,
as a result of which the up, down and strange quarks come in almost
equal numbers resulting in a (near) cancellation of charge. 

The multi-quark hadronic bags are thus not subject to the size restrictions
which are imposed on a ordinary nucleus and can easily fill in the
range of sizes between the nucleus and the neutron star. The possibility
of a strange universe, therefore, cannot be ruled out. As outlined
below, the strangeness can occur at various scales, from forming heavier
than usual isotopes of common elements, to larger strange `nuggets,'
compact stars composed largely (or completely) of strange matter,
to entire 'dark galaxies'. 

The primary motivation for this work is therefore the exploration
of the possible astrophysical consequences of the occurrence of this
other kind of matter and its (close) encounter with the regular nuclear
matter objects, on this planet, as well as elsewhere in the Universe.

\subsection{\label{ssec:sqm}Strange Quark Matter or quasibaryons }

Our usual experience with ordinary matter, suggests that the various
forms of nuclear matter (nucleons, hyperons and baryons) show great
stability. Such \emph{pure} baryonic matter in the universe appears
to be composed entirely of up (u) and down (d) quarks and held in
stable configuration by strong interactions (nuclear forces). The
most stable chemical element with the least energy per baryon 
(\( ^{56}\mathrm{Fe} \))
is therefore the natural choice for the ground state of such forms
of matter. In the absence of any other form of matter in which strong
interactions play the key role, this ground state can also be thought
to be synonymous with the ground state of QCD (Quantum Chromo Dynamics),
since QCD is the framework for the theory of strong interactions. 
In a seminal paper in 1984, however, Edward Witten \cite{wittenci} put forward
a conjecture that a system of 3A up, down, and strange quarks with
the number of u, d and s quarks roughly same, can have a lower energy
per baryon compared to normal nuclear matter objects with mass number
\( A \). This form of \emph{quasibayonic} matter is known in the
literature as Strange Quark matter (SQM). The term quasibaryonic was
coined by ourselves \cite{panic} to imply that the quarks in this case 
would not form individual baryons, but would have wave functions ranging over 
the entire size of the system. It would still be possible to associate mass numbers
with these objects (Color must still be confined, so it is still possible
to talk about baryon number when discussing such a system) in the sense that
a SQM blob of mass number \( A \) is actually a system of \( 3A \) quarks
kept in a color neutral configuration. This is
also a system in which strong interactions play the dominant role
and would therefore represent a new ground state of matter. 

The above statement about the existence of an hitherto unknown form
of matter is known in the literature as the STRANGE MATTER hypothesis.
This assumption lies at the very foundation of the current work.
The hypothesis illustrates that SQM, rather than nuclear matter, can
very well represent the \emph{true} ground state of strongly interacting
matter.

\subsection{\label{ssec:sqmprop}Properties of quasibaryons or SQM}

In ordinary circumstances, normal nuclear matter does not decay to
this true QCD ground state because this would necessary require very
high order weak processes in order to pass down to the novel state,
as strange quarks must be generated in abundance from the u and d
quarks; the timescale for the process being larger than the age of 
the universe \cite{wittenci}. On the other hand, SQM will readily absorb 
neutrons, since their stability is enhanced as they acquire larger 
mass numbers. They can therefore be distinguished from other forms of 
matter in their exceptional stability
and an insatiable appetite for neutrons. It has been found efficient
to divide the spectrum of strange matter into 3 categories arising
mainly from size considerations \cite{mad}.

\begin{figure}[p]
\begin{center}
\includegraphics{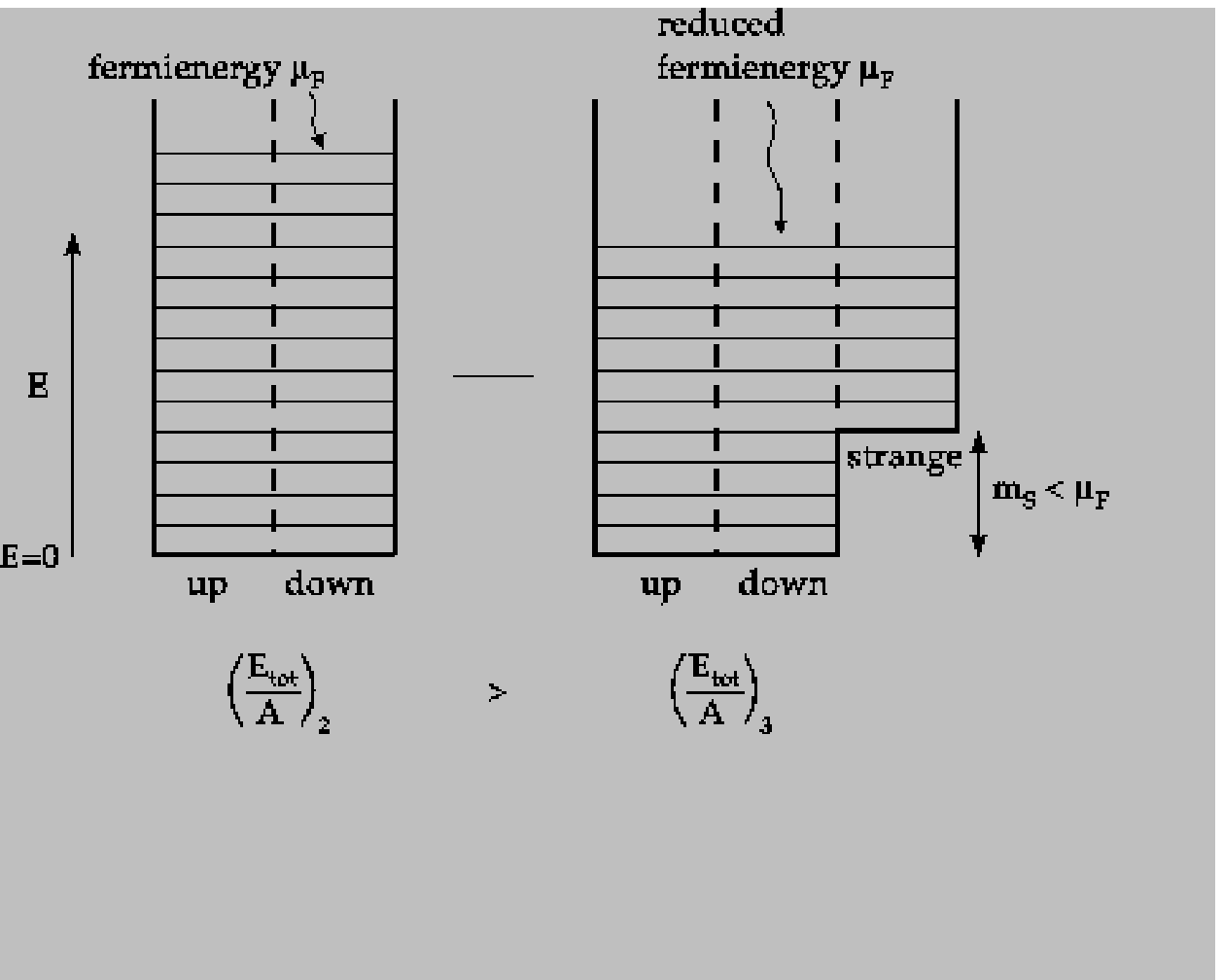}
\end{center}
\caption{\label{fig:fermiwell} The presence of an extra Fermi well reduces the energy of a three flavor
system relative to a two flavour system.}

\end{figure}

\begin{enumerate}
\item \textbf{Bulk Strange Matter}: Bulk strange matter is sufficiently
large (\( \mathrm{A}\sim 10^{44} \) or higher) and surface effects
are usually small so that they may be disregarded in the first approximation.
This is a system of free Fermi gas of 3A u, d and s quarks held in
a quark bag which separates the collection from the vacuum by a phase
boundary. The system has been found to be stable at zero temperature
and pressure. The presence of the strange matter is an essential ingredient,
as a system of u and d quark matter is known to be energetically unstable
relative to the nuclear matter. It is only the presence of an extra
third Fermi well (Fig.\ref{fig:fermiwell})that can actually reduce the energy of a three flavor
system relative to a two flavor system. Ideally, the system is large
enough so that one may not consider the surface effects at all. Bulk
strange matter is also electrically neutral as the number of the quark
flavors are identical and the charge cancellations are near perfect (\( \frac{2}{3}n_{u}-\frac{1}{3}n_{d}-\frac{1}{3}n_{s} \)
vanishes when \( n_{u}=n_{d}=n_{s} \)). The charge cancellations in the above sense 
are actually perfect in the case when masses of the three quark varieties are equal, even when they are not,
electrons ensure local charge neutrality (since the strange quark is heavier than the up and down varieties, there
will be a net positive charge in the absence of electrons). Weak interactions maintain equilibrium in the 
system through flavor conversions like : 
\begin{align*} 
d & \lra u + e^{-} + \bar{\nu_e} \\
s & \lra u + e^{-} + \bar{\nu_e} \\
u + s & \lra d + u
\end{align*} 
The neutrinos generated in such reactions leave the system and are not ascribed any chemical potential.
 The model of bulk strange matter is useful because it can describe the naturally occurring quasibaryonic
systems like strange stars and provides a limit which must hold for
any model of SQM when the dimensions of the system tends to infinity.
The model contains three free parameters, the Bag Constant \( B \)
representing an external pressure that keeps the system bound, the
mass of the strange quark \( m_{S} \) and the strong coupling \( \alpha _{S} \).
The Bag constant \( B \) is essentially identical to the \( B \)
used in the MIT Bag model and is a parametrization of the long range
QCD confinement force, being the difference between the perturbative
and the nonperturbative vacua. The overall conclusion
of such strange matter models is that bulk strange matter is stable
over a certain region in the three dimensional parameter space \( (B,m_{S},,\alpha _{S}) \)
. Strange matter can actually exist in a stable configuration if the
values allowed by the model are consistent with the real world. The
real-world values of the parameters are obtained from bag model fits
to light hadron spectra. The renormalization point of the bag models
is unknown, so neither \( \alpha _{S} \) nor \( m_{S} \) can be
meaningfully compared. While it is not possible to compare the
'windows of stability` for strange matter to known values of the parameters \( \alpha_S, m_S \) and \( B \), 
the windows are quite large. It is therefore extremely likely that
bulk strange quark matter is bound and stable.

\item \textbf{Medium sized strange nuggets:} These nuggets have \( \mathrm{A}\leq 10^{7} \)
and the models used to describe such objects are more detailed and
take finite size effects into accounts. Strange
quark matter in the medium range is still large enough to be treated
as a Fermi gas, but small enough that effects relating to its finite
size must be considered. The radius of such a strangelet is of the
order of few 100's of fm, which is less than the Compton wavelength
of an electron. As a consequence, unlike bulk strange matter, electrons
will not be found within strangelets and the strangelet therefore acquires a net positive charge.
The electrons will now be found `orbiting' the strangelet as in an atom. As a result, coulomb forces within the
strangelet may no longer be neglected.

\begin{figure}[p]
\begin{center}
\includegraphics{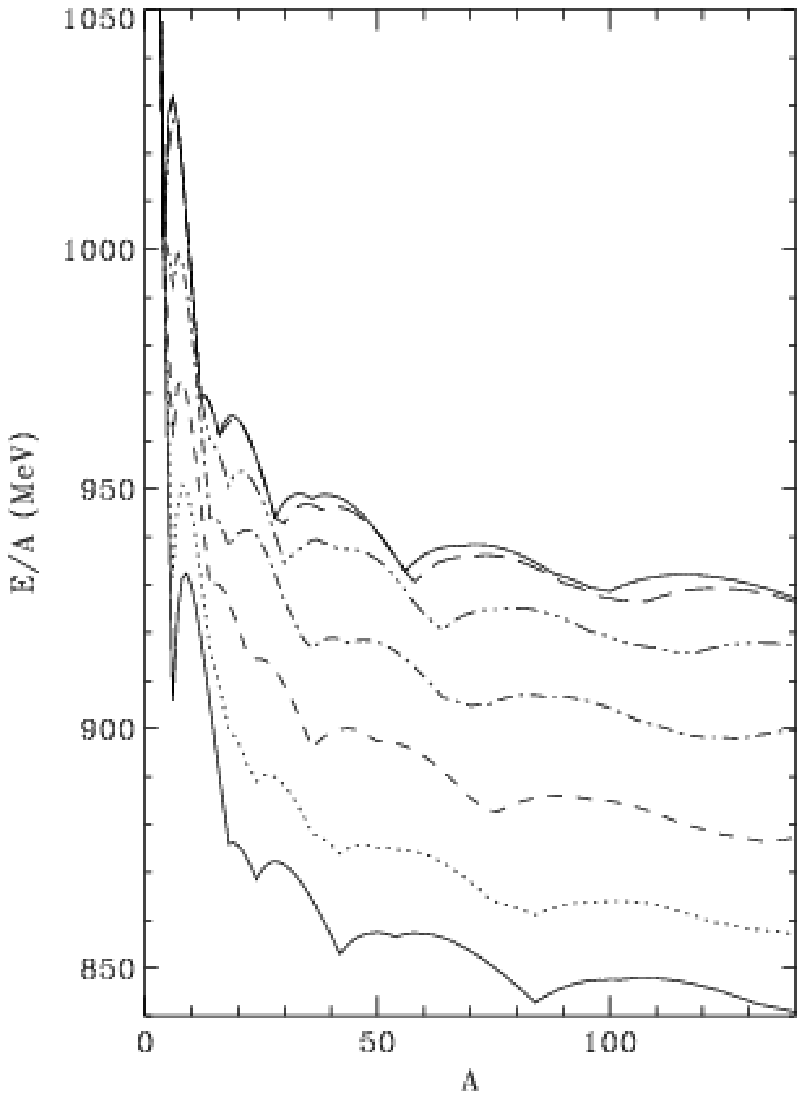}
\end{center}
\caption{\label{fig:madshell} The Energy per baryon Vs. the atomic number $A$ for
strangelets in a shell model calculation with $B = 145^4$ MeV. The different curves represent 
the values of the strange quark mass (0 - 300 MeV). The peaks represent \textit{magic numbers} or the
values of $A$ for which the stability is enhanced with respect to neighbouring atomic numbers \cite{mad}.}

\end{figure}

\item \textbf{Very small strangelets:} These resemble the isotopes of super-heavy
elements in their mass. The study of such small strangelets mainly 
arose from the desire to explore a form of strange matter that can
arise out from the collisions of the heaviest nuclei at the highest
attainable energies. The models of bulk strange matter with surface
effects are not suitable for their description, as one must take into
account the possibility of formation of shell structure in their energy
levels. The model begins by filling the energy levels in a bag, one
quark at a time, minimizing the energy each time with respect to flavor.
The bag radius is adjusted to balance a constant external pressure
\( B \). As the quarks are non-interacting, it
is possible to ignore the perturbative QCD corrections. Ignoring coulomb
interactions is also acceptable, since Z will typically be very small
for these small A systems. This approach yields a relationship between
the energy per baryon (number) vs the baryon number, and reveals
a shell structure analogous to nuclear structure. At first the quark
bag begins to get filled by massless non-strange quarks. Soon the Fermi
energy of the system becomes large enough compared to the strange quark mass
so that it becomes energetically favorable to add massive strange quarks.  
At some future point the system will again become favorable for the acceptance 
of the non-strange quarks and the differential variation of the strange and non-strange
levels with the radius of the bag produces level crossings at some
\( A \) where the strange quarks transform into strange quarks.
There are also some values of \( A \) where the \( \varepsilon  \)
drops rapidly : at these points the decrease in energy from emitting
a baryon is insufficient to offset the energy needed to climb out
of the dip. The increased stability is a signature of the shell closure
at these values of \( A \). It is, in fact, possible to identify a sequence of 
magic numbers (See Fig.\ref{fig:madshell}) at \( A = 6,18,24,42,54,60,84,102 \) 
etc for low \( s \) quark masses, something strongly reminiscent of nuclear physics. For larger 
\( s \) quark masses, it becomes more favorable to use \( u \) and \( d \) quarks 
instead of strange quarks, and these magic numbers change. The energy per baryon approaches the 
bulk limit when \( A \rightarrow \infty \).

The vulnerability of such
a small strangelet depends on their mass (related to their mode of
production) and type of interaction in which they participate : to
pass from one stable configuration to another, the time scales required
are of the order of the weak interaction time scale and, in general
larger \( A \) implies greater stability (provided these objects
are not big enough to have become medium sized nuggets). Thus, for
example, stability of strangelets is of much concern in Heavy Ion experiments
(being governed by strong interaction time scale), while it is almost
guaranteed for collisions of strangelets in the terrestrial atmosphere
(The time delay between successive collisions is large enough so
that they have enough time to switch over to one of such stable configurations).
\end{enumerate}

\section{\label{sec:sqmiss}SQM and the missing matter}

\begin{figure}[p]
\begin{center}
\scalebox{0.6}{\includegraphics{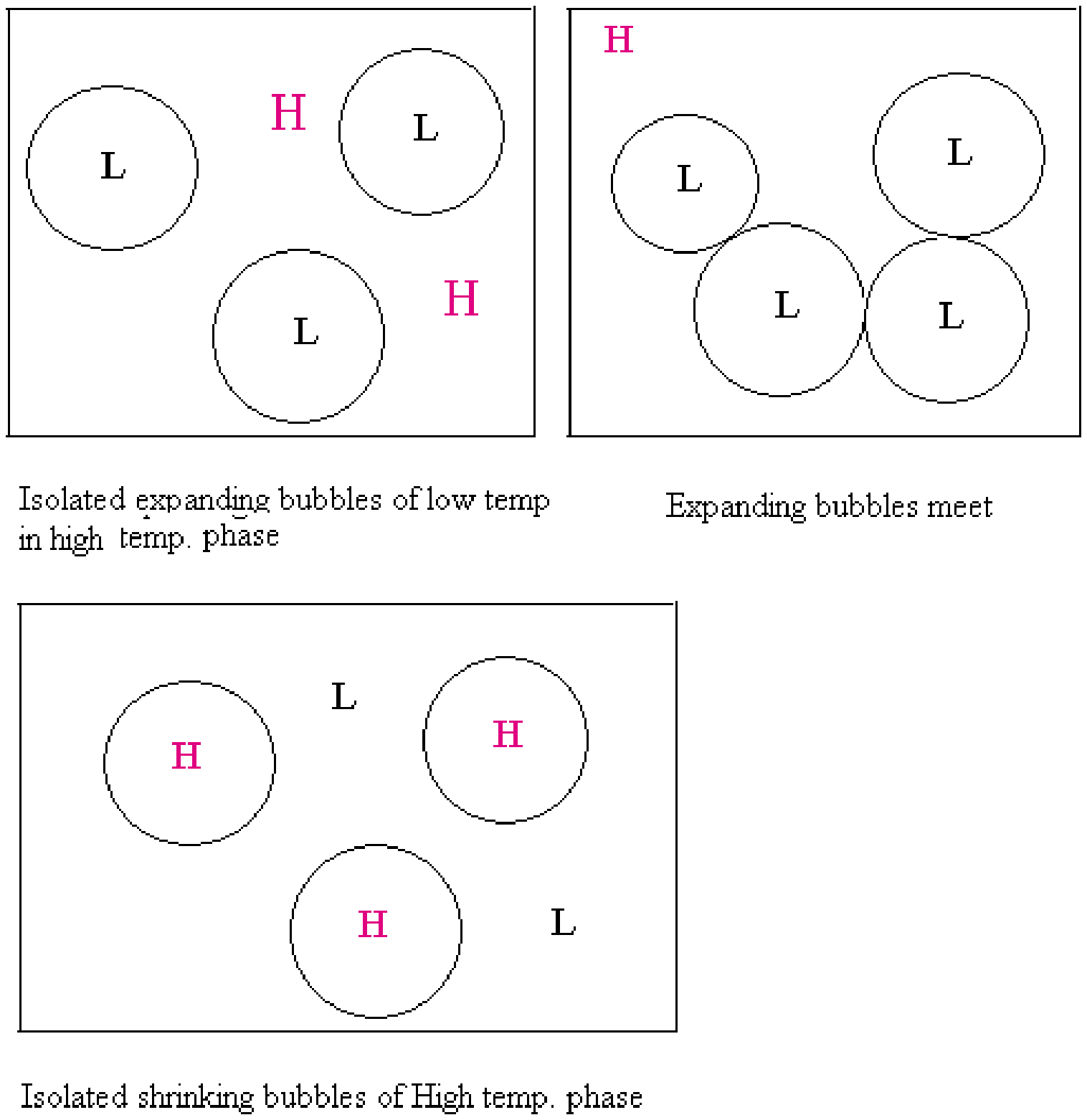}}
\end{center}
\caption{\label{fig:bubbles} The evolution of hadrons through bubble nucleation in the early universe
according to Witten \cite{wittenci}}
\end{figure}

The naturally occurring strangelets (\( \mathrm{A}\sim10 ^{44} \))
could have been formed in the early evolutionary phase of the universe.
In the very same paper \cite{wittenci} where Witten discussed the possible existence of 
stable multi-quark bags stabilized by strangeness, he had raised the possibility that the missing mass
of the universe can be accounted for by SQM. In the following we present a brief synopsis of the 
original work -- it must be mentioned here that although the essence of the agreement remain unaltered,
later works have examined various related aspects in greater length and depth and the numbers quoted here are 
purely of historical significance. We discuss some more recent works in Chap.~\ref{chap:macho}.

Witten's description of the process begins in the very early universe (row 3, table : \ref{tab:universeevents}) when
the universe undergoes a a first order phase transition in the early
Universe from a high temperature state of quasi-free \footnote{in the sense
that they are free to roam inside the bag, without having to be confined in individual
hadronic bags} light quarks to a state of hadronic matter (The cosmic QCD Phase Transition).
The Latent heat released during the first order phase transition holds the temperature to the 
steady value of \( T_\mathrm{c} \) allowing low temperature bubbles of hadronic matter (Fig.\ref{fig:bubbles}) to form.
As the Universe expands, these bubbles grow as they absorb energy from their surroundings. At some point
the lower energy `bubbles' will percolate, and soon after, it is the
high energy regions that form bubbles. As the Universe continues to
expand, the high energy free quark bubbles continue to give off energy.
This release of energy may take two forms. If it comes from evaporation,
i.e. the release of hadrons into the low temperature region, then
the bubbles will continue to shrink until they disappear. If instead
they lose energy via neutrino emission, the baryon number inside
the bubble will remain constant while energy is released. The bubbles
will continue to shrink in size, increasing the baryon density. Eventually
the excess baryons inside will produce a pressure to resist further
contraction. These lumps can now accommodate between 80\% - 99\% of all
the baryon excess of the universe, but are only about \( 10^{-6} \)
cm - 5 cm in radius and their mass lies somewhere between \( 10^{9}-10^{18} \)
gms \footnote{Recent theories suggest larger sizes but hardly alters any of the following conclusions.}. 
In order for these lumps \footnote{Henceforth referred to as SQNs or
Strange Quark Nuggets.}
to survive they could not be composed
of normal matter (quark matter is unstable without strange quarks),
but instead would be composed of strange quark matter. But since these
nuggets are so small, they would scatter very little light and would
be impossible to observe directly. They would be exceptionally bright
candidates for dark matter, being non participants in nucleosynthesis.
After the pioneering work by Witten, E. Farhi and R.L Jaffe \cite{farhij} have
shown that the chunks of the SQM could be stable for a much larger
range of sizes than predicted by Witten. R Riisagar and J. Madsen
\cite{riismad}
found that the primordial quark nuggets had to be made of more than
\( 10^{23} \) quarks if their existence were to be consistent with
both the calculated amount of missing dark matter and the observed
abundance of light isotopes.

\subsection{\label{ssec:sqsec}Strange Quark Nuggets from secondary sources}

The collisions of primordial nuggets and the mergers resulting from
such events might also lead to strangelets of larger size. It is also
possible that deep inside the neutron star, extreme densities and
pressure might have triggered the conversion of nuclear matter to
strange quark matter - it works in the following way. As mentioned
above, non strange quark matter is unstable at zero pressure. The
pressure inside a neutron star can make it possible to make the energies
lower by few 10's of MeV per nucleon at which point stable quark matter
begins to form. However as soon
as a small core of stable quark matter is formed, it can become more
stable by converting some of the light quarks to strange quarks and
grow by absorbing the surrounding nucleons, facing no Coulomb barriers
in their way. In such cases the quark cores should be able to convert
the whole star into a (strange) quark star, providing a contemporary
source of strange matter. Even a droplet of SQM falling into a neutron
star can initiate this conversion and can convert it into a quark star.
The resulting star would be much more compact since it will be bound
by intrinsic quark forces. It is imperative, in this situation, to
figure out what the resulting size of the strange star is going to
be. Quantitative estimates to this query have been obtained using
the solution to the TOV equations of hydrodynamics with an assumed
equation of state. We, on the other hand, tried to develop an analytic
procedure based on the quark mass density dependent model (QMDD).
The QMDD model was first proposed by Fowler, Raha and Weiner \cite{qmdd} to provide
a dynamical description of confinement. In this model the quality
of confinement is mimicked through the requirement that the mass of
the quark becomes infinitely large as the volume increases to infinity,
holding the energy density constant and is summarized in the effect
\[
\begin{array}{ccc}
m_{q} & = & \frac{B}{3n_{B}}\\
m_{s} & = & m_{s_{0}}+\frac{B}{3n_{B}}
\end{array}\]
where \( m_{q} \) is the (density-dependent) mass of the u and d
quarks, \( m_{s} \) is the (density-dependent) mass of the strange
quark, \( m_{s_{0}} \) is the (nonzero) free mass of the strange
quark, \( B \) is the Bag constant or the vacuum energy density within
the bag and \( n_{B} \) is the baryon number density. With this model,
the standard technique of balancing the gravitational attraction to
the quark degeneracy pressure remains applicable even in the case
of quark stars composed of massless u and d quarks, as these quarks
are able to get their masses from the density effects. We discuss
the application of these ideas in chapter \ref{chap:macho}

\subsection{\label{ssec:sqexpt}Experimental searches for strange matter.}

The search for quasibaryons, the living fossils of the early universe,
is an important activity in the field
of QCD and astrophysics. It necessarily has strong theoretical and
experimental perspectives and the search for them is currently an
active experimental pursuit that spans over diverse zones like the
Milky Way Halo, The Earth's crust, Geophysical specimens (meteorites),
High altitude stations, Weather balloons and accelerators for heavy
ion collisions. It is unfortunately a nontrivial task to distinguish
between SQM and normal hadronic dust, since SQM constitute a new form
of matter and not a specific type of particle having a definite mass
; the sole criterion for distinction in this case being the extremely
small charge to mass ratio of the strangelets in comparison to nuclear
particles.

To find evidence for strange quark matter there
are mainly four places of choice.

\begin{enumerate}
\item The Earth's crust and rock samples: The idea that the SQM would have
left their trace in the crust material was first advocated by A.D
Rujula and S.L Glashow \cite{rujula} . According to them there can be three
possible consequences, depending on the size of the strangelets
\footnote{See also Sect.~\ref{sec:exptdust}}:

\begin{enumerate}
\item \( 10^7<3A<10^{14} \) : These particles (nuclearites) would be slowed down and stopped
by earth and could reveal themselves as 

\begin{enumerate}
\item Unusual meteoritic events caused by nuclearites traveling with tremendous
velocities compared to usual meteorites and penetrating large depths of 
the atmosphere without burning off, making their way to the ground. 

\item Earthquakes with special signatures characteristic of a point source and
almost total absence of surface waves (since most of the energy loss is supposed
to take place in the mantle rather than the crust, since the nuclearite traverses
the Earth in less than a minute).
\item peculiar tracks in ancient mica
\end{enumerate}
\item \( 3A>10^{23} \) : They would pass through the earth leaving no traces.
\item \( 3A<10^{7} \) : They might remain embedded in meteoritic or crustal
material. 
\end{enumerate}
\item Heavy Ion Collisions : We have already discussed the prospect of production
of strange quark blobs in experiments \cite{frank} involving massive ion beams. 
Such experiments however, to date have no success to report. It seems 
that the main difficulty associated with the production of strangelets   
is that the strangelets produced in  Heavy Ion Collisions 
will have very little time (governed by the time scale of strong interactions) to 
settle down on a stable configuration through their interaction. \footnote{In contrast, the strange nuggets which appear
in the early history of the universe gets enough opportunity in the cosmological scale,to stabilize through weak interactions.}
Their sizes would necessarily be very small (depending on the share of the incident energy density it will receive).
However If  they are detected, they can be resolved and discerned from other particles  through a mass spectrometer 
on the basis of their extremely small charge-to-mass ratio. 
\item The Atmosphere : Strangelets, generated from collisions
between strange `neutron' stars can also be set off in motion to
be (possibly) revealed as exotic cosmic ray events with extremely
small charge to mass ratios (\( Z/A<<1 \)). Numerical
simulations of head-on collisions of neutrons stars suggest that as
much as 13\% of the total mass of the system might be ejected \cite{nstardum}. These
potentially relativistic strangelets would eventually be impingent
upon the Earth's atmosphere. Even if such collisions are rare, binary
systems may eject some mass during mass transfer; they can also result
from the decay of binary systems. These strangelets would have to
overcome the effect of the magnetic field imposed by the earth in
order to come down to the level of mountain altitudes, where they
can be detected using a ground based large array of passive solid
state detectors. It is also possible to find the evidence for strangelets
in balloon borne experiments using active / passive detectors, but
since the incidence rate of these objects is very small, the probability
of detection must be extremely small. Nevertheless the balloon experiments
are the first of these kind of experiments which showed the signature
of exotic particles which match the charge to mass signature of the
strangelets.
\item The Outskirts of our galaxy: Gravitational microlensing technique
has arrived as a powerful tool for exploring the structure of our
galaxy. The idea of microlensing rests upon the fact that the light
from a distant star is lensed and form a ring around a massive object
that is near and lies along along the line of sight of the observer (Fig.~\ref{fig:lens}).

\begin{figure}[p]
\begin{center}
\scalebox{0.5}{\includegraphics{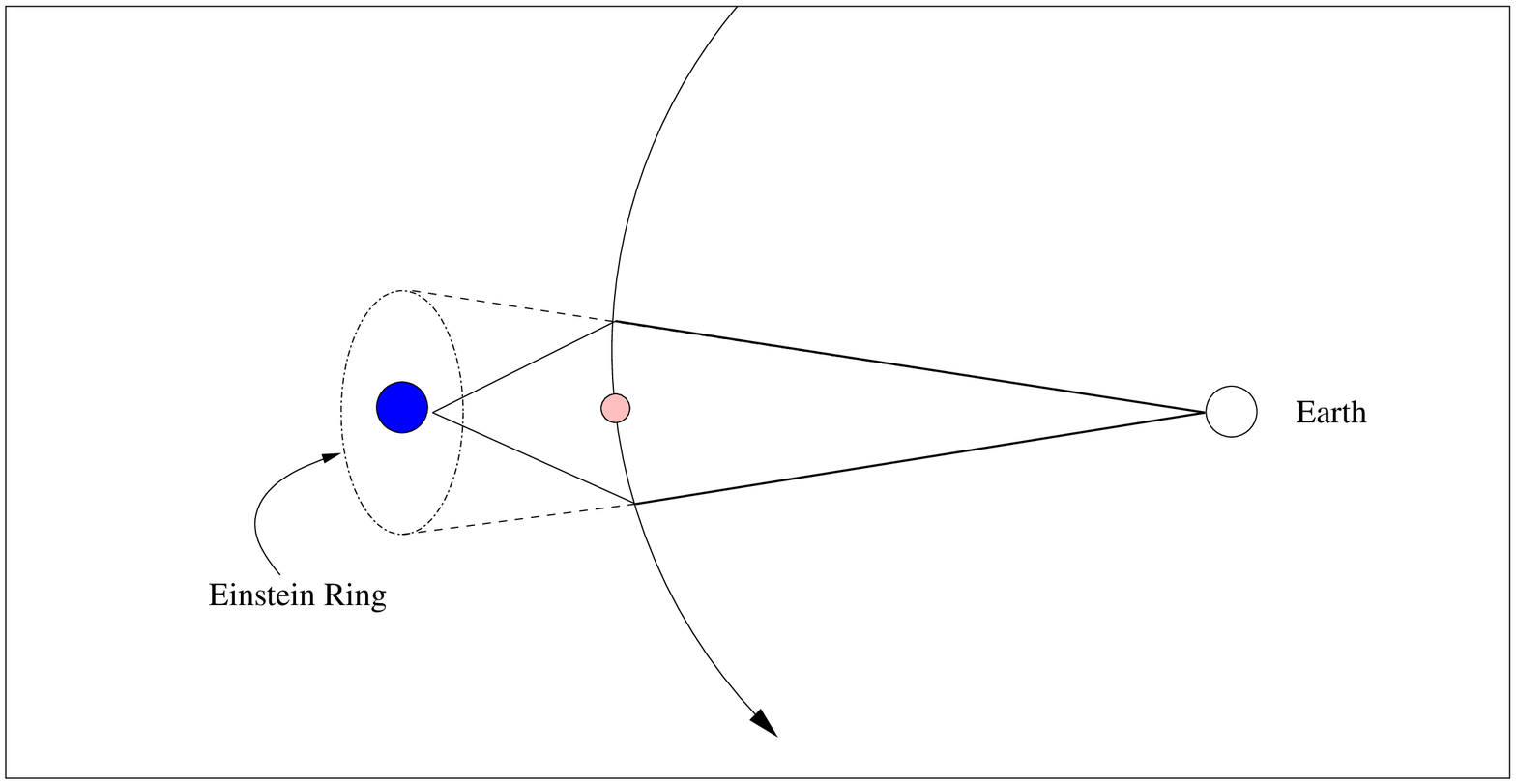}}
\end{center}
\caption{\label{fig:lens} Since Light rays are bent when they pass close to a massive object, light from a distant source may be focussed by a closer object to producing a sudden brightening. If the smaller object's path takes it precisely in front of the other one, the image formed by the "gravitational lens" is a circular ring, referred to as an "Einstein Ring.
}
\end{figure}

In the more likely case in which the massive object is slightly displaced
from the line of sight, it will form a double image of the distant
star, separated by a small angle. In the usual case this angular shift is too small to 
be resolved since the lens masses are in the stellar mass range and distances to the lenses are the 
order of galactic lengths. In this case, however, the image of the
star will appear to brighten up, since light from both images will
pour into it. These observations reveal the size and the mass (Experimentally found
to be \( \sim  M_\odot \)) of the
lenses, which can then be used to judge between possible physical candidates for the lens objects.
The experiments have the potential to reveal objects which are not
available for luminous interception and can reveal the nature of dark
matter present in our galaxy.
\end{enumerate}
\clearpage
\section{\label{sec:thesis}The outline of this thesis}
In this work the content is arranged in the following manner. In Chap.~\ref{chap:dust} we propose 
a basic model of propagation of small lumps of strange matter through the terrestrial atmosphere. The
model relies strongly on the peculiar properties of SQM. On the basis of this model we discuss
the origin of a class of \textit{exotic} cosmic ray events (characterized by very low charge to mass ratios) 
and show that it is possible to relate these events to the passage of small strangelets through 
the terrestrial atmosphere. The elementary model introduced in this chapter is extended in several ways 
in Chap.~\ref{chap:newmodel} in order to deal with the problem in a more satisfactory way and leads to 
results which appear to be in close agreement to the few experimental data available in this field. In 
Chap.~\ref{chap:chandra} we examine the (possible) existence of a maximum mass limit for quark stars from
an analytical standpoint on the basis of the density dependent quark mass model. Finally, in 
Chap.~\ref{chap:macho} we try to unfold a connection between the existence of massive gravitational lenses  
in the halo of our galaxy to the local density of dark matter by asserting that the lenses are made of 
coalesced quasibaryonic matter.   

\chapter{\label{chap:dust}Strange dust in cosmic rays}

In this chapter we will first consider a few cases in which exotic events were
registered in cosmic ray experiments (\ref{sec:exptdust}). It is always
difficult to accommodate any of these events within a conventional framework 
of propagation of cosmic nuclei through the terrestrial atmosphere : we 
discuss these problems in Sec.~\ref{sec:problemdust} and finally, in 
Sec.~\ref{sec:strletdust} examine a new model of strangelet propagation, 
proposed recently by us \cite{jpg@1}. 

\section{\label{sec:exptdust} Existing reports of exotic fragments}
There have been several reports of exotic nuclear fragments, with highly
unusual charge to mass ratio, in cosmic ray experiments. In the following,
several such events are listed, from the least to the most recent.
In 1978, an event (Price's Event, \cite{ex3}) was identified with 
Z \(\sim\) 46 and A \(\sim\) 1000 and was looked upon as a possible 
candidate for a magnetic monopole at that time. Around this time, the 
Centauro cosmic ray events \cite{centauro} detected in emulsion 
exposures taken at Mt. Chacaltaya by the Brazil-Japan collaboration, 
raised great interest. These events were observed at atmospheric 
depths \(\sim\) 500 gm / cm$^2$ and accompanied by hundreds of baryons 
and almost no \( \pi^0 \) or \( \gamma \).

\begin{figure}[p]
\begin{center}
\includegraphics{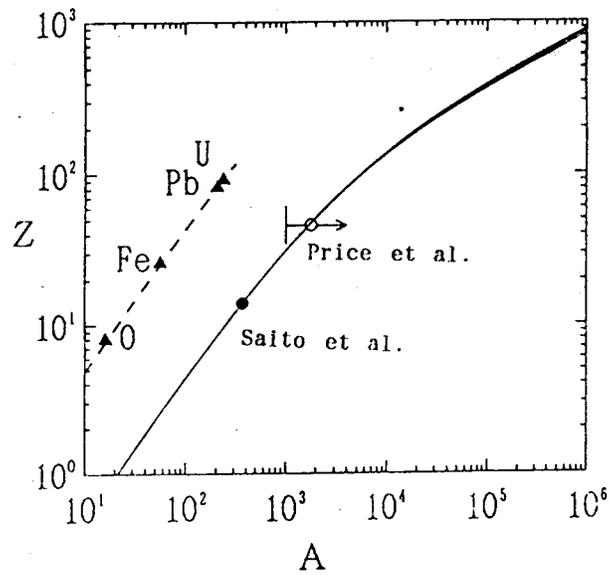}
\end{center}
\caption{\label{fig:price} Z vs A relationship for SQM as given in
\cite{kasu} . The Z-A relationship for normal nuclei is shown
for comparison.}
\end{figure}

In 1990 Saito et. al \cite{saitofukuda} analyzed the data of an 1981 
balloon borne experiment which carried Cerenkov and Scintillation 
counters and claimed to have identified two events which were consistent with 
A \( \sim \) 370 and Z \( \sim \) 14 and could not be explained 
within conventional premises. In Fig.~\ref{fig:price} these two events
are shown in a Z vs A plot. In order to accentuate the highly \emph{unusual}
characteristics of the deviants, the two events are shown alongwith for 
few normal nuclei like Fe, Pb and U.

In 1984, De R\'ujula and Glashow \cite{DG} considered the possibility of
detecting large lumps of SQM, called "nuclearites", of $ A < $ 10$^{15}$
and $Z$ "well beyond any published periodic table". Their main conclusions
have been presented briefly in the introduction (Sec.~\ref{ssec:sqexpt}). 
Among the indications of these events, they considered the possibility of 
observing visible light produced through ionization of the atmosphere as 
well as epilinear seismographic events, along with the possibility of   
finding visible signatures of their tracks on ancient mica. It is 
interesting to note that Anderson \textit{et al} \cite{seismic}
analysed over 1 million seismic data recorded during the period 1990-1993,
in search for an passage of a nuclearite through the mantle of the Earth 
and have tentatively identified one of them as a possible candidate event.

In 1993 Ichimura \textit{et al} \cite{ex2} reported an event called the 
`exotic track' event with Z \(\sim\) 20 and A \( \sim \) 460. The 
report was based on an analysis of a 1989 balloon borne experiment using
solid state nuclear track detectors (CR39).

There have been several other reports of events with $A \sim $ 350 - 500 and 
$Z \sim $ 10 - 20 in cosmic ray experiments \cite{ex3,ex1,ex2,ex4,ex5}, the so-called
exotic cosmic ray events. In the following table (Tab.~\ref{tab:exotica}) we 
present a synopsis of the events in this range of charge and mass. All these
events carry the signature of small charge to mass ratios (\( Z/A<<1 \)) 
characteristic of SQM (\ref{ssec:sqexpt}).

Although these observations come from different groups, the existence of
such objects cannot yet be taken as confirmed, due to various experimental
uncertainties like  switch between gondolas, ambiguities associated with
the calibration of Cerenkov counter output, detector noises, dead time etc,
in the different experiments. These events, thus, are, at best, candidate 
events, although, in spite of that, it is important to understand what they
could be, if they are eventually confirmed.
\vspace{0.3in}
\begin{table}[h]
{\centering \begin{tabular}{|l|c|c|}
\hline 
Event & Mass& Charge\\
\hline 
\hline 
Counter Experiments (\cite{ex1})&
350-450&
14\\
\hline 
Exotic Track(\cite{ex2})&
460&
20\\
\hline 
Price's Event(\cite{ex3})&
1000&
46\\
\hline 
Balloon Experiments(\cite{ex4}-\cite{ex5})&
370&
14\\
\hline 
\end{tabular}\par}
\caption{\label{tab:exotica}Summary of some exotic cosmic ray events}
\end{table}
\vspace{0,3in}

\section{\label{sec:problemdust} Problems related to propagation}
\begin{figure}[p]
\begin{center}
\begin{minipage}[t]{5in}
\centering
\scalebox{0.5}{\includegraphics{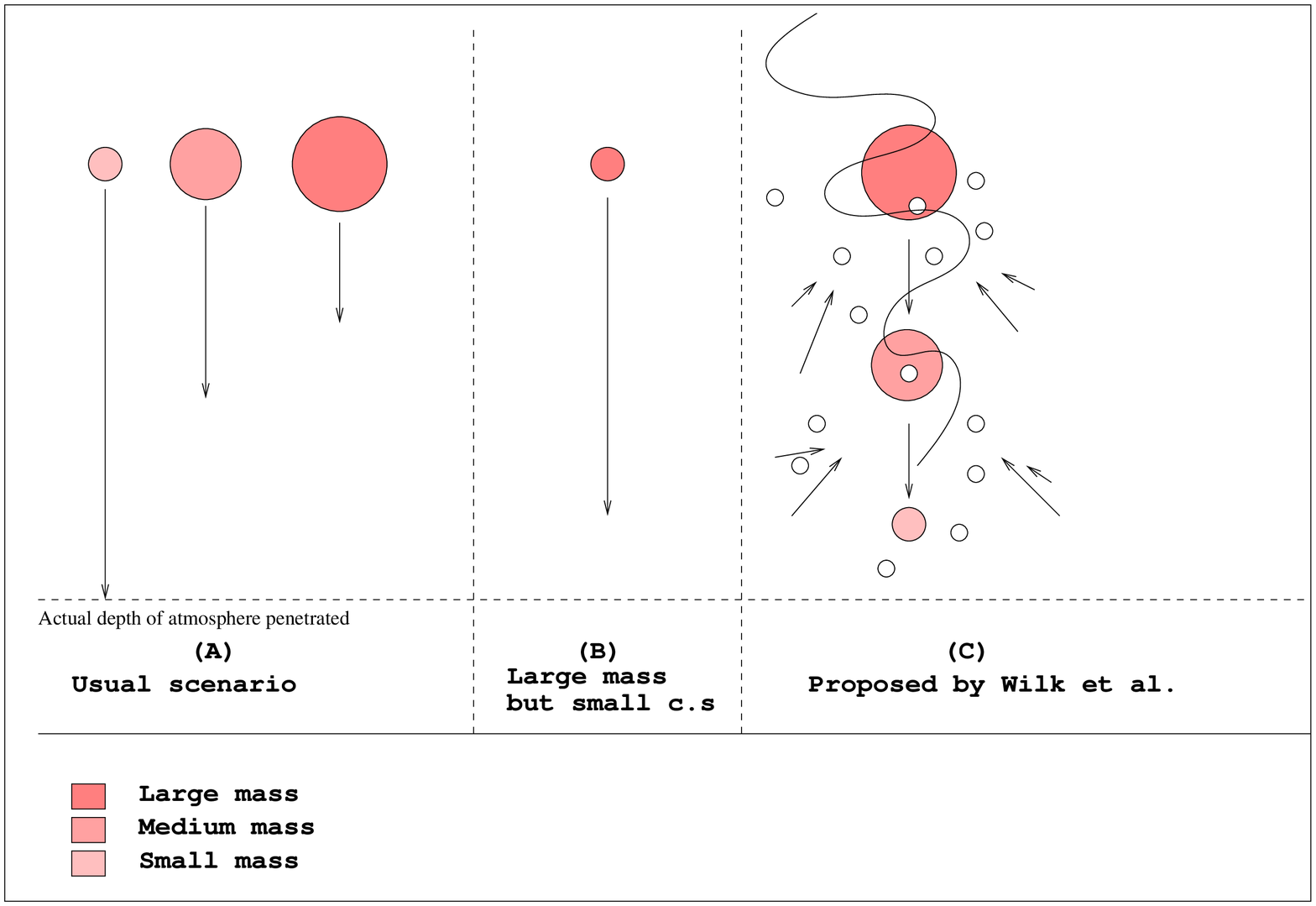}}
\caption{\label{fig:la} Picture to illustrate the difficulty in the 
propagation of strangelets. Panel C shows the solution proposed by Wilk \etal}
\end{minipage}\\[20pt]
\begin{minipage}[b]{5in}
\centering
\scalebox{0.8}{\includegraphics{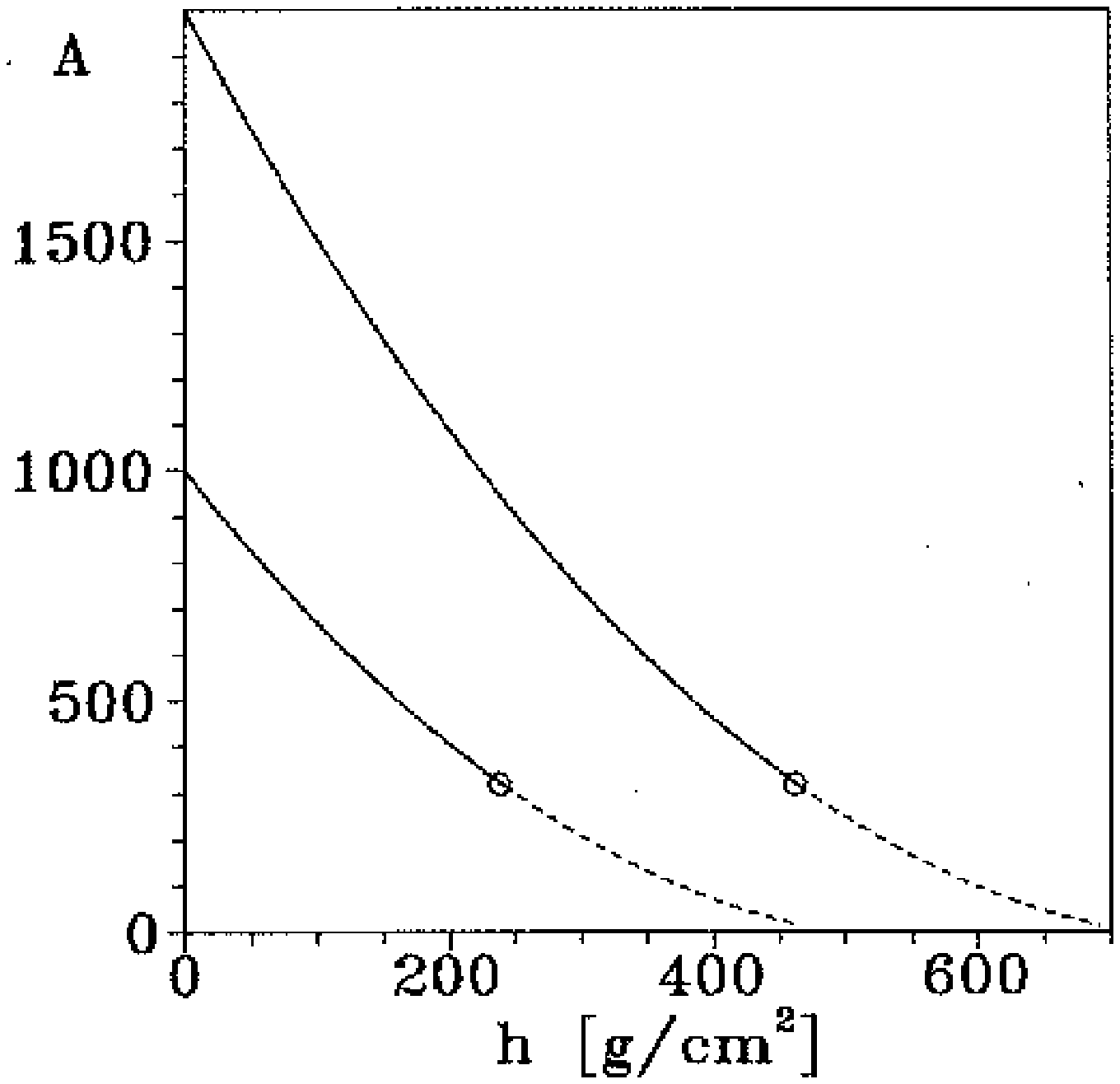}}
\caption{\label{fig:wilk} The variation of the mass of the strangelet 
with atmospheric depth penetrated, as proposed by Wilk \etal\ and illustrated in
Panel C of Fig.~\ref{fig:la}}
\end{minipage}
\end{center}
\end{figure}

In all the events discussed in the above section (\ref{sec:exptdust})
the primary difficulty seems to the extent of penetration of these seemingly 
heavy nuclei in the terrestrial atmosphere, since all the events were observed
near (or slightly higher than) mountain altitudes. The cross sections
of normal nuclides in this mass range would be too large for escaping the 
fate of catastrophic collision with the air nuclei before being intercepted
unhampered into a detector (Fig.~\ref{fig:la},panel A). In the following we 
discuss the evolution of the ideas leading to the interpretation of these 
objects as strangelets. 

In \cite{bj} Bjorken and McLerran ruled out the possibility of a high Z
primary nuclei as the source of the Centauro event mainly on the above ground
(unusually high penetration) and the fact that the mean transverse momentum of the 
secondaries were much higher than the value typical of nuclear fragmentation. In order
for an object to reach comparable altitudes they assumed the object to be glob of nuclear 
matter of an unusual type with density  \( \sim \) 30 - 100 times that of ordinary nuclear 
matter and radius 3 - 5 times smaller than ordinary nuclei. The cross-section of 
the object would thus be much smaller (Fig.~\ref{fig:la}, panel B) compared 
to usual nuclides in this mass range and would face little difficulty in reaching 
the atmospheric depths at which they had been observed. This will be favored all 
the more if the binding energies of the components of this peculiar nuclei were 
larger compared to conventional nuclear matter.

Compressed to such high densities, the constituents of the glob would be in a 
quark matter phase (Sec.\ref{sec:struniv}). In their model of propagation, the 
glob, on its way down the atmosphere, collides with air nuclei and gets 
heated up. It subsequently tries to get rid of the additional energy so 
acquired, either by the radiation of mesons or evaporation of the baryons. 
The fate of the glob depends crucially on its region of metastability : it 
explodes if the region of metastability extends only to a baryon number 
\( N_\mathrm{crit} \). On the other hand, the glob can propagate 
all the way down until it has fully evaporated if it happens to be stable right upto 
\( N_\mathrm{crit} \rightarrow 1 \). In this case if the binding energy per nucleon
decreases with decreasing N (due to the aforesaid radiation / evaporation)
then the nuclei can explode if a central collision with an air nucleus imparts 
sufficient energy to the glob which is getting more and more loosely bound with time. 
However, if the energy per nucleon increases on evaporation, the glob can 
land safely on the terrestrial soil. It is clear that the most essential 
ingredient in the above model is the region of metastability of the globs. 
The quark matter globs were speculated to be (meta)stabilized by the 
presence of a fractionally charged free quark which might help compress the baryons 
to the required high density. This led to an unacceptable flux of quarks at 
the sea level together with an unacceptable rate for horizontal air
showers. A slight variant of this model is capable of reducing the flux of globs 
at the sea level compared with the previous model but remains incompatible with
the rate of horizontal air showers.

The above discussion serves to highlight the difficulties encountered in the
interpretation of the exotic cosmic ray events along conventional lines. Another
issue is that, unlike the above event, most other exotic events of the type 
given in table \ref{tab:exotica} were non-explosive and could not be dealt with properly
within a fireball scenario.
The requirement of metastability was also a 
stringent one, and this has led to various speculations on the composition of 
these quark blobs; e.g Chin and Kerman \cite{chin} proposed the existence of 
metastable multiquark states of large strangeness within the framework of the 
MIT bag model \cite{bag}. The existence of such objects has been 
postulated by other authors \cite{terazawa} too, but the seminal work of Witten
\cite{witten} in 1984 provided the theoretical basis for the study of 
SQM (referred to in the following as strangelets) within the framework of QCD. 
We have already discussed part of this framework in the introduction (\ref{ssec:sqm})
leading to a classification of the SQM in three mass groups (\ref{ssec:sqmprop}). 
In the light of the prior discussion it seems natural to associate some of the events in 
Table.~\ref{tab:exotica} with strangelets since the rather unusual \( e/m << 1\) 
ratios which appear in the table seems to be well correlated 
with the theoretical estimates for \( e/m \) of strangelets \cite{madsen1}.

In spite of this, no consensus has yet emerged primarily because of ambiguities 
related to the mechanism of propagation of these objects through the terrestrial 
atmosphere. For example if a strangelet with baryon number A \(\sim\) 1000 
appears at the top of the atmosphere, there would be a serious problem with 
its penetrability through the atmosphere, as the exotic events are observed 
at quite \emph{low} altitudes. One can assume arbitrarily that their 
geometric cross sections are quite small (as in above). This conclusion seems to 
be rather artificial because there is no compelling theoretical reason to believe 
that the mass-radius relation for SQN's to be much different from normal 
nuclides -- at least not as dramatic as suggested in \cite{ex4}--\cite{ex5}
since the density of SQM is believed to be not much large than ordinary nuclear matter 
\cite{witten}.

As a way out of the situation Wilk et al. \cite{wlk1,wlk2,wlk3} and 
others (e.g \cite{wol}) proposed a mechanism by which the strangelets will be able
to cover great atmospheric depths without having to rely on an atypically
small cross section. They explored the fate of strangelets of 
initial masses of the order of \(10^3\) a.m.u. incident on the upper layers
of the atmosphere\footnote{%
These are medium sized strangelets, according to the classification given 
in \ref{ssec:sqmprop}}. The main assumption was that the mass and hence the 
cross sections of  such strangelets \emph{decrease} rapidly due to their 
collisions with air molecules in their downward journey. In particular, they 
assumed that a mass equal to that of the nucleus of an atmospheric atom 
(A $\sim$ 14.5) is ripped off from the strangelet in every such encounter, as if 
the atmospheric atom drills a bore through the strangelet (See Fig.~\ref{fig:la},
panel C). 
This model is characterized by a critical mass \( m_{\mathrm{crit}}\) such that 
when the mass of the strangelet evolving out of an initially large strangelet
drops below the above critical limit, it simply evaporates into neutrons; 
this happens when the separation energy \( dE/dA \) becomes larger than the 
mass of a baryon. In other words, the condition 
\[ \left.\frac{dE}{dA}\right|_{m_{\textrm{crit}}} > m_{n} \]
would fix the lower limit of the altitude upto which a strangelet would be 
able to penetrate (Fig.~\ref{fig:wilk}).

Let us reiterate the basic conclusions that can be derived from the
earlier works (Fig.\ref{fig:la}). Firstly, strangelets observed at the 
mountain altitudes typically have masses around 300 to 450 and charge between 10 to 20.
But the experimental results obtained till date are inconclusive and hence
they do not impose a strict bound on the mass and charge of strangelets that
can be observed in future experiments. Secondly, although the correlation
between penetrability and geometric cross sections is usually valid for
ordinary nuclei, the same cannot be easily extrapolated to the case of
strangelets since these massive objects are very tightly bound and are not
expected to break up as a result of nuclear collisions. Indeed, in a typical
interaction between a strangelet and the nucleus of an atmospheric atom, it
is more probable for the strangelet to absorb neutrons so that the
colliding nucleus, and not the strangelet, is likely to break up most of the
time. Hence the scheme proposed in \cite{wlk1}, namely that the mass of a
strangelet decreases in every encounter, seems to be unrealistic. In a realistic
model of propagation one also has to consider the effect of the geomagnetic
field which can act on a charged strangelet. Specifically, for medium to
small sized strangelets (see Sec.~\ref{ssec:sqmprop}) the charge on the strangelet will make the 
strangelet travel in twisted paths, increasing the effective length of the path 
and making the globs disappear long before mountain altitudes are reached if 
the cross-sections decrease according to the above prescription.    

In the next section an alternative scheme will be introduced which attempts to 
include the above factors in consideration, specially for small sized strangelets.

\section{\label{sec:strletdust} A new model for propagation}
\begin{figure}[p]
\begin{center}
\begin{minipage}[t]{5in}
\centering
\scalebox{0.5}{\includegraphics{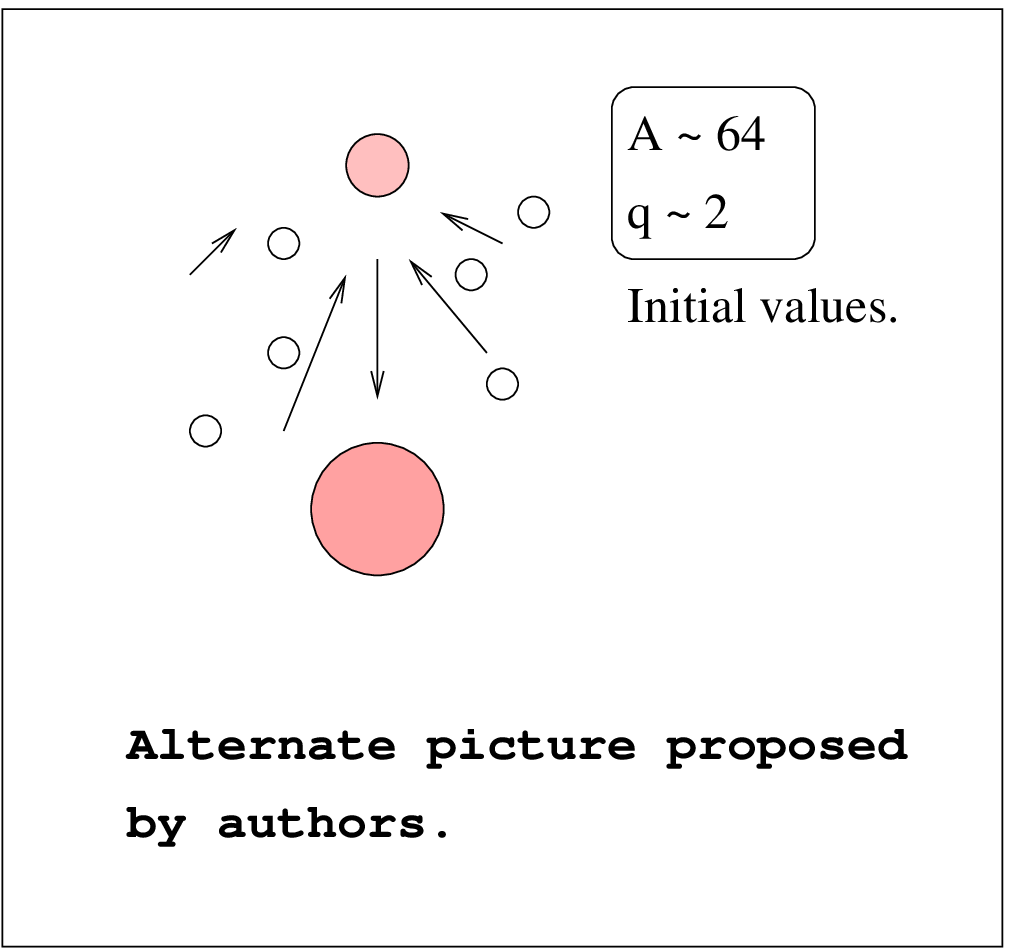}}
\caption{\label{fig:la2} An alternate scheme, proposed by us, for the 
propagation of very small strangelets.}
\end{minipage}\\[20pt]
\begin{minipage}[b]{5in}
\centering
\includegraphics{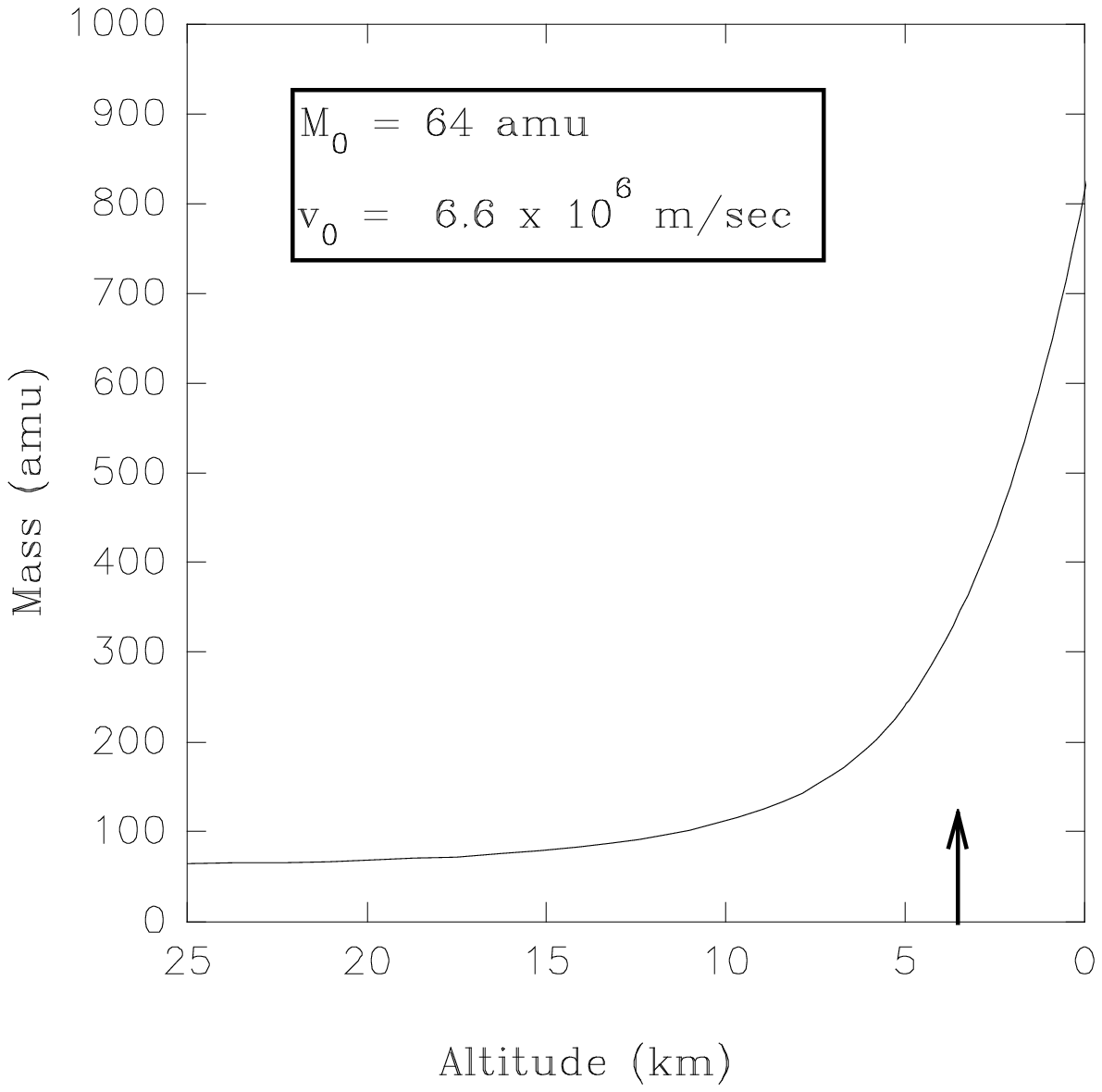}
\caption{\label{fig:mz} Variation of mass of the strangelet with altitude, according 
to our scheme.The arrow corresponds to\ an
altitude of 3.6 km from the sea level.}
\end{minipage}
\end{center}
\end{figure}

The alternative scheme is based on the following premises :
\begin{enumerate}
\item The collision of a lump of SQM with ordinary matter results in the
\emph{absorption} of the neutrons from the colliding nucleus, as a result of
which the mass of the strangelet increases in every collision and it becomes
more tightly bound (Fig.\ref{fig:la2}).
\item The initial masses of the strangelets are assumed to be small in
order to obtain final baryon numbers which are nearly equal to the observed
ones at mountain altitudes. The discussion in the preceding chapter indicates 
that it is quite possible to have stable lumps of SQM with low mass numbers
\footnote{These would fall in the class of very 
small strangelets (\ref{ssec:sqmprop})}.
This would also facilitate a somewhat larger flux in the cosmic rays.
\item The speed, and hence the kinetic energy of these particles, must be
such that they would arrive at a distance of 25 km above the sea level,
surmounting the geomagnetic effects. We start with such an altitude since 
the atmospheric density above 25 km is low enough to be neglected. The charge
of the strangelet is also fixed by this assumption, corresponding to a
certain strangeness fraction.
\end{enumerate}

The simple assumptions proposed above give a picture more or less in accord
with the observation of the propagation of the exotic nuclei in the
atmosphere, which can give useful indications of the type of things to be
expected in an actual experiment. The description of the model is given next.
We consider a situation in which a strangelet with a low baryon number
enters the upper layers ($\sim$ 25 Km  from the sea level) of the atmosphere.
To arrive at this point, a charged particle must possess a speed determined
by the formula (see, \textit{e.g}, \cite{menzel}).
\begin{equation}
\label{cutoffspeed}
\frac {pc}{Ze} \geq \frac{M}{r_{o}^{2}} \frac{cos^{4}\vartheta}{\left
( \sqrt{1+cos^{3}\vartheta}+1\right )^2}
\end{equation}
where $M$ is the magnetic dipole moment of the Earth, $r_o$ the radius of
the Earth and $\vartheta$ is the (geomagnetic) latitude of the point of
observation ($\sim 30^o$, which might represent a location in north eastern
India). $p$ and $Ze$ represent the momentum and charge, respectively, of the
particle. The magnetic field of the Earth is taken to be equivalent to that
due to a magnetic dipole of moment $M=8.1\times 10^{22} J/T$, located near the
centre of the earth, the dipole axis pointing North-South. We have fixed the mass,
initial speed and charge to be 64 amu, $6.6\times10^{7}$m per sec and 2
(electron charge), respectively, at the initial altitude of 25 km. 
\par
In the course of its journey, the strangelet comes in contact with
air molecules, mainly $N_2$. During such collisions, the strangelet absorbs
neutrons from some of these molecules, as a result of which it becomes more
massive. The effect of such encounters is summarized in the formula

\begin{figure}[p]
\begin{center}
\includegraphics{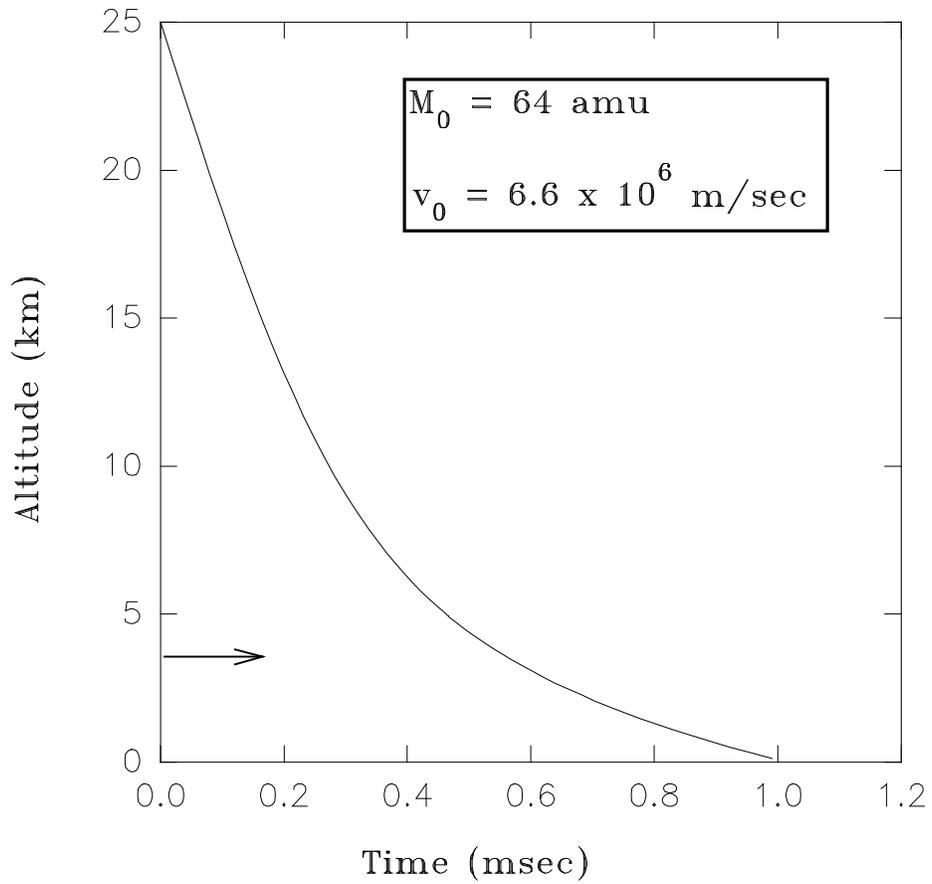}
\end{center}
\caption{\label{fig:at} Variation of altitude with time. The arrow corresponds to an
altitude of 3.6 km from the sea level.}
\end{figure}

\begin{equation}
\label{eatthem}
\frac{dm_S}{dh}= \frac{f\times m_{N}}{\lambda}
\end{equation}
where $m_S$ is the mass of the strangelet, $m_{N}$ the total mass of the
neutrons in the atmospheric atom, $\lambda$  the mean free path of the
strangelet in the atmosphere and $h$ the path length traversed. (It should be
emphasized here that the strangelets would preferentially absorb neutrons,
as protons would be coulomb repelled. Nonetheless the strangelet can absorb 
some protons in the initial phase of the descent, when the relative velocity 
between the strangelet and the air molecule is large. Thus, in this phase, both
the mass and the charge of the strangelet will increase, while in the later 
phase the charge absorption is expected to become strongly inhibited. We do 
not address this issue in the present chapter
\footnote{We do address this issue in the next chapter}
for the sake of simplicity, although it can be readily seen that the rate of
increase in mass would obviously be faster than that in charge.) In the
above equation, $\lambda$ depends both on $h$, which determines the density
of air molecules and the instantaneous mass of the strangelet, which relates
to the interaction cross section. The mean free path  decreases as lower
altitudes are reached since the atmosphere becomes more dense and the
collision frequency increases. Finally the factor $f$ determines the fraction
of neutrons that are actually absorbed out of the total number of neutrons in
the colliding nucleus. The expression for this factor has been determined by
geometric considerations \cite{gosset} and is given by
\begin{equation}
\label{overlap}
f=\frac{3}{4}{\left(1-\nu\right)}^{1/2}{\left(\frac{1-\mu}{\nu}\right)}^{2}
-\frac{1}{3}{\left[3{\left(1-\nu\right)}^{1/2}-1\right]}
{\left(\frac{1-\mu}{\nu}\right)}^{3}
\end{equation}

\begin{figure}[p]
\begin{center}
\includegraphics{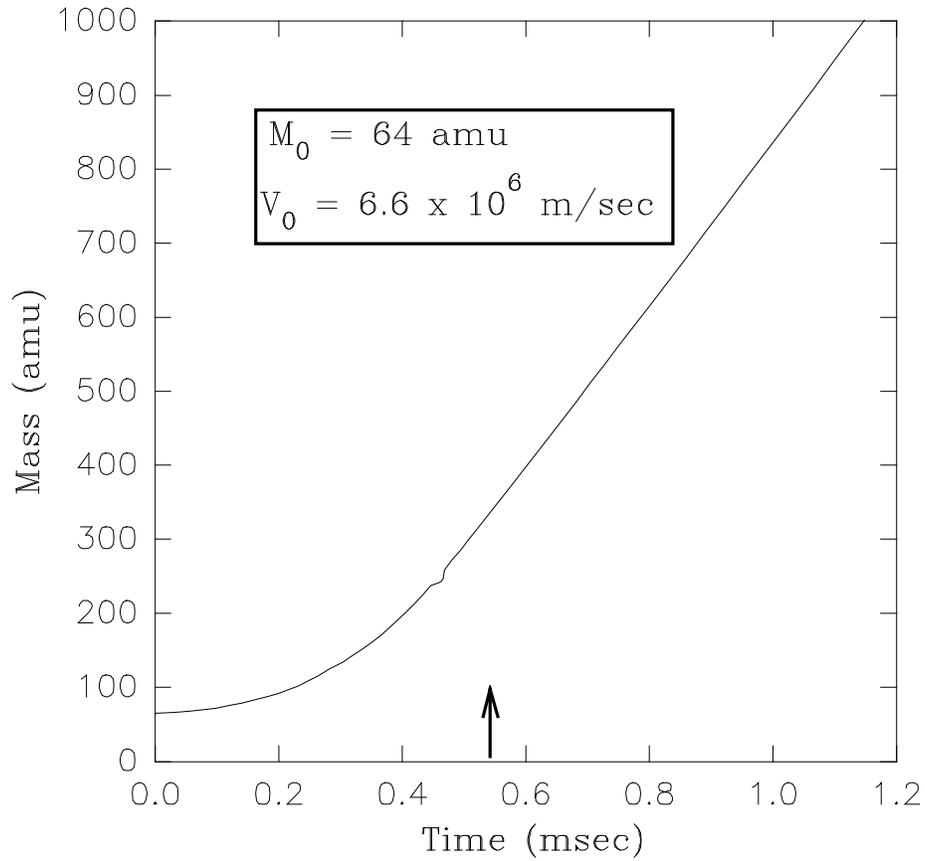}
\end{center}
\caption{\label{fig:mt} Variation of mass with time. The time indicated by the arrow is the
time taken to reach an altitude 3.6 km from the sea level, starting from
25 km.}
\end{figure}

In eqn(\ref{overlap}), $\mu=\frac{b}{R_1+R_2}$ and $\nu=\frac{R_1}{R_1+R_2}$ where
$b$ is the impact parameter. $R_1$ and $R_2$ are the radii of the strangelet
and the nucleus of the atmospheric atom, respectively.
$f$ is initially small but grows larger and reaches the limiting value 1
when the strangelet grows more massive.

\par
The above considerations lead us to a set of differential equations of the
form
\begin{equation}
\label{diffeq1}
\frac{d\vec{v}}{dt}=-\vec{g}+\frac{q}{m_S}\left ( \vec{v} \times \vec{B}
\right ) - \frac{\vec{v}}{m_S}\frac{dm_S}{dt}
\end{equation}
In eqn(\ref{diffeq1}), $-\vec{g}$ represents the acceleration due to gravity, $\vec{B}$
is the terrestrial magnetic field, $q=Ze$ and $\vec{v}$ represents the
velocity of the strangelet. These equations were solved by the $4^{th}$
order Runge Kutta Method for the set of initial conditions described above.

\begin{figure}[p]
\begin{center}
\includegraphics{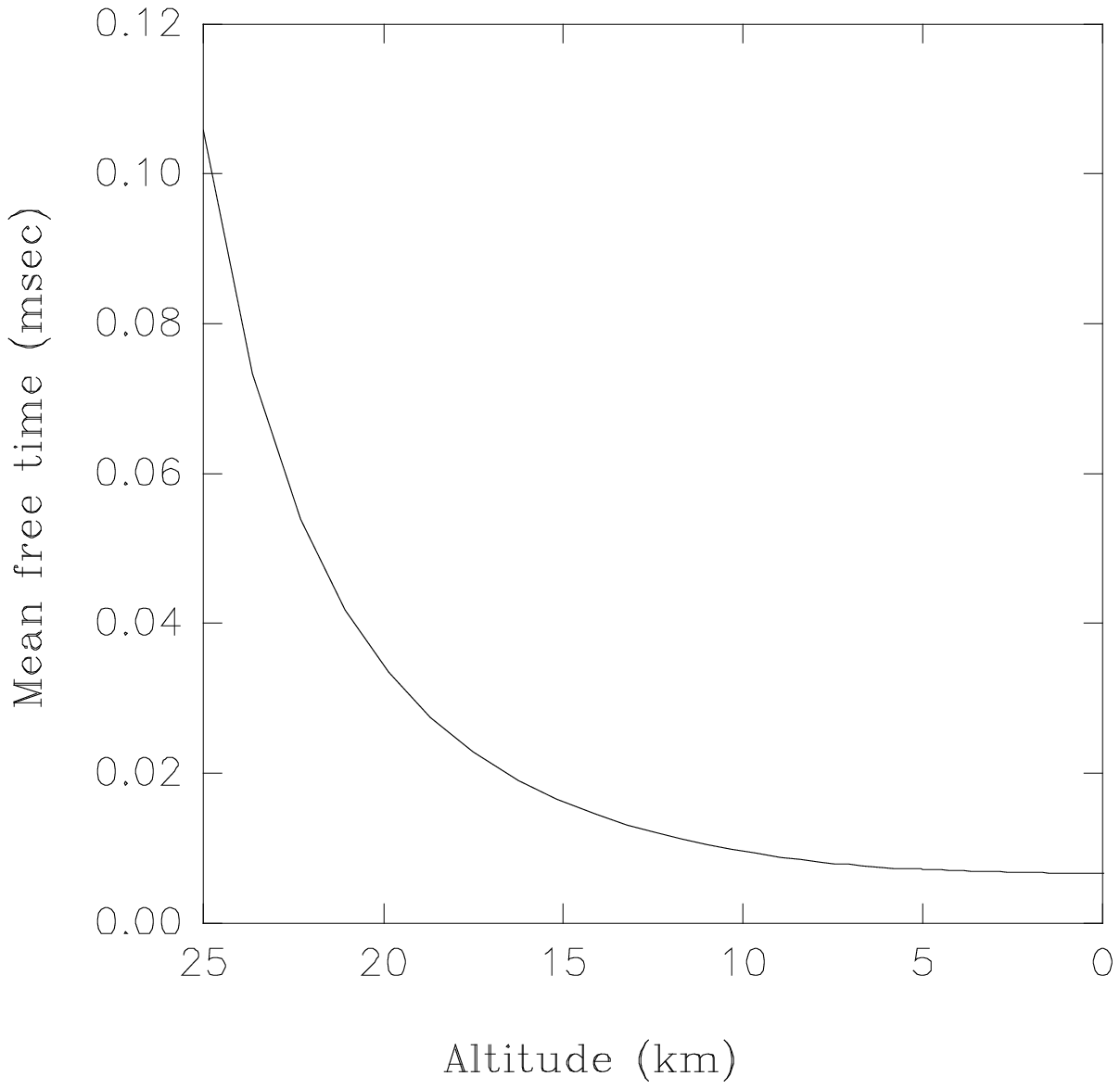}
\end{center}
\caption{\label{fig:ftm} Change of mean free time with altitude.}
\end{figure}

\begin{figure}[p]
\begin{center}
\includegraphics{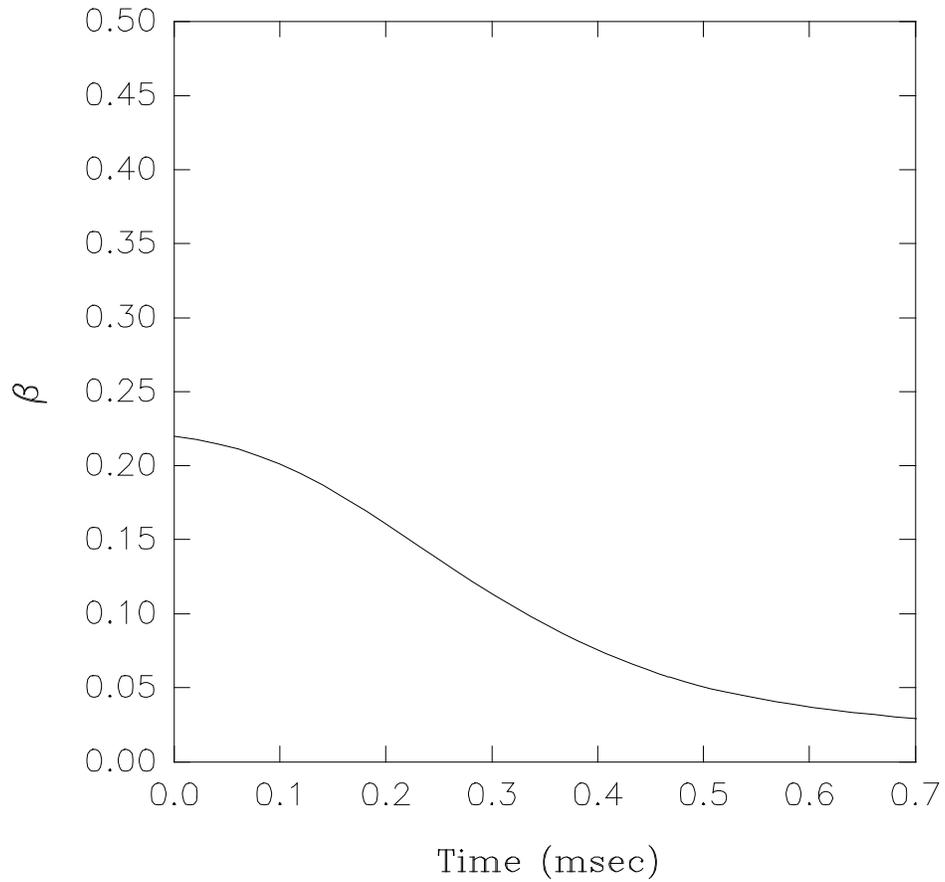}
\end{center}
\caption{\label{fig:btm} Change of $\beta=v/c$ of the strangelet with time.}
\end{figure}

The results are shown in figs.~ \ref{fig:mz} -- \ref{fig:btm}. 
Figure \ref{fig:at} shows the variation of
Altitude with time, the zero of time being at 25 km. The time required to
reach a place which is about 3.6 km above the sea level (height of
a typical north east Indian peak like `Sandakphu', where an experiment to
detect strangelets in cosmic rays using a large detector array is being set
up \cite{saha}) is indicated in the figure.
The next figure (Fig.\ref{fig:mt}) shows the change of the mass of the strangelet
with time and figure \ref{fig:mz} shows the growth of the strangelet mass with
altitude. It can be seen from the figures that the expected mass at the
aforementioned altitude comes out to be about 340 amu or so. Figure \ref{fig:ftm} shows
the variation of the mean free time with altitude. The mean free time at all
positions are more than the time scale for weak interactions ($10^{-8}$ sec) so that
the strangelet gets enough time to stabilize and adjust itself to new baryon number 
configurations. Finally, figure \ref{fig:btm} shows the variation of $\beta=v/c$ 
with time, showing that the speed of the strangelets decrease as they grow more massive.

In this chapter we have advocated a dynamical model of the propagation of strangelets
through the terrestrial atmosphere. It is based strongly on the characteristic property
of SQM which is responsible for regarding objects made of similar stuff as 
the ground state of QCD. It is realistic enough to include the difference in the 
interaction process which is expected when SQM and not ordinary nuclei, collide with 
atmospheric nuclei. The effects of Earth's gravitational and magnetic fields are 
included in the equations of motion so that it is possible to derive meaningful 
information directly from the resulting trajectory. The main conclusion of the 
model is that the exotic cosmic ray events with very small Z / A ratios at 
mountain altitudes  could result from SQM droplets which need not be large
initially. Thus the flux of the strangelets may be appreciable enough to make their
detection by a large area detector at mountain altitudes a real possibility. 

\chapter{\label{chap:newmodel}Energy loss of fast strangelets}
\newcommand{\ppf}{\ensuremath{\vec{P}_f}}
\newcommand{\ppi}{\ensuremath{\vec{P}_i}}
\newcommand{\g}{\ensuremath{\gamma}}
\newcommand{\dg}{\ensuremath{\Delta\gamma}}
\newcommand{\Dt}{\ensuremath{\Delta t}}
\newcommand{\DM}{\ensuremath{\Delta M}}
\newcommand{\ddt}[1]{\ensuremath{\frac{d#1}{dt}}}
\newcommand{\DDt}[1]{\ensuremath{\frac{\Delta #1}{\Delta t}}}
\newcommand{\oof}[1]{\ensuremath{\mathcal{O}(#1)}}
\newcommand{\vel}{\ensuremath{\vec{v}}}
\newcommand{\tcs}{\ensuremath{\theta_s^C}}
\newcommand{\tcn}{\ensuremath{\theta_N^C}}
\newcommand{\tls}{\ensuremath{\theta_s^L}}
\newcommand{\tln}{\ensuremath{\theta_N^L}}
\newcommand{\tlcm}{\ensuremath{\theta_\text{CM}^L}}

In the previous chapter we have proposed a dynamical model for the propagation
of strangelets of low mass numbers through the terrestrial atmosphere, where
the stability of SQM plays a very important role. In that model the mass
of the strangelet increases when it undergoes a collision with atmospheric
atoms during the course of the journey. Using straightforward geometrical 
considerations, it has been shown \cite{jpg@23} that the strangelet can grow from 
$A$ = 64 amu to $A \sim$ 340 amu by the time it reaches an altitude  
$\sim$ 3.5 km, the altitude of a typical mountain peak with adequate 
accessibility for setting up a large detector array. This remarkable 
possibility makes it imperative to explore the consequences of this novel
mechanism with greater care.

The basic model proposed earlier is summarized in the equation of motion
eqn.~\ref{diffeq1}. While the first two terms in this equation have obvious
significance, the third term accounts for the deceleration of
the strangelet due to its peculiar interaction with the air molecules; 
strangelets can readily absorb matter and become more strongly bound, unlike 
the normal nuclear fragments which tend to break up \cite{wit@23}. In this chapter 
the basic formalism will be extended \cite{asp,prl@33} in several ways, elaborated in the following
section.

\section{\label{sec:refine} A more refined model}
In the earlier model, the strangelet acquired mass only through the absorption 
of neutrons from the air molecule, since it was assumed that the repulsive 
Coulomb barrier on the surface of the strangelet will keep the protons off the strangelet.
However, if we want to study initially fast moving strangelets, it appears quite 
possible that they absorb a few protons in the initial phase of their journey, when the 
relative velocity between the strangelet and the air molecule is large.
In the previous chapter, the charge of the strangelet played only an indirect role,
since it entered the equation of motion eqn.~\ref{diffeq1} only through
the coupling with the geomagnetic field. Thus, although it was instrumental 
in steering the course of the strangelet it didn't affect the strangelet speed. 
Therefore, although the presence of the charge increased the length of the trajectory 
before mountain altitudes were reached and led to larger mass increments compared 
to that for a neutral strangelet, it did not directly affect the instantaneous mass accumulation 
rate of the strangelet (eqn.~\ref{eatthem}). In this chapter the formalism takes care of 
the accretion of charge of fast moving strangelets by formulating the problem 
in a relativistic setting. As a consequence of proton absorption, the issue of loss 
of energy of the strangelet through ionization of the surrounding media cannot be ignored any more.
The accumulated charge thus will be able to influence the speed 
of the strangelet directly and it will be seen that the ionization losses become quite 
significant at comparatively low altitudes (where the atmosphere is dense and the 
strangelet is sluggish) and provide a lower limit to the height at which the 
strangelets can be detected successfully. In the previous analysis we have traced 
the fate of strangelets whose initial speeds at the upper layer of the atmosphere 
were only slightly higher than the lower bound on such velocities imposed by the 
geomagnetic field -- however a larger spectrum of initial velocities can only be 
examined within a consistent relativistic framework. Although our model of propagation 
of strangelets rely heavily on the stability of small lumps of SQM 
\cite{jaffe@32,car@32,an@32,jurg@32,mador@32}, the probability of fission like 
fragmentation of such lumps cannot be ruled out, specially
for highly energetic collisions of the strangelet with the air molecules, which can 
now occur at relativistic speeds. In this work we disregard the possibility of such 
events by providing a upper limit to the initial speed of the strangelet, above which 
it may no longer be practicable to evade the possibility of fragmentation. We have 
estimated that for our case (initial A larger than 40 for which the stability appears
to be more robust due to an underlying shell like structure \cite{mador@32,mad@32}),
this upper limit on the velocity comes out to be slightly above 0.7c
(see App.~\ref{app:cmenergy}). 

In order to incorporate the above effects, we first modify the term responsible for the
absorption of neutrons in eqn.~\ref{diffeq1} to include
the rate for proton absorption. It should be noted that absorption of
neutrons would lead only to mass increase while that of protons would
increase both the mass and charge of the strangelets.

In order to relate the proton absorption cross-section to the neutron absorption cross section, 
we adopt the following simple model. The shell structure of \( \mathrm{N}_{2} \)
dictates that only the single proton belonging to the outermost shell
can be considered to be sufficiently free for the consideration of
absorption by the strangelet. We describe the classical motion of the proton of
energy E in the vicinity of the strangelet after the model of the motion of a
free particle of unit charge in the repulsive Coulomb field of the strangelet. 
The total energy of the proton at time t is given by

\begin{equation}
\label{eprot}
E=\frac{1}{2}m(\dot{r}^{2}+r^{2}\dot{\phi }^{2})+U(r)=\mathrm{const}
\end{equation}
where \( U(r) \) represents the Coulomb energy of the proton and other
quantities have their usual significance. The effective energy for the
equivalent one dimensional problem is 
\[U_{\mathrm{eff}}(r,L)=\frac{L^{2}}{2mr^{2}}+U(r)\] where \(L=m\dot{r}^{2}\dot{\phi }\)
is the angular momentum of the proton. The angular momentum \( L \) also
equals \( mv_{0}b \) where \( v_{0} \) is the relative speed with which the 
\( \mathrm{N}_{2} \) nuclei (and hence, its constituent protons) approach the 
strangelet and \( b \) is the impact parameter. The minimum separation along 
the trajectory occurs when - 
\begin{equation}
\label{minsep}
\frac{(mv_{0}b)^{2}}{2mr_{\mathrm{min}}}+U(r_{\mathrm{min}})=E=\frac{1}{2}mv_{0}^{2}
\end{equation}
Assuming that charge transfer can take place when \( r_{\mathrm{min}}\le R_{s} \) (the
radius of the strangelet), the corresponding value of b ($\equiv b_c$) for which this 
occurs can be solved by substituting \( b=b_{c} \) and \( r_{\mathrm{min}}=R_{s} \) 
in eqn. \ref{minsep}. This yields 

\[b_{c}^{2}=R_{s}^{2}(1-U(R_{s})/(\frac{1}{2}mv_{0}^{2}))\]
The above assumption is equivalent to saying that all protons for
which \( b>b_{c} \) misses the strangelet, while those with
\( b_{c}\leq b_{c} \) are captured. Thus we can write the capture cross
section for protons \( \sigma _{p} \) as 
\begin{equation}
\label{sigmap}
\sigma _{p}=\pi b_{c}^{2}=\pi R_{s}^{2}\left[ 1-\frac{Z_{S}e^{2}}{4\pi \varepsilon _{0}R_{S}}\left/ \frac{1}{2}mv_{0}^{2}\right. \right] 
\end{equation}
In contrast, the absorption cross section for neutrons \( \sigma _{n} \) is 
just \( \pi (R_{n}+R_{s})^{2} \) and hence we can infer that the accretion 
to the strangelet due to its interaction with neutrons \( \frac{dm_{s_{n}}}{dt} \)
is related to that due to protons \( \frac{dm_{s_{p}}}{dt} \) by
\begin{equation}
\label{fpndef}
\frac{dm_{s_{p}}}{dt}=\frac{\sigma _{p}}{\sigma _{n}}\frac{dm_{s_{n}}}{dt}\equiv fpn\frac{dm_{s_{n}}}{dt}
\end{equation}
whence, 
\begin{equation}
\label{fpn}
f_{pn}=\frac{R_{s}^2}{(R_{n}+R_{s})^2}\left[ 1-\frac{Z_{s}e^{2}}{4\pi \varepsilon _{0}R_{s}}\left/ \frac{1}{2}mv_{0}^{2}\right. \right] 
\end{equation}
Thus, $f_{pn}$ determines the relative probability for a proton to undergo
the above process vis-a-vis a neutron, and is less than one, on account of
the coulomb barrier present at the surface of the strangelet.

We carry on the analysis by extending the formalism slightly to accommodate relativistic 
speeds. The equation of motion eqn.~\ref{diffeq1}, generalized to a relativistic 
form leads to eqn.~\ref{newneweq} in a straightforward manner (see App.~\ref{app:releq})

\begin{equation}
\label{newneweq}
\gamma m_{s}\frac{d\vec v }{dt}=-m_{s}\vec g +q(\vec v \times \vec B )-
\gamma \vec v \left(\frac{dm_{s_{n}}}{dt}+\frac{dm_{s_{p}}}{dt}\right)
-m_{s}\vec v \frac{d\gamma}{dt}-\frac{f\, (v)}{\sqrt{3}}\vec v
\end{equation}

In the above equation the mass transfer rates for the proton and the neutron are related
by an equation (eqn.~\ref{newfpn}) similar to eqn.\ref{fpn} but where 
\( \frac{1}{2}mv_{0}^{2} \) is replaced by it's relativistic equivalent \( E \):

\begin{equation}
\label{newfpn}
f_{pn}=\frac{R_{s}^{2}}{(r_{n}+R_{s})^{2}}\left(1-\frac{1}{E}
\frac{Z_{s}e^{2}}{4\pi \epsilon _{0}R_{s}}\right)
\end{equation}

Finally, the last term of equation (\ref{newneweq}) accounts for the ionization
loss. The expression for \( f(v) \) is given by \cite{ion1@32} 
\begin{equation}
\label{ionloss}
f\, (v)=-\frac{dE}{dx}=
\frac{Z_s^{2}e^{4}nZ_{med}}{4\pi \epsilon ^{2}_{0}m_{e}v^{2}}ln\left(
\frac{b_{max}}{b_{min}}\right)  
\end{equation}

Here,  \( n \) represents the number density of the atmospheric atoms at a
particular altitude, \( Z_{med} \) is the number of electrons per atom of
\( N_{2} \) which can be ionized, \( m_{e} \) is the mass of the electron
and \( b_{max} \) and \( b_{min} \) are the maximum and minimum values of
the impact parameter. At large velocities, expression (\ref{ionloss})
reduces to, with \( I \) denoting the average ionizing energy,
\begin{equation}
\label{ilossrel}
f\, (v)=
\frac{Z_s^{2}e^{4}nZ_{med}}{4\pi \epsilon ^{2}_{0}m_{e}v^{2}}ln\left( \gamma
^{2}\frac{2m_{s}v^{2}}{I}-\beta ^{2}\right)  
\end{equation}

However, when the velocity of the strangelet falls below a critical value
\( v\leq 2Z_{s}v_{0} \)( \( v_{0}=2.2\times 10^{6}m/s \)
is the speed of the electron in the first Bohr orbit), electron capture
becomes significant which can be accounted for by the replacement
\( Z_{s}\rightarrow Z_{s}^{\frac{1}{3}} \frac{v}{v_{o}} \) \cite{ion1@32,ion2@32}. 

Equation (\ref{newneweq}) was solved by the \( 4^{th} \) order Runge-Kutta
method with different sets of initial mass, charge and \( \beta \). It may
be mentioned at this point that the first term in eqn (\ref{newneweq}) is not
important in magnitude, as is to be expected. We have nonetheless included
it for numerical stability. This serves to define the downward vertical
direction in the vector algorithm, especially for very small initial
velocities.

\section{\label{ssec:c3result}Results}

\begin{figure}[p]
\centering
\includegraphics{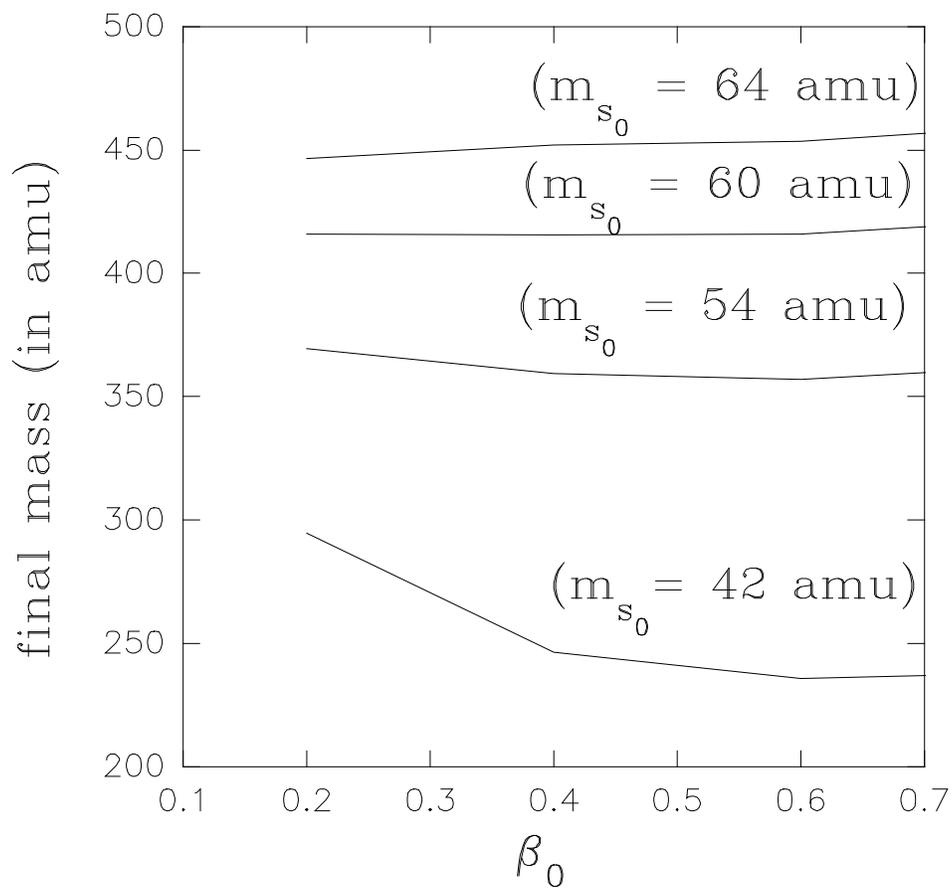}
\caption{\label{fig:fmb00}Variation of final masses with initial \( \beta (\beta_0) \) for different
initial masses of the incident strangelet.}
\end{figure}

Figure \ref{fig:fmb00} shows the final masses (for initial masses 42, 54, 60 and 64
\( amu \) and a fixed initial charge 2) as a function of initial
\( \beta \). It is seen, especially for smaller initial masses, that the
final mass decreases at first with increasing initial \( \beta \)
and then begins to increase again after a critical \( \beta \) is reached
and this critical value of \( \beta \) shifts to the left for larger
initial mass.  Although mathematically delicate (it can be 
seen from eqn.~\ref{newneweq} that a higher value of speed leads to an increasing 
value of the mass increment, which in turn slows down the particle), a qualitative 
explanation for this feature might be given as follows. One can think of the total 
region, through which the strangelet travels, being divided into two not-too-distinct 
subregions. In subregion I, corresponding to higher altitudes, the number 
of atmospheric particles is small, while this number is considerably larger in 
subregion II, corresponding to lower altitudes. For a strangelet of small initial mass 
(smaller size), the strangelet has a greater chance to escape subregion I if \( \beta \) 
is higher, so that it will pick up lesser mass from this region. On the other hand if 
\( \beta \) is very high, the volume that the strangelet sees will be contracted 
(the twisted tube through which it travels will be constricted), as a result of which
it will interact with a greater number of atmospheric particles whence it will pick 
up a larger number of nucleons. It is clear that for an initially bigger (more massive)
strangelet, this critical \( \beta \) will be lower, as it will be able to sweep
through a larger number of atmospheric particles right from the start. In a nutshell,
this effect can thus be ascribed to higher speeds leading to larger mass increments, 
whose effect would be more pronounced for lower initial masses.

\begin{figure}[p]
\centering
\includegraphics{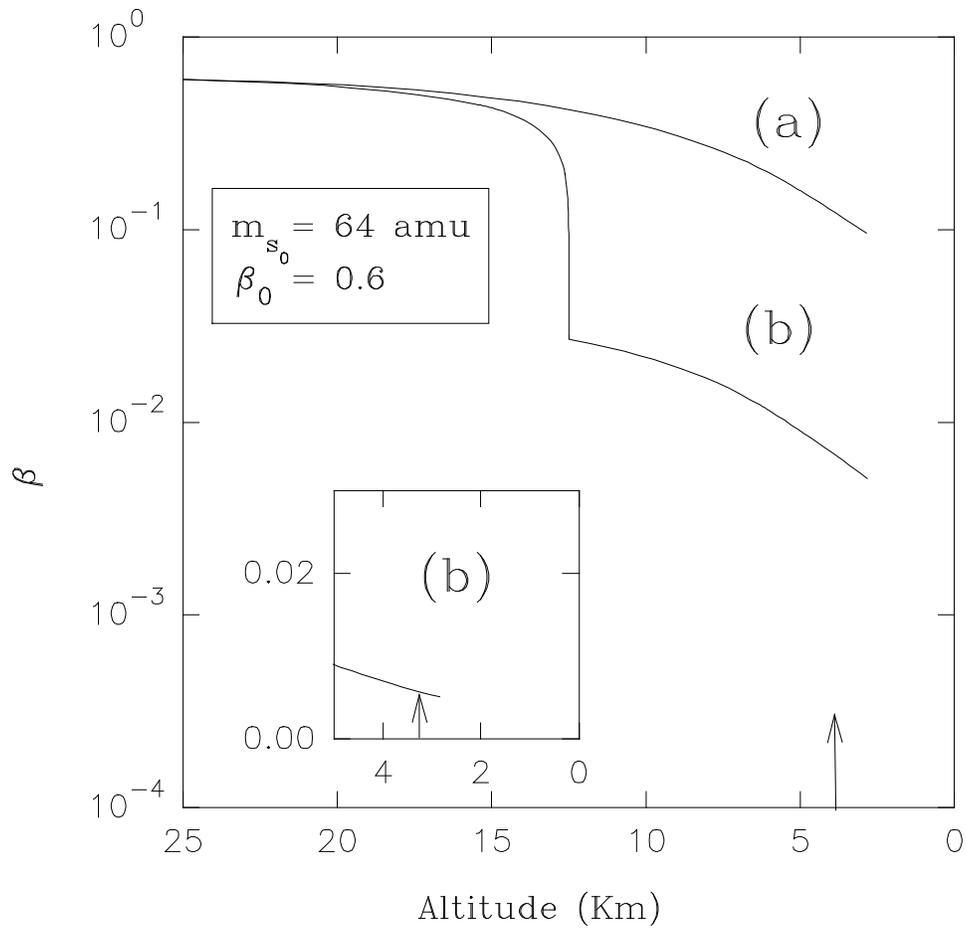}
\caption{\label{fig:bh3final} Variation of final \( \beta \) with altitude (a) for constant charge
and without ionisation loss and (b) including proton absorption as well as
ionisation loss}
\end{figure}

\begin{figure}[p]
\centering
\includegraphics{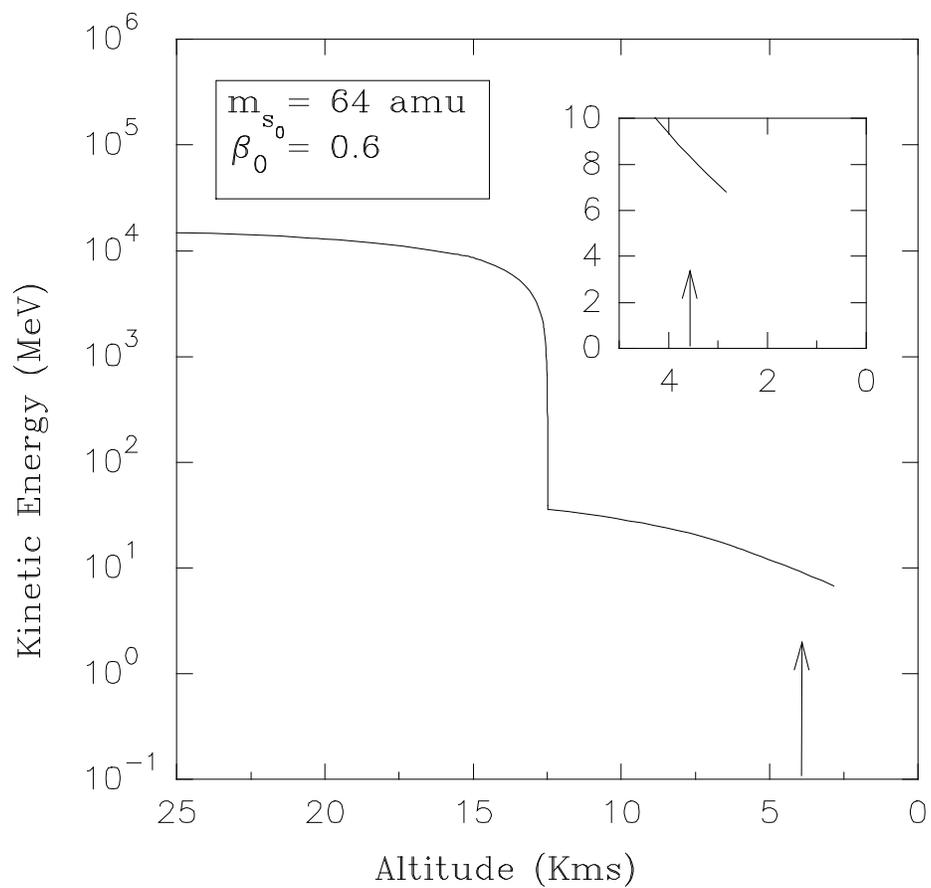}
\caption{\label{fig:kh3} Variation of kinetic energy of the strangelet with altitude}
\end{figure}

Let us now consider a representative set of data with initial mass 64
\( amu \) and charge 2 for detailed discussion. The results for \( \beta _{0}=0.6 \) 
are shown in figures \ref{fig:bh3final} and \ref{fig:kh3},where the variation of 
speed (\( \beta  \)) and the energy of the strangelet with altitude are depicted. The 
sharp change seen at \( \sim \) 13 km corresponds to the onset of electron capture, which
is handled phenomenologically through the effective \( Z_s \).
The insets of figures \ref{fig:bh3final} and \ref{fig:kh3} show a zoomed-up view of the
respective quantities near the endpoint of the journey. It is apparent from
the figures that the ionization term reduces the overall energy and speed
considerably from the non-dissipative situation \cite{jpg@23}. However, the
zoomed-up insets in figs.\ref{fig:bh3final} and \ref{fig:kh3} show that the strangelets 
may have enough energy to be detectable at an altitude of 3.6 km from the sea level. 
For example, for the values of the initial quantities \( m_{s_{o}} \) and
\( \beta_{o} \) shown here, the strangelet is left with a kinetic energy
\( \sim  \)8.5 MeV (corresponding to \( \frac{dE}{dx} \sim 2.35\, MeV/mg/cm^{2} \)
in a Solid State Nuclear Track Detector (SSNTD) like CR-39), which, although
small, is just above the threshold of detection 
\( (\frac{dE}{dx})_{crit}\sim \, 1- 2 MeV/mg/cm^{2} \) for \( \beta < 10^{-2} \) 
in CR-39 for the present configuration. Below this height, the possibility of 
their detection with passive detectors like SSNTD reduces to almost zero.

\begin{table}
{\centering \begin{tabular}{|c|c|c|c|c|c|}
\hline 
\( \beta _{0} \)&
\( m_{s_{0}} \)&
\( m_{l} \) (amu)&
\( q_{l} \)&
\( \beta _{l}\times (10^{-3}) \)&
\( e_{l} \) (MeV)\\
\hline 
&
42 & 294.7 & 3 & 2.8 & 1.05 \\
0.2&
54 & 369.4 & 4 & 3.0 & 1.55 \\
&
60 & 415.8 & 4 & 3.0 & 1.80 \\
&
64 & 446.5 & 5 & 3.1 & 1.98 \\
\hline 
&
42 & 246.4 & 6 & 4.9 & 2.84 \\
0.4&
54 & 359.5 & 8 & 4.7 & 3.73 \\
&
60 & 415.6 & 8 & 4.7 & 4.25 \\
&
64 & 452.0 & 9 & 4.6 & 4.63 \\
\hline 
&
42 & 235.8 & 10 & 7.4 & 5.97 \\
0.6&
54 & 357.1 & 12 & 6.6 & 7.15 \\
&
60 & 416.0 & 13 & 6.4 & 7.87 \\
&
64 & 453.6 & 14 & 6.3 & 8.39 \\
\hline 
&
42 & 236.4 & 12 & 8.6 & 8.16 \\
0.7&
54 & 359.1 & 14 & 7.6 & 9.59 \\
&
60 & 418.3 & 15 & 7.3 & 10.46 \\
&
64 & 456.3 & 16 & 7.2 & 11.11 \\
\hline 
\end{tabular}}
{\centering
\caption{\label{tab:compvalues}The final values, denoted with suffix \( l \), are tabulated along
with initial \( \beta \) (\( \beta_0 \))}}
\end{table}

Table \ref{tab:compvalues} lists the final values of the mass, the charge, 
\( \beta  \), and the energy of the strangelet at the end of the journey for different
initial velocities. A comparison between table \ref{tab:exotica}, first mentioned in 
the beginning of Chap.~\ref{chap:dust} and Tab.~\ref{tab:compvalues} shows that the final masses and charges are very similar to the ones found in cosmic ray events.

The experimental verification of SQM in cosmic ray flux (and the mechanism
of their propagation through the earth's atmosphere) is thus possible with a
suitable ground based detector set up at high altitudes of about 3 to 5 km.
At such altitudes, the predicted energy range of the resulting penetrating
particles with mass $M$ between 300 and 400 and $Z$  between 10 and 15
should lie between 5 to 50 MeV. (This estimate corresponds to an averaging
over all angles of incidence at the top of the atmosphere, taken to be 25
km here, as mentioned above.) 
A suitable locality for such observations at an altitude of about 3.5 km
above the sea level has been identified at Sandakphu, in the middle ranges
of the eastern Himalayas, with adequate accessibility and climatic
conditions. Continuous exposure for months or years at a stretch of a
detector assembly with stacks of SSNTDs like CR-39, covering a total area
of about 400 m$^2$, is planned there. (The number of events due to
strangelets may be as few as 5 - 10 per  100 m$^2$ per year, according to
our approximate estimates (see Chap.~\ref{chap:strabund}.) The major considerations in this respect are cost, 
structural simplicity, and long time stability of the detection sensitivity against
temperature fluctuations of several tens of Celsius degrees between summer
and winter months and the ruggedness of the passive detectors. Regarding all
these aspects, commercially available CR-39 appears to be the most suitable
choice, which has been shown in NASA SKYLAB experiments \cite{chan@23,beaujean@23}
to be capable of detecting heavy ions with energies upto 43 MeV/u. The
signatures produced in such detectors in terms of mass, charge and energy of
detectable strangelets can be evaluated in the expected $dE/dx$ range by
measurements of track dimensions. For this purpose, additional calibration
experiments, exposing CR-39 samples to heavy ions with variable charge
states at almost similar energy ranges, are necessary which can be made at 
several existing heavy ion accelerator facilities. With efficient etching
and automated track measurements, backgrounds of low energy secondary
radiation with lower charge or mass are not expected to pose any serious
problems. Due to specific inherent technical problems like "fading" of
thermo-luminescent substances over a long interval of time, this kind of
material do not seem to be practical in our experimental conditions. CR-39 has an additional
advantage over the other types of passive semi-conductor detectors using
co-polymers like SR6, CN85 or Lexan; a large amount of characteristic
experimental data are already available for CR-39 in the existing literature.
As alternatives, Mica or Overhead Transparency Foils may also be considered
and calibration experiments using these materials will be conducted at
accelerator facilities to judge their suitability. Other active detectors
and devices do not appear to be suitable for installation at proposed
mountain heights for stand-alone operation over long periods and are
therefore not being considered at present.

In conclusion, we have presented a model for the propagation of 
cosmic strangelets of none-too-large size through the terrestrial 
atmosphere and shown that when proper account of charge and mass 
transfer as well as ionization loss is taken, they may indeed 
reach mountain altitudes, so that a ground based large detector 
experiment would have a good chance of detecting them.

\begin{subappendices}
\section{Relativistic equation of motion}
\label{app:releq}

\begin{figure}[ht]
\centering
\psfrag{D}{\scalebox{2}{$M$}}
\psfrag{E}{\scalebox{2}{$M-\Delta M$}}
\psfrag{C}{\scalebox{2}{$M^*$}}
\psfrag{T}{\scalebox{2}{\fbox{$t$}}}
\psfrag{S}{\scalebox{2}{\fbox{$t+\Delta t$}}}
\psfrag{V}{\scalebox{2}{$\vec{v}$}}
\psfrag{W}{\scalebox{2}{$\vec{v}+d\vec{v}$}}
\psfrag{X}{\scalebox{2}{$\vec{u}$}}
\psfrag{G}{\scalebox{2}{$\gamma$}}
\psfrag{H}{\scalebox{2}{$\gamma+\Delta\gamma$}}
\psfrag{I}{\scalebox{2}{$\gamma_u$}}
\scalebox{0.7}{\includegraphics{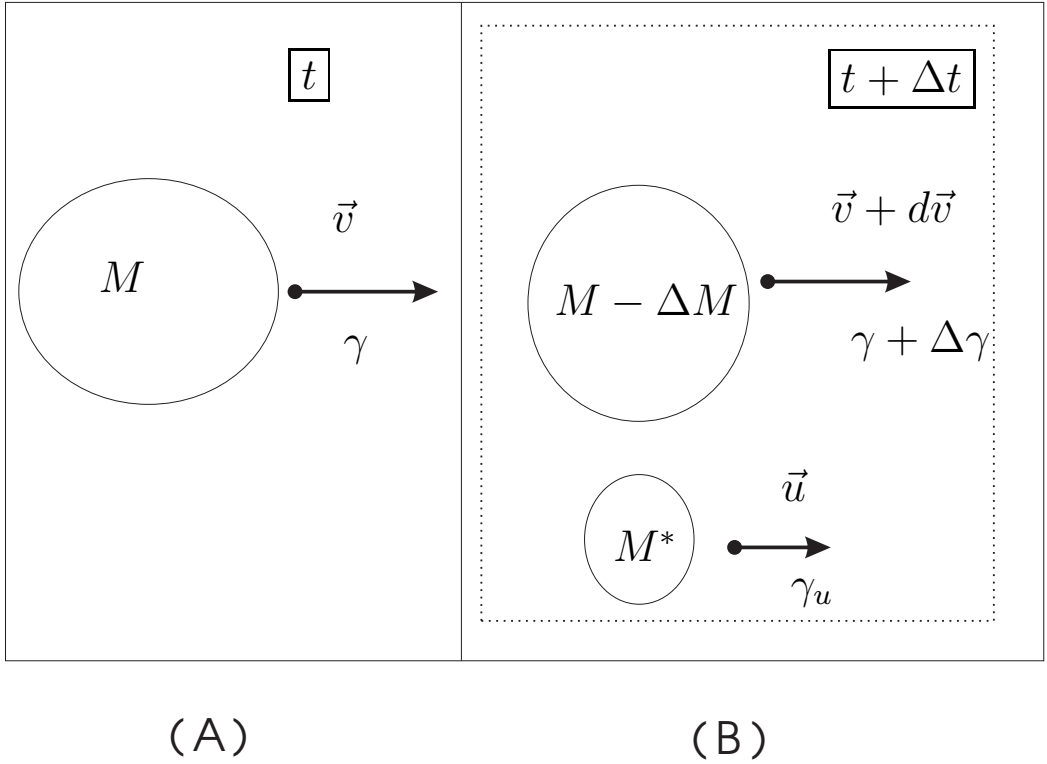}}
\caption{\label{fig:releq} Relativistic rocket motion : (A) The rocket at time t, (B) The rocket
and it's eject at time $t+\Delta t$}
\end{figure}

Starting with a relativistic form of the equation of rocket motion,
we derive the equation of motion applicable for our case of a relativistic snowball. The equation of motion 
eqn.~\ref{diffeq1} is generalized to a relativistic form by interpreting \( \vec{p} \) as 
the relativistic three momentum. We consider a system of variable (\emph{proper}) mass M at the instant 
\( t \) which changes to one with mass \( M - dM \) at the instant \( t+dt \) (see Fig.~\ref{fig:releq}). The mass of the ejecta depicted by \( \delta M^{*} \) is assumed to be different from \DM since it moves with a different 
velocity \( \vec{u} \). The velocities at the two points of time are denoted by \( \vec{v}(\gamma)\) and \( \vec{v}(\gamma+d\gamma) \) respectively, where \( \g \) are the respective Lorentz factors. For the system
enclosed in the dotted boundary,

\begin{align*}
& \vec{F}_\mathrm{ext} \simeq \frac{\ppf - \ppi}{\Dt} \\
       & =     ((\g+\dg)(M-\DM)(\vel+d\vel) + \vec{u}\g_u\DM^* - \g M\vel) / \Dt \\
& \text{In our case the particle is simply absorbed, so one can set}\; \vec{u}\rightarrow 0\\
& \text{and after discarding any terms}\; \oof{\Delta^2}\\
& = \g M \DDt{\vel} - \g \vel\:\DDt{M} + \DDt{\g} M \vel 
\end{align*}
Replacing \( \DDt{M} \rightarrow -\ddt{M} \) and disregarding terms \( \oof{\beta^2} \)
we get,
\begin{equation}
\label{prenewneweq}
\g M \ddt{\vel} = \vec{F}_\mathrm{ext} - \g\vel\:\ddt{M} - M\vel\:\ddt{\g} \tag{$*$}
\end{equation}
From (\ref{prenewneweq}) we recover the equation of motion \ref{newneweq} appropriate 
for the situation. It may be noted that for rocket motion the \ddt{M} is negative, while it 
is positive in the case for relativistic snowballs.

\section{Threshold speed of strangelet for breakup}
\label{app:cmenergy}

\begin{figure}[ht]
\centering
\psfrag{X}{\scalebox{2}{$\theta_s^L$}}
\psfrag{Y}{\scalebox{2}{$\theta_s^c$}}
\psfrag{Z}{\scalebox{2}{$\theta_N^c$}}
\psfrag{W}{\scalebox{2}{$\theta_{\text{CM}}^L$}}
\scalebox{0.7}{\includegraphics{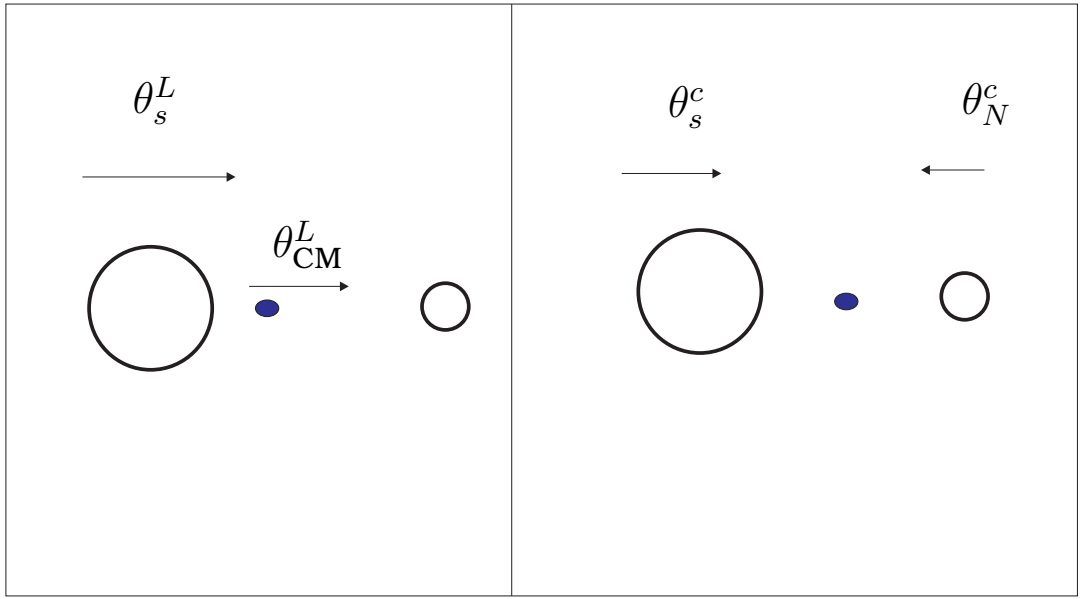}}
\caption{\label{fig:scoll} Colission of strangelet and air nuclei as seen 
from the (A) Lab frame (B) Center of mass frame. The small filled circle
represents the center of mass of the colliding particles.}
\end{figure}

In this appendix we outline the calculations leading to the estimate 
of the upper limit of strangelet velocity. Roughly speaking, the 
reduced mass \(\mu\) of the two body system of the strangelet (\(m_s=64\) GeV) 
and an average air nucleus (\(m_N=14\) GeV) is \(\frac{m_s m_N}{m_s+m_N} \sim 11.48\)
GeV. A crude estimate can be obtained by examining the kinetic energy \(\mu(\g-1)\) 
available in this system.  Assuming that this is enough to break up the strangelet 
completely\footnote{%
we have taken the representative value of the binding energy per baryon of
a strangelet from \cite{mad@32}}, one obtains a \(\g \sim 1.28 \) corresponding to a 
\(\beta \sim 0.72 \). 
To obtain a more careful estimate of the available energy we examine the  
energy \( E_\text{CM}^\text{avl}\) available in the center of mass frame 
of the air molecule suffix $N$) and the strangelet, (suffix $s$) (see Fig.~\ref{fig:scoll}) which may be effective 
in breaking the strangelet up, 
\begin{equation}
\label{ecm}
E_\text{CM}^\text{avl}=m_s(\g_s^C-1) + m_N(\g_N^C - 1)
\end{equation}
The center of mass quantities (superfix $C$) can be expressed in terms of 
l aboratory quantities (superfix $L$) as follows 
In the CM frame the strangelet and the air molecule approach each other with equal magnitude 
of three momentum, i.e
\begin{align}
\label{a1}
m_s\beta_s^C\g_s^C &= m_N\beta_N^C\g_N^C \quad \text{or,} \notag \\
m_s\sinh\tcs & = m_N\sinh\tcn
\end{align}
On the other hand \( \tcn=-\tlcm \) and \(\tcs = \tls - \tlcm = \tls + \tcn \). 
Since \(\tcn\) is a negative quantity, redefining \( \tcn \rightarrow \abs{\tcn} \), 
one gets, 
\begin{equation}
\label{a2}
\tcs=\tls - \tcn
\end{equation}
From \ref{a1} and \ref{a2} it follows that, 
\begin{align}
\label{a3}
m_s\sinh(\tls-\tcn) & = m_N\sinh\tcn \quad \text{or,} \notag \\
\coth\tcn & =\frac{m_N}{m_s}\csch\tls + \coth\tls
\end{align}
Finally, from \ref{a3} and using \( \tls = \tanh^{-1}\beta \), \(\tcn\)
can be expressed in terms of \(\beta\) alone and the quantities 
\( \beta_{s|N}^C = \tanh\theta_{s|N}^C \) can be similarly evaluated leading
to an evaluation of \(E_\text{CM}^\text{avl}\) in terms of the incident speed.
\end{subappendices}

\chapter{\label{chap:strabund} Strangelet event rates and abundances}
In the earlier chapters (Chap.~\ref{chap:dust}-\ref{chap:newmodel}) we have discussed 
some of the problems associated with the penetration of small lumps of SQM through the 
terrestrial atmosphere. In order to handle some of the quite unusual properties of SQM with 
respect to ordinary cosmic ray particles in a proper fashion, we have also put forward 
a dynamical model for the propagation of these objects through the terrestrial atmosphere. 
Working within the framework of the above model, we have been able to obtain the expected 
charge and mass range \footnote{These values compare reasonably well with the few experimental 
values quoted in Tab.~\ref{tab:exotica}} with which these objects can arrive on the surface 
of the Earth. 
Within the scope of these chapters, we have also discussed some of the possible ground based 
experimental set-ups which can be decisive about the nature of the exotic cosmic ray events 
of the above type. In this chapter we, therefore try to provide an estimate of the expected 
flux of the very small strangelets which can be intercepted by similar ground-based experiments. 
As a further consequence, we also try to estimate the relative abundance of the accumulated strangelets
in the Earth's crust. 

\section{\label{sec:galflux}Flux of galactic strangelets}
In Chap.~\ref{chap:intro} we have indicated various ways by which small lumps of SQM may be formed.
These include both galactic as well as extragalactic sources. However, most of the strangelets 
are expected to arrive from the local, galactic source of dark matter, whose density 
\( \rho \sim 10^{-24} \) gm / cm$^3$. From this, one can obtain an upper limit of the flux in the 
following way \cite{jholden@4}.  Assuming that all the dark matter consists of (spherical) strangelets of a 
certain radius \( r_s \), the number density \( n \) of such strangelets will be about 
\[ n = \frac{3}{4\pi r_s^3} \frac{\rho}{\rho_n} \] where \( \rho_n \) is the typical density of nuclear matter objects. 
These objects move about randomly with a speed \( v \) determined by the depth of the gravitational 
potential of the galaxy, which is  about \( 10^7 \) cm/sec. This results in a current 
density \( j \sim n v \) \footnote{The expression for isotropic current density is actually \( 1/4\ n v \), but dropping 
the numerical prefactor should not introduce any significant errors, which are already quite large, in view 
of the large approximations in effect.} 
 of strangelets in all possible directions. From this information 
one can find the number of such events expected during a year which may be registered 
by a passive detector array of a given size set up at mountain altitudes. In 
table ~\ref{tab:flux} we list the event rate corresponding to strangelets of various 
sizes according to this scheme.  

\vspace{0.30cm}
\begin{table}
{\centering \begin{tabular}{|c|c|c|c|c|}
\hline
\( r \) & \( A \) & \( n \) & Flux & Flux$_{400}$   \\
\hline\hline
.01 & \( 5.8\times 10^{32} \)  & \( 1.03\times 10^{-33} \) & \( 1.03\times 10^{-26} \) & \( 1.29\times 10^{-12} \)  \\
10  & \( 5.78\times 10^{41} \) & \( 1.03\times 10^{-42} \) & \( 1.03\times 10^{-35} \) & \( 1.29\times 10^{-21} \)  \\
\( 4.8\times 10^{-13} \) & 64 & \( 9.35\times 10^{-3} \)  & \( 9.3 \times 10^{3} \) & \( 1.17 \times 10^{18} \) \\
\( 5.56\times 10^{5}  \) & \( 1.0 \times 10^{56} \) & \( 5.9 \times 10^{-57} \) & \( 5.9 \times 10^{-50} \) & \( 7.5 \times 10^{-36} \) \\
\hline 
\end{tabular}\par}
\caption{\label{tab:flux} Estimated flux of strangelets. Column 1, radius of strangelets in 
cm, column 2, mass number, column 3, $n$ - the number density of strangelets, column 4, Flux of strangelets 
per cm$^2$ per second, column 5, Flux on 400 m$^2$ per year.}
\end{table}
\vspace{0.30cm}

The procedure illustrated in table~\ref{tab:flux}  yields an extreme overestimate for very small 
strangelets, since, in that case, the whole local dark matter density in the galaxy gets assigned to 
very small strangelets (e.g row 3, column 2) of the same size. On the contrary, it is a well known 
fact (first mentioned in Sec.~\ref{ssec:sqexpt}, item 4) that the primary constituents of the dark halo 
of the Milky Way are objects in the mass range of \( \sim \) \( M_{\odot} \), as indicated by 
microlensing experiments. In a subsequent chapter \ref{chap:macho} we propose that there are reasons to 
believe that entire dark halo is composed of quasibaryonic lenses 
of \( A \sim 10^{55-56} \). If this is true, then the very small strangelets most probably originate from 
the collisions of these dark lenses (Sec.~\ref{ssec:sqexpt}, item 3). In this section we present two 
estimates of the strangelet flux based on the above hypothesis. 

The first estimate is based on the values of flux already given in table.~\ref{tab:flux}, but using a modified
version of the procedure given above. If one envisages the collisions between 
two dark compact objects as being similar to two colliding nuclei, then a rough estimate of how much of the ejecta results in small strangelets can be obtained by using the fact that in a multifragmentation process \cite{stock,cwill,mashen} 
the mass yield approximately obeys a power law behavior \( \propto A^{-\tau} \) with \(\tau \sim 2/3 \). From row 4 of 
table.~\ref{tab:flux} we read that the flux for objects of mass \( M_{\odot} \) is \( \sim 7.5 \times 10^{-36} \) for 
a 400 m$^2$ of detector area. However if we assume that this flux is the outcome of a ``reaction'' product of 
the collision of two objects in the Solar mass range, then the number for small fragments of a given mass number 
\( A \) is this number, enhanced by a factor \( \left(\frac{A}{1.0\times 10^{56}}\right)^{-2/3}\). For the typical 
mass of the strangelet assumed in chapter \ref{chap:newmodel}, the factor comes out to be \( \sim 5.9\times 10^{35} \),
so that we get a flux of about 4 - 10 strangelets per year on a 400 m$^2$ passive detector layout. 

The second estimate essentially borrows on the idea given in \cite{madsen}, in which the authors estimated 
the background flux obtainable from stellar collisions. Here the principal assumption is based on the observation 
that since several pulsars are members of binary systems, the two components of a binary are ultimately going to collide. 
If such collisions spread as little as 0.1 \( M_\odot \) of non relativistic strangelets with baryon number A, the 
number of strangelets released in a single collision will be \[ N = \frac{0.1 M_\odot}{A m_n} \] where, \( m_n \) is 
the nucleon mass. If such objects are distributed homogeneously over a halo of radius \( R_h \sim 10 \) kpc, the 
number of particles flowing out isotropically per unit time, per unit area normal to the flow direction will be given by 
\( \frac{3 N}{4\pi R_h^3} v \), where v is the flow velocity. The flux, per unit solid angle is then,
\[ F = \frac{3}{16\pi^2} \frac{0.1 M_\odot}{R_h^3 m_n} A^{-1} v \] or about \( 10^{-6} A^{-1} v_{250} \) 
cm$^{-2}$s$^{-1}$sterad$^{-1}$, where \(v_{250}\) is the speed measured in units of 
250 km s$^-1$, the typical speed of SQM in the galactic halo. The number of binary mergers in the galaxy 
have been estimated \cite{curran} by a detailed computer simulation and can be quoted as 
\( \mathscr{R} \sim 10^{-6} \mathrm{yr}^{-1} \). Thus if one binary coalescence occurs per \(10^6\) years, 
but the disintegration products scatter over a region of radius 10 kpc, then we need to calculate the probability
that one such event takes place within 10 kpc of the Earth. Since there would have been \( \sim  10^3 \) such events 
since the formation of the Milky Way Galaxy, the required probability is 
\[ \mathscr{P} = 10^3 \times 
\frac{\frac{4\pi}{3} R_h^3 }{4 \pi \times R_h \times R^2} \sim 10^{-3}\]
as the Milky Way Galaxy has the shape of a very flattened spheroid of major radius \( R \sim 10\; \mathrm{Mpc} \)
and minor radius \( R_h \lesssim 10\; \mathrm{kpc} \). Then, the anticipated flux of strangelets of \(A \sim 64 \) would be 
\mbox{\( F_{A=64} \times \mathscr{P} \)} or about 5 /  \(( \mathrm{m}^2\; \mathrm{Yr}\; \mathrm{sterad} )\). The upper limit of flux on a 400 sq.m 
detector area, according to this estimate is as Flux$_{400}$ \( \sim 10^2 \) per year. The actual expectation 
should be much lower, in view of the fact that the composition of the ejecta will necessarily 
accommodate strangelets of varying sizes.

\section{\label{sec:relab}Relative abundance of strangelets on the Earth's crust}

In this section we try to estimate the expected relative abundance 
of strangelets on the Earth's surface. In an earlier chapter (Chap~\ref{chap:newmodel}) we have presented
a scheme of propagation of such strangelets through the terrestrial atmosphere
and indicated the expected charge and mass with which these objects can arrive
on the earth's surface. The essential difference between the propagation of
such strange quark balls and normal cosmic rays is the fact that the strangelets
can absorb nucleons (mainly neutrons) and become more stable when they interact
with the atmospheric nuclei. In the course of propagation they lose energy both
due to these collisions and the ionization loss (the strangelets have a small
positive charge to start with, and they pick up some more from the protons in the
initial phases of their journey through collisions with atmospheric nuclei) 
and as a result come down with very small velocities and land on the crust of the 
earth almost like a parachute. Starting with a mass of \( \sim 64 \) they might end 
up with a mass \( \sim 350 \) and charge 14 (See Chap.~\ref{chap:newmodel}, Tab.~\ref{tab:compvalues}). 
It appears that the only way that they can propagate down
the Earth's surface is by the help of water percolation. In this section we make
some not too unreasonable assumptions in order to estimate the relative abundance
of such elements with respect to Silicon, since the charges of the strangelets
are similar to it (Z = 14)\footnote {And hence might be separable using very high precision isotope
separation methods.}, and also because Si is the most abundant material on the Earth's 
crust. 

With the flux values calculated in Sec.~\ref{sec:galflux}, the total number of particles per 
unit area, \( N_{s} \), that have accumulated on the earth's crust, since the time of formation of the 
atmosphere ( which we take to be \( 4.4\times 10^{9} \) years) can be calculated. Out of these particles a 
smaller number can actually percolate within in a rectangular box of area 1 sq. m and depth \( \sim \)10m (the
depth up to which the water can come down) per year. Here we assume that a fraction of the particles
\( y \sim 1 - 10^{-3} \) which have been deposited over a unit area have actually come down through water percolation 
in this box.  The total mass contributed by the strangelets in this way to the box mass, \( m_{s} \)
is then \( N_s A_s m_n y \) , where \( m_n \) is the average nucleon mass and \( A_s \sim 350 \) is
the mass number of the strangelet (we use SI units in these estimates). The mass contributed by Silicon, 
\( m_{Si} \) is roughly \( V \rho _{e}\times r_{Si} \) , where \( \rho _{e} \) is the mean density of 
Earth \cite{earthdata}, \( V \) is the volume of the box and \( r_{Si} \) is the relative abundance 
of Silicon ( 27.7 \%). 
The relative abundance calculated for strangelets (\( r=\frac{m_{s}}{m_{s}+m_{Si}} \)) 
calculated in this way, come out roughly near \( 10^{-19} - 10^{-15} \).

\begin{figure}[p]
\begin{center}
\scalebox{0.6}{\includegraphics{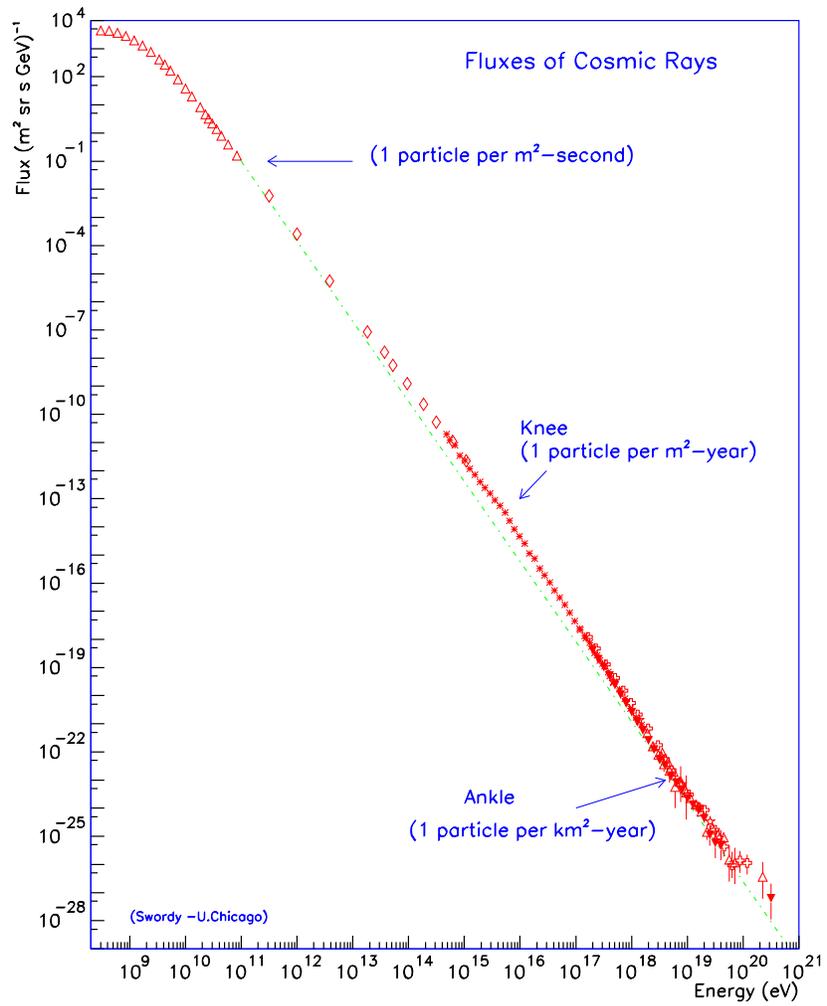}}
\end{center}
\caption{\label{fig:cosmicflux} Flux of cosmic rays at the same total energy
of the incident particles \cite{groom}} 
\end{figure}

In order to compare the values of relative abundances obtained above, we refer to the events analyzed by 
Saito et.al \cite{saitogul} (see also the discussion at the beginning of Sec.~\ref{sec:exptdust}). 
These events have a relative abundance of \( 2.1\times 10^{-5} \) with respect to the total number of 
normal cosmic ray particles observed at the same total energy(including rest energy). Since the total 
energy of the strangelets
lie somewhere in the \( 100 \) to \( 1000 \) GeV (corresponding to \( \beta =0.2-0.7 \))
range we accept the corresponding flux window between \( 10^{-3} \) to \( 10 \)
particles per \( m^{2} \) per steradian per sec per GeV in the primary cosmic ray spectrum (Fig.~\ref{fig:cosmicflux}).
Thereafter, following the same methods as given above, the relative abundance turns out to be 
\( \sim 4.7 \times 10^{-16) - 10^{-9}} \). Thus in this case, the expected relative abundance is considerably 
higher compared to our earlier estimate.

\chapter{\label{chap:chandra}The Chandrasekhar limit for quark stars}
\section{\label{sec:maxmass} Maximum mass of compact objects}
In this chapter we address the issue of the existence of a possible maximum mass limit for quark stars 
from an analytical standpoint. In the first half of the chapter we compare the state of the 
problem for different classes of compact objects with respect to the above limit. 
In the other half we illustrate the analytic procedure leading to an evaluation of the 
maximum mass limit, and finally comment on the nature of the results obtained.

The Quark stars, if they exist, belong to a family of compact objects like the white dwarfs and 
neutron stars. These objects have been a topic of interest for several decades. Compact
objects like the ones mentioned above are produced as the end product of the stellar 
evolution, \emph {i.e.}, when the nuclear fuel of the normal stars has been consumed.
The factor which decides whether a star ends up as a white dwarf, neutron star or a black
hole is primarily the star's mass. White dwarfs originate from light stars with masses 
\( M \lesssim 4 M_\odot \). They no longer burn nuclear fuel, but cool off gradually, 
radiating away the last bit their of thermal energy. The decreasing total energy
leads to a  gradual contraction of the star\footnote{The contraction is a must for 
any gas with a polytropic index \( \Gamma > 4/3 \)} to the point at which the 
density increases so much that the breakdown of Maxwell-Boltzmann relations give way 
to a degenerate electron gas which can exert pressure even at zero temperature 
by the virtue of Pauli exclusion principle. The pressure of the Fermi gas of electrons
acts against the contraction of the star and might be able to bring it to a halt. 
The existence of a maximum mass limit for these objects can be inferred qualitatively (after Landau) 
as follows. 

The total energy at the equilibrium point consists of a positive Fermi energy part which 
\( \sim N^{1/3} \) and a negative contribution due to the gravitational energy  
( \( \sim N \) ), both of which scales as \( 1/R \). The much weaker dependence of the 
Fermi energy on \( N \) relative to the gravitational energy term is in fact the key 
factor responsible for the existence of a maximum mass for this class of objects. 
This implies that although stable equilibrium is possible for small values of N,
larger values will make the gravitational term dominate, making the net energy negative
which increases toward zero as \( R \rightarrow \infty \). Thus, beyond a critical value 
of the mass the star cannot escape the fate of a gravitational collapse. 

Neutron stars originate from stars more massive than the progenitors for the white 
dwarfs. At very high densities the electrons react with the protons to form neutrons 
via inverse beta decay. The incorporation of such effects in white dwarf matter leads 
to an instability which settles down to a stable configuration only when almost all the 
protons and electrons squeeze together to form neutrons and the system can be once 
more supported against the huge gravitationally generated inward pressure by the 
pressure due to degenerate neutron gas. 

Applying essentially similar qualitative arguments one forsees a maximum mass limit for
the neutron stars also and both the collapsed star types turn out to have a maximum mass 
( \( \sim 1.5 M_\odot \) ), beyond which they collapse to black holes 
\cite{chandra,landau,shapiro}. 

In a pioneering work in 1926, S. Chandrasekhar identified the pressure which 
holds up white dwarfs with the electron degeneracy pressure.  Actual white 
dwarf models, taking the effects of relativistic speed of electrons in the 
degenerate electron equation of state were constructed by him in 1930 \cite{chanda,chandb}.
In the course of these studies, Chandrasekhar made the very important discovery
that the maximum mass limit of white dwarf stars have to be \( \sim 1.4 M\odot \),
the exact value depending on the composition of the stellar matter. This maximum mass
limit is called the \emph{Chandrasekhar Limit} in honor of its discoverer. In 1932,
L.D. Landau \cite{landau} presented an elementary explanation of the Chandrasekhar
limit and applied his arguments in a similar manner several months later 
to neutron stars.

Although the Chandrasekhar limit refers strictly to 
white dwarfs, the limiting mass for neutron stars is also loosely called the 
Chandrasekhar limit, primarily because the limits in the two cases turn out to be the
same \cite{landau}; the sizes are however vastly different (white dwarfs have 
\( R \sim 10^{-2}R_\odot \)
\footnote{ where, $R_\odot$ is the Solar Radius, $\sim 6.96 \times 10^{10} cm $ }   
whereas for neutron stars \( R \sim 10^{-5}R_\odot \) \cite{shapiro}, Tab.1.1). 
The maximum mass for the white dwarfs depend essentially on the fundamental constants while
the maximum mass for neutron stars is a sensitive function of the yet-unknown equation
of state of nuclear matter and requires the solution of the TOV equation for 
relativistic hydrodynamics along with an assumed equation of state.

The neutron star, so formed, is essentially a huge nucleus and the nucleons (mainly neutrons) 
which make it up can undergo a hadron - quark phase transition at high density and/or temperature
(see sec.~\ref{sec:struniv}). This is likely, since the central densities in the neutron stars are 
high enough to favour such a transition. In this phase transition, the individual hadronic 
boundaries dissolve and the quarks get trapped within a larger bag whose radius coincides
approximately with the star radius and the neutron star becomes a quark star in the process
\cite{baym,alcock,qsrev}. We have already mentioned that the central hypothesis of the 
present thesis is based on the suggestion of Witten \cite{witten@c} that strange quark matter may be 
the true ground state of the strongly interacting matter. In this circumstance, quark stars, if they are formed, would 
preferably convert to strange stars, comprising {\it u, d} and {\it s} quarks, under weak interaction. 
Several authors ( for example, \cite{jpg,ghosh1} ) have used different models to understand 
the properties of strange stars. For a review, see \cite{qsrev}. For such quark stars, the maximum
mass would indeed be almost the same as that for neutron stars.

In contrast to other compact objects, the strange stars need not be the direct product of stellar evolution. 
This is an unique property which distinguishes them from other compact objects, e.g it is conceivable that 
if a large amount of quark matter exists in the universe as a relic of the cosmological quark-hadron 
phase transition \cite{apj}, it could clump under gravitational interaction (see Chap.~\ref{chap:macho})
and even form invisible quark galaxies \cite{sweden}. The `stars' of such a galaxy would
be strange stars which do not evolve from neutron stars and thus are not governed by the Chandrasekhar 
limit for neutron stars. It is, therefore, highly relevant to inquire if there exists, just like the case of 
ordinary compact stars, an upper limit on the strange stars beyond which they would be gravitationally unstable. 
In the past, the problem has been approached numerically. Starting with the seminal work of Witten
\cite{witten@c}, most authors have concentrated on solving the Tolman-Oppenheimer-Volkov 
(TOV) equation (see, for example, \cite{shapiro}) for the quark matter equation of state. 
While the results show that there does exist a limiting mass for quark stars
(which is very close to that for neutron stars), there is no \textit{a priori}
argument to prove that such a limit should exist or that it should
depend mostly on fundamental constants, as is the case for the ordinary
compact stars \cite{shapiro}. In this chapter we show analytically, \textit{from first principles}, 
that such a limit exists for compact quark stars and that it is mostly determined by
universal constants. 

\section{\label{sec:maxcalc} The Analytical Form of the maximum mass limit}  

In the following treatment we essentially follow Landau \cite{landau} and
apply a general and simple picture of energy balance to a system
of \em massless \em quarks, confined in a large bag \cite{baym} characterized
by a constant energy density \( B \). The above is adopted as a working model 
of a strange star for the present purpose.

As in the case of white dwarfs and neutron stars, the equilibrium should occur 
at a minimum of the total energy per fermion \( e \), where \( e \equiv e_{F}+e_{G} \), 
in which \( e_{F} \) is the Fermi energy  and \( e_{G} \) is the gravitational energy per fermion.
There is a crucial difference in the way the (Newtonian) gravitational energy
can be estimated for quark stars and the ordinary compact stars. The Newtonian 
gravitational energy is a macroscopic quantity, and for ordinary compact stars, this
mass is due almost entirely to the baryons. For quark stars, however, one needs to
identify the total mass as the total ( thermodynamic as well as the confining ) energy in
the star. In order to estimate the gravitational energy per fermion one needs a  
prescription for incorporating both contributions into an effective quark mass. 

A suitable framework capable of handling this issue was formulated quite some time ago \cite{dden} 
in which the authors (The model is briefly discussed in Sec.~\ref{ssec:sqsec}) proposed a dynamical 
model of confinement in a many body system of quarks. This was,  
in turn, motivated by an earlier work by Pati and Salam \cite{patisalam} who 
pictured confinement as the quark having a small mass inside a hadron and 
a very large mass outside. The standard description of 
confinement is provided by the bag model \cite{mit}, which implies that, 
for a many body system, the energy energy density inside the bag, for small 
total quark number density, differs by a positive constant (\( B \)) from that 
of the true vacuum outside. The QMD model, inspired by the \emph{Archimedian
principle} advocated in \cite{patisalam} parametrized confinement through a 
density dependent quark mass which varied so as to agree with the bag model 
limit of constant energy density, i.e 
\begin{equation}
\label{qmdmass}
m_q \sim \frac{B}{n_q}\: \text{as} \: n_q \rightarrow 0
\end{equation}  
where \(m_q\) is the \emph{effective} quark mass and \(n_q\) is the total
quark number density. Thus the mechanism of confinement is mimicked through 
the requirement that the mass of an isolated quark becomes infinitely large so
that the vacuum is unable to support it. The picture given in eqn.~\ref{qmdmass} 
then tells us that for a system of quarks at zero temperature the energy density tends 
to a constant value while the mass tends to infinity as the volume tends to 
infinity or the density tends to zero.

The number density of fermions is related to the chemical potential as
\[ n=\frac{g}{6\pi ^{2}}\mu ^{3}\]
which dictates that 

\begin{equation}
\label{mu-n}
\mu =(\frac{9\pi }{2g})^{\frac{1}{3}}\frac{N^{\frac{1}{3}}}{R}
\end{equation}

In the above relations, \( n \) is the number density, \( N \) the total
number of fermions in a star of radius \( R \), \( g \) the statistical
degeneracy factor and \( \mu  \) is the chemical potential.

The fermion energy density is given by

\begin{equation}
\label{eden}
\varepsilon _{F}=\frac{g}{8\pi ^{2}}\mu ^{4}
\end{equation}

and hence the Fermi energy per particle of the quarks becomes 

\begin{equation}
\label{ef}
e_{F}=\frac{\varepsilon _{F}}{n}
=\frac{3}{4}(\frac{9\pi }{2g})^{\frac{1}{3}}\frac{N^{\frac{1}{3}}}{R}
\end{equation}

The mass \( M \) of the star can be written in terms of \( N \) and
\( B \) (the bag constant), if the density
\( \rho (r) \) in the star is assumed to be roughly constant throughout
the volume of the star. Hence using eq.(\ref{ef}),

\begin{equation}
\label{Mass1}
M=\int _{0}^{R}4\pi r^{2}\rho (r)\, dr
=\frac{4}{3}\pi R^{3} B + e_{F} N
=\frac{3}{4}(\frac{9\pi }{2g})^{\frac{1}{3}}\frac{N^{\frac{4}{3}}}{R}
+\frac{4}{3}\pi BR^{3}
\end{equation}

Extremising the mass \( M \) (eq. \ref{Mass1}) with respect to \( R \) gives,

\begin{equation}
\label{Ef=3b}
(\frac{9\pi }{2g})^{\frac{1}{3}}\frac{N^{\frac{4}{3}}}{R^4}
=\frac{16}{3}\pi B     \nonumber \\
\Rightarrow \\
\varepsilon _{F}=3 B
\end{equation}

Substituting eq.(\ref{Ef=3b}) in the expression for \( M \)
(eq. \ref{Mass1}),
\begin{equation}
\label{Mb}
M = 4 B V = \frac{16}{3} \pi B R^3
\end{equation}

We note that this is very similar to the condition obtained for hadronic bags
\cite{mit}. As a next step we find the \( R \) for which the
total energy per fermion would be maximum.

The gravitational energy per fermion \( e_{G} \) is

\begin{equation}
\label{eg1}
e_{G}=-\frac{GM{m_{eff}}}{R}
\end{equation}
where \( m_{eff} \) is the effective quark mass inside the star. Assuming
that the effective quark mass contributes to the total star mass \( M \),
one can write for a strange star with \( N \) quarks,
\begin{equation}
\label{Mm}
M = N m_{eff} \Rightarrow m_{eff} = \frac{4 B}{n}
\end{equation}

As mentioned above, the effect of confinement in a quark matter system was
shown \cite{dden} to be incorporable in the effective quark mass, which, the
quarks being fermions, coincides with the quark chemical potential . As a
result, one gets, in the limit of vanishing quark density \cite{dden},
\[ \mu = \frac{B}{n} \] .
This, together with the eq.(\ref{Mm}), gives
\begin{equation}
\label{qmas}
m_{eff} = 4 \mu
\end{equation}
where all the energy ( thermodynamic and confining ) is included in the
effective gravitational mass of the quarks inside the strange star.

Using equations (\ref{eg1}), (\ref{Mm}) and (\ref{qmas}) we get 

\begin{equation}
\label{eg2}
e_{G}=-\frac{64}{3}(\frac{9\pi }{2g})^{\frac{1}{3}}G \pi B R {N^{\frac{1}{3}}}
\end{equation}

Minimising the total energy \( e=e_{F}+e_{G} \) with respect to \( N \), we
get the expression for maximum value of \( R \) as

\begin{equation}
\label{rmax}
R_{max}=\frac{3}{16}\frac{1}{\sqrt{\pi GB}}
\end{equation}

It may be observed, that in this case the \( N \) dependence of both \( e_F \), eqn.~\ref{ef}
and \( e_G \), eqn.~\ref{eg2} are 
the same, while the \( R \) dependence is different for the two terms.
Finally, the maximum mass of the strange star is computed by substituting the
value of \( R_{max} \) (from equation \ref{rmax}) in equation ( \ref{Mb}).

\begin{equation}
\label{Mmax}
M_{max}=\frac{16}{3}{\pi B {R_{max}}^3}
\end{equation}

The chemical potential \( \mu \) can be evaluated in terms of \( B \)
using equations (\ref{eden}) and (\ref{Ef=3b}). Substituting this in
eq.(\ref{Mm}) gives the value of \( N_{max} \). The values of \( R_{max} \),
\( M_{max} \) and \( N_{max} \) are tabulated below (Table \ref{tab:qstar}) for various values of
the Bag constant \( B \).

\vspace{0.30cm}
\begin{table}
{\centering \begin{tabular}{|c|c|c|c|}
\hline 
\( B^{1/4}\, (MeV) \)&\( R_{max}\, (Km) \)&\( \frac{M_{max}}{M_{\odot }} \)&\( N_{max} \) \\
\hline 
\hline 
145&12.11&1.54&1.55 \( \times \) 10$^{57}$\\
\hline 
200&6.36&0.81&5.90 \( \times \) 10$^{56}$\\
\hline 
245&4.24&0.54&3.21 \( \times \) 10$^{56}$\\
\hline 
\end{tabular}\par}
\caption{\label{tab:qstar} Computed values of the maximum Mass, Radius and 
Baryon number for quark stars for several values of the Bag constant}
\end{table}
\vspace{0.30cm}

Thus we have demonstrated that there exists a limiting mass (the 
so called Chandrasekhar limit) for compact quark stars, beyond which they would be
gravitationally unstable. As with other classes of compact objects, the maximum
mass depends mostly on universal constants ( \( G \) as well as \( \hbar \) and \( c \), which do not
occur explicitly due to our use of the naturalised units ) and on the bag energy \( B \).
The bag energy is treated as a parameter here, but it is often regarded as a universal 
constant in its own right, since it represents the difference between the non-perturbative 
and perturbative vacua of Quantum Chromodynamics. It can be seen from the above equations 
(\ref{rmax}, \ref{Mmax}) that the physical radius \( R_{max} \), corresponding
to the maximum mass as well as the maximum mass itself, are independent of
the number of quark flavors. Although \( N_{max} \) depends on the statistical 
degeneracy factor \( g \) ( or equivalently, the number of flavors), the dependence
is extremely weak, as can be readily checked from equation (\ref{mu-n}). In fact, we have
verified that there is almost no difference in \( N_{max} \) between the
cases with 2 or 3 flavor quark matter. This, in turn, implies that the assumption of
massless quarks ( even for \( s \) quarks ) does not materially affect these results. While it 
is true that the methods applied in this chapter are pedestrian in nature, the
limits agree well with those found with the help of the numerical solutions of
the TOV equation (see, for example, \cite{witten@c} ). Although, in this work, we have adopted the
simplifying assumption of a constant density profile for the quark star in order  
to have a simple analytic solution, we too get the characteristic scaling behavior \cite{witten@c}
( \( R_{max} \propto B^{-1/2}, \, M_{max} \propto B^{-1/2} \) ) obtained from 
detailed numerical solutions. This show that the simple picture presented here adequately incorporates 
the essential physics of the structure of the quark stars.

\chapter{\label{chap:macho}MACHOs as quasibaryonic dark matter}
\section{\label{sec:cdminhalo} CDM Objects in the halo}
One of the mysteries that persist in the standard cosmological model is the nature of 
dark matter. It has long been conjectured that we live in a nearly critical density 
universe, although there had been no evidence of the required accumulation of matter 
through observations based on the  spectrum. However indirect 
evidences (see \ref{ssec:dm}) do suggest that there is an abundance of matter in 
the universe which is non-luminous, since it either does not interact, or does 
so very weakly, with the all forms of matter, except through gravitational interaction. 
The present consensus (for a short discussion, see \ref{ssec:dm}; for an extended review, 
see \cite{turner1@6,turner2@6}) based on recent experimental data is that 
the universe is flat and that a sizable amount of the dark matter is "cold", 
{\it i.e.} nonrelativistic, at the time of decoupling. For example, The WMAP 
survey data \cite{wmap@6} partitions the total matter-energy content of the 
universe roughly as 73\% smooth dark energy,23\% cold dark matter leaving the 
rest 4\% to luminous matter which goes into the making of bright galaxies and stars.

The nature of the 23\% cold non-luminous matter continues to be a mystery, at 
least within the standard framework of particle interactions, mainly because the required 
accumulation of baryons is clearly unaccountable by existing and extremely 
reliable data on the nucleosynthesis event responsible for the generation of nuclear 
matter. Most of the proposed dark matter candidates, therefore, rely on an extrapolated 
particle interaction model often venturing far into exotic domains. In recent years, 
there has been experimental evidence \cite{alcock1@6,aubourg1@6} for at least 
one form of dark matter - the Massive Astrophysical Compact Halo Objects (MACHO) - 
detected through gravitational microlensing effects proposed by Paczynski \cite{pacz@6} 
some years ago. As of now, there is no clear picture as to what these objects are made of 
in spite of a lot of efforts spent in studying them. Based on about 13 - 17 Milky Way 
halo MACHOs detected in the direction of LMC - the Large Magellanic Cloud (we are 
not considering the events found toward the galactic bulge), the MACHOs are expected to 
be in the mass range (0.15-0.95) \( M_{\odot} \), with the most probable mass being 
in the vicinity of 0.5 \( M_{\odot} \) \cite{sutherland@6,alcock2@6}, substantially 
higher than the fusion threshold of 0.08 \( M_{\odot} \). The MACHO collaboration 
suggests that the lenses are in the galactic halo. Assuming that they are 
subject to the limit on the total baryon number imposed by the 
Big Bang Nucleosynthesis (BBN), there have been suggestions that 
they could be white dwarfs \cite{fields@6,freese@6}. 
It is difficult to reconcile this with the absence of sufficient 
active progenitors of appropriate masses in the galactic halo. Moreover, 
recent studies have shown that these objects are unlikely to be 
white dwarfs, even if they were as faint as blue dwarfs, since this 
will violate some of the very well known results of BBN \cite{freese@6}.
There have also been suggestions \cite{schramm@6,jedam1@6,jedam2@6} that they could be 
primordial black holes (PBHs) ( \( \sim \) 1 \( M_{\odot} \) ), 
arising from horizon scale fluctuations triggered by pre-existing 
density fluctuations during the cosmic quark-hadron phase transition. 
The problem with this suggestion is that the density contrast 
necessary for the formation of PBH is much larger than the 
pre-existing density contrast obtained from the common 
inflationary scenarios. The enhancement contributed by the QCD phase 
transition is not large enough for this purpose. As a result a 
fine tuning of the initial density contrast becomes essential which 
may still not be good enough to produce cosmologically relevant amount 
of PBH \cite{schmid@6}. Alternately, Evans, Gyuk \& Turner \cite{evans@6}
suggested that some of the lenses are stars in the Milky Way disk
which lie along the line of sight to the LMC. 
Gyuk \& Gates \cite{gyuk@6} examined a thick disk model, 
which would lower the lens mass estimate. Aubourg et al. 
\cite{aubourg2@6}  suggested that the events could arise from 
self-lensing of the LMC. Zaritsky \& Lin \cite{zaritsky@6} have argued
that the lenses are probably the evidence of a tidal tail arising from the
interaction of LMC and the Milky Way or even a LMC-SMC (Small Magellanic
Cloud) interaction. These explanations are primarily motivated by the
difficulty of reconciling the existence of MACHOs with the known 
populations of low mass stars in the galactic disks.

\section{\label{sec:stablenugs} Stability of quark nuggets}
In this chapter we accept the standpoint that the lensing MACHOs are indeed in the 
Milky Way halo and propose a theory which relates them to the quark nuggets 
which could have been formed in a first order cosmic quark - hadron phase
transition, at a temperature of  \( \sim \) 100 MeV during the microsecond era
of the early universe. In our picture, the MACHOS evolved out of these 
primordial quark nuggets. A few statements on the aforesaid quark-hadron phase 
transition may not be irrelevant here. 
The order of \textit{any} phase transition carries the most significant bit 
of information about the phenomenon; however the order of the deconfinement phase 
transition is an unsettled issue till now. It is generally believed that a 
true second order phase transition is inconceivable in cosmological scenarios
since nature does not provide an exact chiral symmetry \cite{witten@6}. In a pure ({\it i.e.} only gluons) 
\( SU(3) \) gauge theory, the phase transition is of first order, as suggested 
by Lattice gauge theory. However, there exist no unequivocal approaches 
in the case when dynamical quarks are also present on the lattice, and instead
of studying the deconfinement transition, one investigates the chiral transition. 
Although these two phase transitions are commonly treated to be equivalent,  
there is no definite reason why they should be simultaneous or of the same 
order \cite{alam1@6}. The order of the chiral phase transition depends critically
on the strange quark mass. Although the chiral phase transition is probably of 
first order for large strange quark mass, it may be of second order for 
lighter strange quarks. The situation remains controversial since the strange quark 
mass is of the order of the QCD scale \cite{blazo@6}. In addition, the finite size 
effects of the lattice may tend to mask the true order of the transition. We are
however, more concerned about the deconfinement transition and if it is really
of the first order, the masking effect associated with it would be negligibly 
small in the early universe. In such circumstances, Witten (1984) argued, in a 
seminal paper \cite{witten@6}, that strange quark matter could be the {\it true} 
ground state of {\it Quantum Chromodynamics} (QCD) and that a substantial amount of 
baryon number could be trapped in the quark phase which could evolve into strange 
quark nuggets (SQNs) through weak interactions. (For a brief review of the formation 
of SQNs, see Alam, Raha \& Sinha \cite{alam3@6}.) 
At this point, the most important question is whether the nuggets, so formed 
can be stable on cosmological time scales. The first study on this issue was 
addressed by Alcock and Farhi in 1985 \cite{alkockfarhi@6}. They argued 
that a SQN can evaporate neutrons from its surface at \( T \geq I_N \) where 
\( I_N \sim 20 - 80 \) MeV is the binding energy per baryon in SQM at zero
temperature from the neutron mass. According to their calculations 
QN's must have a baryon number in excess of \( 10^{51-53} \) in order to 
survive on cosmological time scales.  This number is larger by a few orders pf magnitude
compared to the total baryon number of the universe at the 
aforesaid temperature and hence they concluded that 
it was apparently impossible to have any QN surviving till the present time. This 
conclusion was reexamined by Madsen et al. in 1986 \cite{madsen86@6}, who pointed
out that the neutron evaporation was a surface process in which the surface of a QN 
got gradually depleted of u and d quarks, as more and more neutrons were eliminated. 
In this process the flavor chemical equilibrium between the u,d and s quarks on the surface 
was lost and further evaporation would be suppressed till some s quark converts back to a u and 
d quark or there was some transport of u and d quarks from the core to the surface 
through convective process. Since both these processes are slow enough, the critical 
size of the nuggets which can survive is effectively lowered to \( \sim 10^{46} \). In a 
later work \cite{alcockolinto@6} it was suggested that the nuggets can also annihilate by 
boiling off hadronic bubbles from the bulk of the QN's, but this was shown to have a 
rate  much smaller than that for surface evaporation \cite{madsen89@6}. All of the above studies 
used idealized thermodynamic and binding energy arguments to calculate the baryon evaporation 
rate, with strong assumptions like geometrical cross sections, surface transparency of the QN to baryons
etc. Later, studies using QCD - motivated dynamical models (like chromoelectric flux tube model)  
of baryon evaporation from SQNs have established \cite{pijush1@6,sumi@6} that primordial 
SQNs with baryon numbers above \(\sim\) 10\(^{40 - 42}\) would be cosmologically stable. 

In a previous work  \cite{alam3@6} by Alam, Raha and Sinha, it was
shown that without much fine tuning, these 
stable SQNs could provide even the entire closure density (\(\Omega \sim \) 1) 
\cite{alam3@6}. Thus, the entire cold dark matter (CDM) 
( \( \Omega_{\mathrm{CDM}} \sim \) 0.3-0.35 ) could easily be explained by stable SQNs.
\par
We can estimate the size of the SQNs formed in the {\it first order} cosmic QCD 
transition in the manner prescribed by Kodama, Sasaki and Sato \cite{kodama@6} in 
the context of the GUT phase transition. For the sake of brevity, let us recapitulate 
very briefly the salient points here; for details, please see \cite{alam3@6} and 
\cite{bhatta2@6}.  
Describing the cosmological scale factor \( R \) and the coordinate radius \( X \) in 
the Robertson-Walker metric through the relation 
\begin{eqnarray} 
ds^2 &=& -dt^2 + R^2 dx^2 \nonumber \\
&=& -dt^2 + R^2\{dX^2 + X^2(sin^2 \theta d\phi^2 
+ d\theta^2)\}, 
\end{eqnarray}
one can solve for the evolution of the scale factor \( R(t) \) in the mixed phase of 
the first order transition. In a bubble nucleation description of the QCD transition, 
hadronic matter starts to appear as individual bubbles in the quark-gluon phase. With 
progressing time, they expand, more and more bubbles appear, coalesce and finally, 
when a critical fraction of the total volume is occupied by the hadronic phase, a 
continuous network of hadronic bubbles form (percolation) in which the quark bubbles 
get trapped, eventually evolving to SQNs. 
The time at which the trapping of the false vacuum (quark phase) happens 
is the percolation time \( t_p \), whereas the time when the phase transition starts 
is denoted by \( t_i \). Then, the probability that a 
spherical\footnote{For the QCD bubbles, there is a sizable 
surface tension which would facilitate spherical bubbles.} region of 
co-coordinate radius \( X \) lies entirely within the quark bubbles would obviously 
depend on the nucleation rate of the bubbles as well as the coordinate radius 
\( X(t_p,t_i) \) of bubbles which nucleated at \( t_i \) 
and grew till \( t_p \). For a nucleation rate \( I(t) \), this 
probability \( P(X,t_p) \) is given by 
\begin{equation}
P(X,t_p) = exp \left[-\frac{4 \pi}{3}
\int_{t_i}^{t_p} dt I(t) R^3(t) [X + X(t_p,t_i)]^3\right].
\end{equation}
After some algebra \cite{bhatta2@6}, it can be shown that if all the 
cold dark matter (CDM) 
is believed to arise from SQNs, then their size distribution peaks, for 
reasonable nucleation rates, at baryon number \( \sim \) 10\(^{42 - 44}\), 
evidently in the stable sector. It was also seen that there were almost 
no SQNs with baryon number exceeding 10\(^{46-47}\), comfortably lower 
than the horizon limit of \( \sim 10^{50} \) baryons at that time. 
Since  \(\Omega_{\mathrm{B}}\) is only about 0.04 from BBN, 
\( \Omega_{\mathrm{CDM}} \) in the form of SQNs would correspond to
\( \sim 10^{51} \) baryons so that there should be 
10\(^{7-9}\) such nuggets within the horizon limit at the microsecond 
epoch, just after the QCD phase transition \cite{alam2@6,alam3@6}. We shall 
return to this issue later on.
\par
It is therefore most relevant to investigate the fate of these SQNs.
Since the number distribution of the SQNs is sharply peaked 
\cite{bhatta2@6}, we shall assume, for our present purpose, that all 
the SQNs have the same baryon number. 
\par

\section{\label{sec:nugsstick} Coalescence of the primordial QN's}

The SQNs formed during the cosmic QCD phase transition at 
T \(\sim 100 \) MeV have high masses (\( \sim 10^{44} \)GeV) and 
sizes (\( R_N \sim 1 \)m) compared to the other particles (like 
the usual baryons or leptons) which inhabit this primeval universe. 
These other particles cannot form structures until the temperature 
of the ambient universe falls below a certain 
critical temperature characteristic of such particles; till then, 
they remain in thermodynamic equilibrium with the radiation and 
other species of particles. This characteristic temperature is 
called the freezeout temperature for the corresponding particle.
Obviously the freezeout occurs earlier for massive particles for the
same interaction strength. In the context of 
cosmological expansion of the universe this has important 
implications ; the 'frozen' objects can form structures. These 
structures do not participate in the expansion in the sense that 
the distance between the subparts do not increase with the scale
size and only their number increases due to the cosmological scale 
factor.  
\par
For the SQNs, however, the story is especially interesting. Even if they 
continue to be in kinetic equilibrium due to the radiation pressure 
(photons and neutrinos) acting on them, their velocity would be extremely 
non relativistic. Also their mutual separation would be considerably 
larger than their radii; for example, at T \(\sim 100 \)MeV, the mutual 
separation between the SQNs (of size \( \sim 10^{44} \) baryons) is 
estimated to be around \( \sim 300 \)m. It is then obvious that the 
SQNs do not lend themselves to be treated in a hydrodynamical framework; 
they behave rather like discrete bodies in the background of the 
radiation fluid. Due to their large surface area they experience quite 
substantial radiation pressure, in addition to gravitational forces 
due to the other SQNs.
\par
In such a situation, one might be tempted to assume that since the SQNs 
are distributed sparsely in space and interact only feebly with the other
SQNs through gravitational interaction, they might as well 
remain forever in that state. This, in fact, is quite wrong, as we
demonstrate below.
\par
The fact that the nuggets remain almost static is hardly an issue 
which requires justification. The two kinds of motion that they can 
have are random thermal motion and the 
motion in the gravitational well provided by the other SQNs. This 
other kind of motion is typically estimated using the virial theorem, 
treating the SQNs as a system of particles moving under mutual 
gravitational interaction \cite{bhat@6,peebles@6}. The kinetic energy 
( \( K \) ) and potential
energy \( V \) of the nuggets at temperature T = 100 MeV can be 
estimated as, 
\begin{eqnarray}
K= \frac{3}{2} N k_{b} T  \nonumber \\
V=\sum_{i,j} G\frac{M_{i} M_{j}}{R_{i,j}}=
\frac{GM^{2} N^{2}}{2 R_{av}} 
\end{eqnarray}
where \( k_b \) is the Boltzmann constant, \( M_{i}, M_{j} \) are 
the masses of the ith and jth nugget, \( R_{i,j} \) is the distance 
between them and \( R_{av} \) is the average inter-nugget distance. 
Substituting the number of nuggets \( N = 10^7 \), the baryon number of 
each nugget to be \( 10^{44} \) and \( R_{av} = 300 m \), one gets 
\( K = 2.4 \times 10^{-4} \) and \( V = 3.09 \times 10^{35} \) 
(in MKS units) so that the ratio of \( K \) and \( \frac{V}{2} \) 
becomes \( \sim 10^{-39} \). Thus it is impossible for these 
objects to form stable systems, orbiting round each other. On the other
hand the smallness of the kinetic energy shows that gravitational 
collapse might be a possible fate. 
\par
Such, of course, would not be the case for any other massive 
particles like baryons; their masses being much smaller than SQN, 
the kinetic energy would continue to be very large 
till very low temperatures. More seriously, the Virial theorem can be
applied only to systems whose motion is sustained. 
For SQNs, a notable property is that they become more and more bound 
if they grow in size (see Chap.~\ref{chap:dust}). Thus SQNs would absorb baryons impinging on
them and grow in size. Also, if two SQNs collide, they would naturally 
tend to merge. In all such cases, they would lose kinetic energy,
making the Virial theorem inapplicable. 
\par
One can argue that the mutual interaction between uniformly 
dispersed particles would prevent these particles from forming a 
collapsed structure, but that argument holds only in a static and 
infinite universe, which we know our universe is not. Also a perfectly
uniform distribution of discrete bodies is an unrealistic idealization 
and there must exist some net gravitational attraction on each SQN.
The only agent that can prevent a collapse under this gravitational pull 
is the radiation pressure, and indeed its effect remains 
quite substantial until the drop in the temperature of the ambient 
universe weakens the radiation pressure below a certain critical value. 
In what follows, we try to obtain an estimate for the point of time at 
which this can happen. 
\par
It should be mentioned at this juncture that for the system of discrete 
SQNs suspended in the radiation fluid, a detailed numerical simulation 
would be essential before any definite conclusion about their temporal 
evolution can be arrived at. This is a quite involved problem, 
especially since the number of SQNs within the event horizon, as 
also their mutual separation, keeps increasing with time. Our 
purpose in the present work is to examine whether such an effort 
would indeed be justified.  
\par
Let us now consider the possibility of two nuggets coalescing together 
under gravity, overcoming the radiation pressure. The mean separation 
of these nuggets and hence their gravitational interaction are 
determined by the temperature of the universe. If the entire CDM
comes from SQNs, the total baryon number contained in 
them within the horizon at the QCD transition temperature 
( \( \sim \) 100 MeV) would be  \( \sim 10^{51} \) (see above). 
For SQNs of baryon number \( b_{N} \) each, the number of SQNs 
within the horizon at that time would be just 
\( \left( 10^{51} /b_{N}\right) \). 
Now, in the radiation dominated era the temperature dependence of density 
\( n_{N} \sim T^3 \), horizon volume \( V_{H} \) varies with time as 
\( t^3 \), { \it{i.e.} } \( V _{H} \sim T^{-6} \) and hence the 
variation of the total number inside the horizon volume will be 
\(N_{N} \sim T^{-3} \). So at any later time, the number of SQNs within 
the horizon ( \( N_{N} \) ) and their density ( \(n_{N}\) ) as a 
function of temperature would be given by : 
\begin{eqnarray}
N_{N}(T) & \cong & {\frac{ 10^{51}} {b_{N}}}
\left(\frac{100\mathrm{MeV}}{\mathrm{T}}\right)^3 \label{Nn} \\
n_{N}(T) & = & \frac{N_{N}}{V_{H}}=
\frac{3N_{N}}{4 \pi (2t)^{3}}\label{nn} 
\end{eqnarray}
where the time \( t \) and the temperature \( T \) are related in the 
radiation dominated era by the relation :
\begin{equation}
t = 0.3 g_{*}^{-1/2}\frac {m_{pl}}{T^{2}} \label{tT}
\end{equation}
with \( g_{*} \) being \( \sim \) 17.25 after the QCD transition \cite{alam3@6}.
\par
\begin{figure}[p]
\scalebox{0.6}{\includegraphics{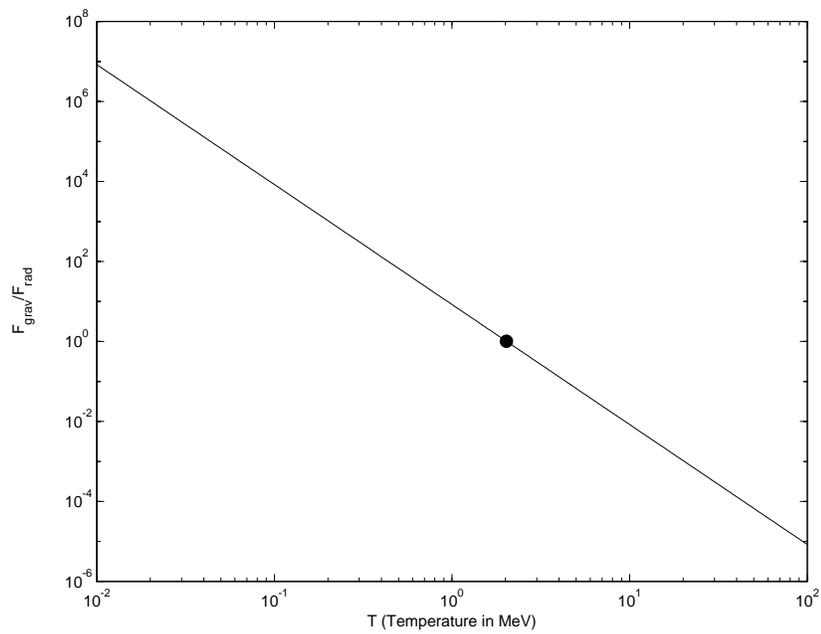}}
\caption{\label{fig:xx} Variation of the ratio 
\( F_{\mathrm{grav}} / F_{\mathrm{rad}} \)
with temperature. The dot represents the point 
where the ratio assumes the value 1.}
\end{figure}

From the above, it is obvious that the density of SQNs decreases as \( 
t^{-3/2} \) so that their mutual separation increases as \( t^{1/2} 
\). Therefore, the force of their mutual gravitational pull will 
decrease as \( t^{-1} \). On the other hand, the force due to the 
radiation pressure (photons and neutrinos) resisting motion under 
gravity would be proportional to the radiation energy density, which 
decreases as \( T^{4} \) or \( t^{-2} \). It is thus reasonable to 
expect that at some time, not too distant, the gravitational pull 
would win over the radiation pressure, causing the SQNs to coalesce 
under their mutual gravitational pull.
The expression for the gravitational force as a function of temperature
T can written as :
\begin{equation}
F_{\mathrm{grav}}=\frac {G {b_N}^2 {m_n}^2} {{\bar{r}_{nn}(T)}^2}
\end{equation}
where \( b_N \) is the baryon number of each SQN and \( m_n \) is the 
baryon mass. \( \bar{r}_{nn}(T) \) is the mean separation between 
two nuggets and is given by the cube root of the ratio \( \kappa \) of 
total volume available and the total number of nuggets. One can roughly
estimate how \( \kappa \) varies with temperature in the following way.
The time-temperature relation eqn.~\ref{tT} can be written in the form 
\[ t = \frac{.8324}{T^2} \]
where t is in seconds and T is the temperature in MeV. Using this,
the horizon radius \( R_H = 2 c t \), expressed in conventional units (m)
is given by \[R_H = \frac{1.665 c}{T^2} \] and the horizon volume 
\(V_H = \frac{19.328\ c^3}{T^6}\). Finally, writing \(b_N \simeq 10^x \)
one can get the following approximate expression for \( n_N \)
\[n_N(T) = 10^{51-x} \left( \frac{100}{T}\right)^3 \] In the following
eqn.~\ref{kappa}, we have substituted the value \( x \rightarrow 44 \) 
(or, \( b_N = 10^{44} \)) in the ratio \( V_H / n_N \) to get an estimate 
for \( \kappa \).

\begin{equation}
\label{kappa}
\kappa = \frac {1.114 \times 10^{-12}c^3}{T^3}
\end{equation}
The force due to the radiation pressure on the nuggets may be roughly
estimated as follows. We consider two objects (of the size of a typical SQN) 
approaching each other due to gravitational interaction, overcoming the 
resistance due to the radiation pressure. The usual isotropic radiation 
pressure is \(  \frac{1}{3} \rho c^2 \), where \( \rho \) is the total
energy density, including all relativistic species. The nuggets will have 
to overcome an additional pressure resisting their mutual motion, which is
given by \( \frac{1}{3} \rho c^2 (\gamma - 1) \); the additional pressure 
arises from a compression of the radiation fluid due to the motion of the 
SQN. The moving SQN would become a oblate spheroid (with its minor axis 
in the direction of motion due to Lorentz contraction), whose surface area 
is given by 
\[ S = 2\pi a^2 + \frac{2\pi a b \sin^{-1} \epsilon}{\epsilon} \]
where a is the length of the major axes perpendicular to the direction of 
motion, b is the length of the smaller axis \& where the ellipticity \( \epsilon \)
is given by \[ \epsilon = \sqrt{1-\frac{b^2}{a^2}} \] In this case 
\( a \rightarrow R_N \) \& \( b \rightarrow R_N / \gamma \), where \( \gamma \) 
is the Lorentz factor corresponding to the moving SQM. With these substitutions
\( \epsilon = \sqrt{\frac{\gamma^2 - 1}{\gamma}} \) and 
\[ S = 2\pi R_N^2 \left( 1 + \frac{\sin^{-1}\epsilon}{\gamma\epsilon} \right) \]
for small values of \( \epsilon \) (small \( \gamma) \), \( sin^{-1} \epsilon \sim \epsilon \), so
that the surface area becomes \(2 \pi R_N^2 \frac{\gamma + 1} { \gamma} \). 
Thus the total radiation force resisting the motion of SQNs is
\begin{equation}
F_{\mathrm{rad}}=\frac{1}{3} \rho_{\mathrm{rad}} c v_{\mathrm{fall}}
(\pi R_N^2) \beta \gamma
\end{equation}
where \( \rho_{\mathrm{rad}} \) is the total energy density at 
temperature \( T \), \( v_{\mathrm{fall}} \) or \(\beta c\) is the
velocity of SQNs determined by mutual gravitational field and \( \gamma \) 
is \( 1 / \sqrt{1-\beta^2} \). The quantities  \( F_{\mathrm{rad}} \),
\( \beta \) and \( \gamma \) all depend on the temperature of the epoch
under consideration. (It is worth mentioning at this point that the \( 
t \) dependence of \( F_{\mathrm{rad}} \) is actually \( t^{-5/2} \), 
sharper than the \( t^{-2} \) estimated above, because of the \( 
v_{\mathrm{fall}} \), which goes as \( t^{-1/4} \).)
The ratio of these two forces is plotted against temperature in 
figure \ref{fig:xx} for two SQNs with initial baryon number \( 10^{42} \) each. 
It is obvious 
from the figure that ratio \( F_{\mathrm{grav}}/F_{\mathrm{rad}} \)
is very small initially. As a result, the nuggets will remain 
separated due to the radiation pressure. For temperatures lower than 
a critical value \( T_{\mathrm{cl}} \), the gravitational force starts 
dominating, facilitating the coalescence of the SQNs under mutual 
gravity.
\par
Let us now estimate the mass of the clumped SQNs, assuming that 
all of them within the horizon at the critical temperature will 
coalesce together. This is in fact a conservative estimate, since the
SQNs, although starting to move toward one another at 
\( T_{\mathrm{cl}} \), will take a finite time to actually coalesce, 
during which interval more SQNs will arrive within the horizon.
\par
In table \ref{tab:macho}, we show the values of \( T_{\mathrm{cl}} \) for SQNs 
of different initial baryon numbers along with the final masses of 
the clumped SQNs under the conservative assumption mentioned above.

\begin{table}
\begin{tabular}{|c|c|c|c|}
\hline 
\( b_{N} \) &\( T_{\mathrm{cl}} \) (MeV) & \( N_{N} \) &\( M/{M\odot} \) \\
\hline
& & & \\  
\( 10^{42} \) & 1.6  & \( 2.44 \times 10^{14} \) & 0.24  \\
& & & \\
\( 10^{44} \) & 4.45 & \( 1.13 \times 10^{11} \) & 0.01  \\
& & & \\
\( 10^{46} \) & 20.6 & \( 1.1 \times 10^7 \)   & 0.0001 \\
& & & \\
\hline
\end{tabular} 
\caption{\label{tab:macho} Critical temperatures ( \( T_{\mathrm{cl}}\) ) of SQNs of 
different initial sizes \( b_{N} \), the total number \( N_{N} \) 
of SQNs that coalesce together and their total final mass in 
solar mass units.}

\end{table}
\par
It is obvious that there can be no further clumping of these already
clumped SQNs; the density of such objects would be  too small within the
horizon for further clumping. Thus these objects would
survive till today and perhaps manifest themselves as MACHOs. 
It is to be reiterated that the masses of the clumped SQNs given in 
table \ref{tab:macho} are the lower limits and the final masses of these 
MACHO candidates will be larger. (The case for \( b_{N} = 10^{46} \) 
is not of much interest, especially since such high values of 
\( b_{N} \) are unlikely for the reasonable nucleation rates 
\cite{bhatta1@6,bhatta2@6}; we therefore restrict ourselves to the other 
cases in table \ref{tab:macho} in what follows.) 
A more detailed estimate of the masses will require a
detailed simulation, but very preliminary estimates indicate that 
they could be 2-3 times bigger than the values quoted in table \ref{tab:macho}.
\par
The total number of such clumped SQNs ( \( N_{\mathrm{macho}} \) ) 
within the horizon today is evaluated in the following way. With the
temperature \( \sim 3 {^0}K \) and time \( \sim 4 \times 10^{17} \)
seconds, the total amount of visible baryons within the horizon volume 
can be evaluated using photon to baryon 
ratio \( \eta \sim  10^{-10} \). The amount of baryons in the CDM will
be \( \frac{\Omega_{\mathrm{CDM}}}{\Omega_{\mathrm{B}}} \) times the
total number of visible baryons. This comes out to be \( \sim 1.6 \times
10^{79} \), \( \Omega_{\mathrm{CDM}} \) and \( \Omega_{\mathrm{B}} \)
being 0.3 and 0.01\footnote{%
The visible baryons occur in two forms. As visible stars they make up
about 0.3 - 0.6 \% and as hot intergalactic gas, they contribute about 
0.5 \% of the total density, i.e about 0.8 - 0.11 \% in all.
}
 respectively. The total number of baryons in a MACHO
is \( b_{N} \times N_{N} \) i.e. \( 2.44 \times 10^{56} \) and
\( 1.13 \times 10^{55} \) for initial nugget sizes \( 10^{42} \) and
\( 10^{44} \) respectively. The quantities \( b_{N} \) and
\( N_{N} \) are taken from the Table \ref{tab:macho}. So dividing the total number 
of baryons in CDM by that in a MACHO,
the \( N_{\mathrm{macho}} \) comes out to be in the range 
\( \sim 10^{23-24} \). 
\par
We can also mention here that 
if the MACHOs are indeed made up of quark matter, then they cannot 
grow to arbitrarily large sizes. Within the (phenomenological) Bag model 
picture \cite{chodos@6} of QCD confinement, where a constant vacuum 
energy density (called the Bag constant) in a cavity containing 
the quarks serves to keep them confined within the cavity, we have 
earlier investigated \cite{ban@6} the upper limit on the mass of 
astrophysical compact quark matter objects. It was found that for a 
canonical Bag constant {\bf B} of (145 MeV) \(^4 \), this limit 
comes out to be 1.4 \( M_\odot \) (see Chap.~\ref{chap:chandra}). The 
collapsed SQNs are safely below this limit. (It should be remarked here 
that although the value of {\bf B} in the original MIT bag model is taken 
to be {\bf B} \(^{1/4} \) = 145 MeV from the low mass
hadronic spectrum, there exist other variants of the Bag model 
\cite{hasen@6} , where higher values of {\bf B} are required. Even 
for {\bf B} \(^{1/4} \) = 245 MeV, this limit comes down 
to 0.54  \( M_\odot \) \cite{ban@6}, which would still 
admit such SQN.
\par
\begin{figure}[p]
\begin{center}
\includegraphics{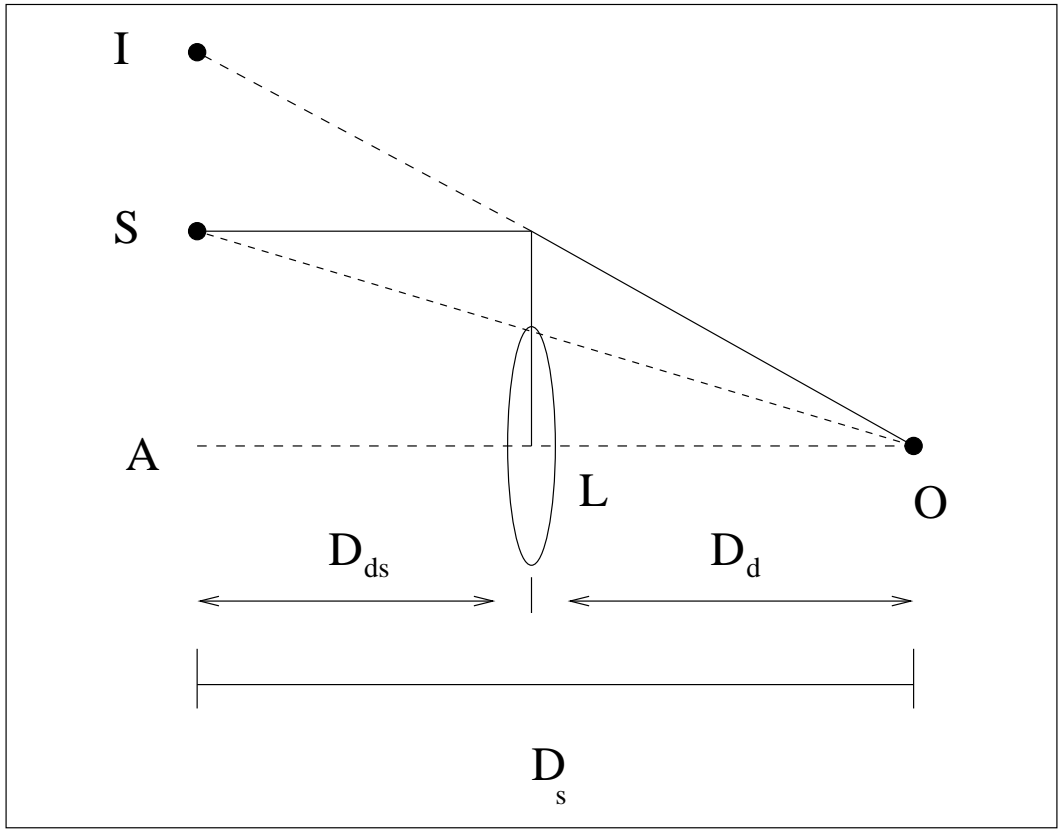}
\end{center}
\caption{\label{fig:lensn} Geometry of a gravitational lens system.
The light ray propagates from the source S, and while passing the
lens, gets deflected and reaches the observer at O. I is the image of
the source S . The distances between the observer and the source,
the observer and the lens and the lens and the source are
$D_s$,$D_d$,$D_{ds}$ respectively. OA is the optic axis.
\texttt{The figure has been adapted from \cite{narayan@6}}
}
\end{figure}

As a consistency check, we can perform a theoretical estimate of the
abundance of such MACHOs in the galactic halo which is 
conventionally given by the optical 
depth. The optical depth is the probability that at any instant of time
a given star is within an angle \( \theta_{E} \) of a lens, the lens 
being the massive body (in our case MACHO) which causes the deflection 
of light. In other words, optical depth is the integral over the number 
density of lenses times the area enclosed by the Einstein ring of 
each lens. The expression for optical depth can be written
as \cite{narayan@6}:
\begin{equation}
\tau = \frac{4 \pi G}{c^2} D_{s}^{2}\int \rho(x) x (1-x) dx 
\end{equation}
where \( D_s \) is the distance between the observer and the source,
\( G \) is the gravitational constant and \( x=D_{d} {D_{s}}^{-1} \),
\( D_d \) being the distance between the observer and the lens (Fig.~\ref{fig:lensn}). In
particular \( \rho \) is the mass-density of the MACHOs, which is of the
form \( \rho = \rho_0 \frac{1}{r^2} \) in the naive spherical halo model,
which we have adopted in our calculations. In the present case 
\( \rho_0 \) is given by 
\begin{equation}
\rho_0 = \frac{ M_{\mathrm{macho}} \times N_{\mathrm{macho}}}{4\pi R } 
\end{equation}
where 
\( R= \sqrt{D_{e}^2 + D_{s}^2 + 2 D_{e} D_{s} \cos\phi} \), 
\( \phi \) and \( D_{e} \) being the inclination of the LMC and the
distance of observer (earth) from the Galactic center respectively. 
\( M_{\mathrm{macho}} \) 
and \( N_{\mathrm{macho}} \) are the mass of a MACHO and 
the total number of MACHOs in the Milky Way halo.  
\par

The total visible mass of the Milky Way 
( \( \sim  1.6 \times 10^{11} M_{\odot} \) ) 
is equivalent to the mass of \( \sim 2\times 10^{68} \) baryons. 
This corresponds to  a factor of  \( \sim 2 \times 10^{-9} \) of all the 
visible baryons 
within the present horizon. Scaling the number of clumped SQNs within the 
horizon by the same factor yields a total number of  MACHOs,
\( N_{\mathrm{macho}} \sim 10^{13-14} \) in the Milky Way halo 
for the range of  baryon number of 
initial nuggets \( b_{N} = 10^{42-44} \). The value of  
\( D_{e} \) and \( D_{s} \) are taken to be 10 and 50 kpc, respectively.
The value of the inclination angle used here is 40 degrees. Using these 
values for a naive inverse square spherical model comprising such 
objects upto the LMC, we obtain an optical depth of  
\( \sim 10^{-6}-10^{-7} \).  
The uncertainty in this value is mainly governed by the value of  
\( \eta \), \( \Omega_{\mathrm{CDM}} \) and \( \Omega_{\mathrm{B}} \), 
and to a lesser 
extent by the specific halo model. This value compares reasonably well 
with the observed value \cite{sutherland@6,alcock2@6} and may be taken as a measure of reliability in 
the proposed model.
\par
As an interesting corollary, let us mention that the scenario presented
here could have other important astrophysical significance. The origin of
cosmic rays of ultra-high energy  \( \ge 10^{20} \) eV continues to be a 
puzzle. One of the proposed mechanisms \cite{pijush2@6} envisages a 
top-down scenario which does not require an acceleration mechanism and 
could indeed originate within our galactic halo. For our picture, such 
situations could easily arise from the merger of two or more such MACHOs, 
which would shed the extra matter so as to remain within the upper mass 
limit mentioned above. 
\par
We thus conclude that gravitational clumping of the primordial SQNs
formed in a first order cosmic quark - hadron phase transition appears to
be a plausible and natural explanation for the observed halo MACHOs. It
is quite remarkable that we obtain quantitative agreement with the
experimental values without having to introduce any adjustable parameters
or any fine-tuning whatsoever. 

\newpage
\rhead{Conclusions}
\chapter*{}
\addcontentsline{toc}{chapter}{Conclusions}
\vspace{-1.77in}
\begin{center} 
\vspace{0.2in}
\begin{center}
\includegraphics[scale=0.4]{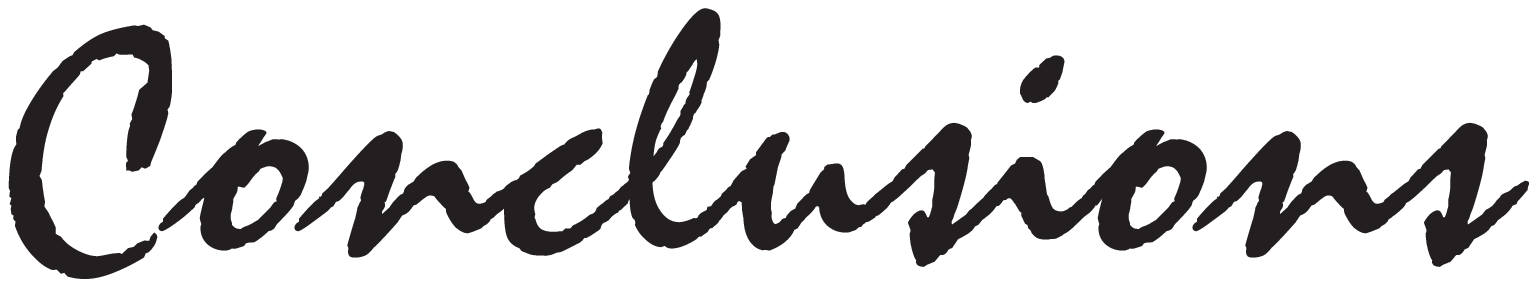}
\end{center}
\end{center}
\vspace{0.2in}
The present work is based on our quest for understanding the possible
role played by strange quark matter in the dark matter content of
the universe, which is believed to comprise more than 90 \% of the
total material content of the universe. The viewpoint which has emerged
out of this effort can be summarized as follows. 

At the age of a few microseconds, the universe underwent a possibly
first order phase transition. This phase transition is responsible
for generating the hadrons, which, before the transition existed in
the form of Quark-Gluon Plasma; the baryon number of the universe
was therefore contained in the quarks prior to this phase transition.
These hadrons, namely the baryons among them, would eventually form the lighter elements
through nucleosynthesis at an Universal age of $\sim180$ seconds (Tab.~\ref{tab:universeevents}).
Part of these light elements would be resynthesized by stars to heavier
varieties and some of it would spill all over the world due to supernovae
explosions. Some of these matter would be luminous and the rest of
it would exist in the form of dark gas -- but the total contribution
to the matter sector, due to all such forms of matter, collectively
called nuclear matter, would not be able to account for the fact the
fact that we live in a nearly closure density universe. The \emph{missing
matter} can however be very well accounted by Quasibaryonic matter
if one takes into account the fact that the bulk of the baryon number
content of the universe gets concealed within SQN's which also form
out of the aforementioned phase transition. It may be mentioned at
this point that although all of what has been said before remains
valid irrespective of whether the phase transition proceeds through
a second order (or even continuous) process, but the formation of SQN's
require the phase transition to be first order. The SQN's (The 
Acronym SQN first appear in \ref{sec:sqmiss}), thus formed are massive objects which tend to clump
together under mutual gravitational attraction, but prevented by the
radiation pressure during the radiation dominated phase of the universe.
As the universe expands, it cools down and as shown in Chap.\ref{chap:macho} , these
objects could coalesce together forming objects in the half solar mass range.
This happens when the Universe has cooled down to $\sim1-10$ MeV,
the exact time depending on the initial size of the SQN's. The objects,
so formed, would be more or less uniformly distributed through the
volume of the universe and have properties which are characteristic
for dark matter candidates. We have calculated the expected number
of such candidates at the current time residing in the Milky way halo
and estimated their optical depth for gravitational microlensing experiments
which look for dark lenses in the galactic halo. The value of the
optical depth, so obtained, compare reasonably well with the observed
values and may be taken to be a measure of the reliability of the
dark matter model proposed by us. It should be emphasized that a definite
conclusion can only be reached after a detailed simulation is carried
out. The central value of the mass range of the dark halo lenses are
typically in the 0.5 $M_{\odot}$ range. The masses of the clumped
SQN's can however be more or less than this value, depending on the
time when they clump. The results obtained in Chap.\ref{chap:chandra} however indicates
that the dark SQM lenses cannot be much heavier than this, since the
maximum mass range of compact quark matter objects is in the same
range. This so called \emph{Chandrasekhar Limit} does not depend strongly
on the number of quark flavors that goes into the making of the stars
and is also applicable to the lens SQN's. Thus although gravitational
clumping might produce over-sized SQN's, these would have to disintegrate,
in order to maintain stability to sizes $\sim0.5\, M_{\odot}$ or
less. The halo SQN's are therefore suitable candidates for the MACHO's
found in the gravitational microlensing experiments. In the course
of evaluation of the maximum mass limit for a configuration of massless
quarks through an analytic procedure, we have adopted the Landau picture
of energy balance through the density dependent quark mass model of
confinement which was appropriate for the situation. 

According to our picture, the local source of dark matter arises mainly
from SQM blobs in the half solar mass range. It is therefore quite
possible that occasional collisions of such objects can release bursts
of small, atomic sized \emph{strangelets} in every possible directions,
and a few of them may be intercepted by our planet as well. In Chap.\ref{chap:strabund}
we examine the probability of observing such particles in Earth based
experiments by various means. It has been emphasized earlier in this
work ( Chap.\ref{chap:dust} and Chap.\ref{chap:newmodel}) that the detection of strangelets
is crucial both from the standpoint of astrophysics as well as strong
interaction physics. For astrophysics it can provide confirmation
for the nature of dark matter - it's detection will have the implication
that dark matter is quasibaryonic in form. This will also help the
\emph{Nuclear desert} to be filled up with intermediate baryon number
objects between atomic species to nuclear stars. For QCD it will provide
experimental justification for Witten's conjecture that the true ground
state of strong interaction physics is SQM rather than the ordinary nuclear matter. 

In Chap.\ref{chap:dust} and and Chap.\ref{chap:newmodel} we have therefore examined
thoroughly the problem of strangelet propagation through the terrestrial
atmosphere. In order for the model to be consistent with the hypothesis
of stability of strange matter with respect to ordinary nuclear matter,
the strangelets have been invested with the extraordinary property
of absorbing a fraction of atmospheric particles which are incident
on it. This property is in stark contrast with the passage of
an ordinary ( heavy) cosmic ray particle which usually breaks up under such
impact. In this model the strangelet grows like a snowball, absorbing
mass and also some charge from the atmospheric particles; however
it's energy decreases due largely to the ionization loss of the surrounding
media and partly due to the impacts. In fact the energy decreases
so much that they go beyond the range of detectability of passive
Solid State Nuclear Track Detector's below typical mountain altitudes.
The study reveals two important aspects : for one, it reproduces the
observed pattern of several \emph{exotic} cosmic ray events (very
small $e/m$ ratio and detection at atmospheric depths much higher
than ordinary cosmic ray particles, but not lower than typical mountain
altitudes, as well as the value of the charges and masses found at
those altitudes.) fairly well, suggesting that these events, previously
unclassified, can now be associated, quite justifiably, with the passage
of strangelets through the atmosphere. Secondly, the estimated energy
deposition of the particles in SSNTD's like CR-39 show that these
are just above the threshold of detection at mountain altitudes, indicating
that a ground based large area detector array of SSNTD's might be
quite capable of picking up strangelet signals. 

Throughout our work we argue that strange matter is an essential component
of the dark matter forms present in the Universe. The whole work,
save for the part that deals with the maximum mass limit of quark
stars, depend rather crucially on two factors :

\begin{enumerate}
\item The strange matter hypothesis.
\item The existence of the first order Quark-Hadron deconfining phase transition in
the early universe.
\end{enumerate}
The general agreement is that  both are well founded hypothesis;
however, none of them, so far, has passed the test of experimental
verification. It thus appears that, once again, cosmological experiments
will be able to disentangle issues which accelarator based experiments
may not be able to address.

In the course of this work, several threads have emerged, which can
extend the ideas developed in several ways. In Chap.\ref{chap:macho}, as
well as earlier in this chapter, we have mentioned that a detailed
simulation is needed before one can reach a conclusion about the formation
of half solar mass objects by the coalescence of the SQN's. We propose
to undertake an extensive numerical study of this case in near future.
The idea behind the model of neutron absorption by SQM, used in Chap.
\ref{chap:dust}-\ref{chap:newmodel} is also applicable in a cosmological setting, since the depletion
of baryons near SQN's can cause local baryon inhomogeneities and affect
the nucleosynthesis in unknown ways. The network of SQN's, in some
ways, form the first instances of structure in the early Universe.
Inspired by the applicability of the density dependent quark mass
model to quark star systems, we also propose to study the issue of
the stability of SQM in the context of this model. In the present
work we have not examined the role of SQM in the dark energy content
of the universe, but there do exist some theoretical hints to assume
that they might play an important role in providing for the acceleration
of the universal expansion rate. Some work in this direction is already
in progress, but it already raises some intriguing questions which
seem to be tied to the foundations of quantum mechanics; in particular,
the effect of quantum entanglement in relativistic many body systems
need to be explored much further before a definite commitment can
be made. 

In a nutshell, then, we have argued that the standard model of particle
interactions and the strange matter hypothesis together can account
for the cosmological dark matter problem, without having to resort
to exotic reformulations of the physics of particle interactions. 

\vspace{0.2in}
\begin{center}
\includegraphics[scale=0.4]{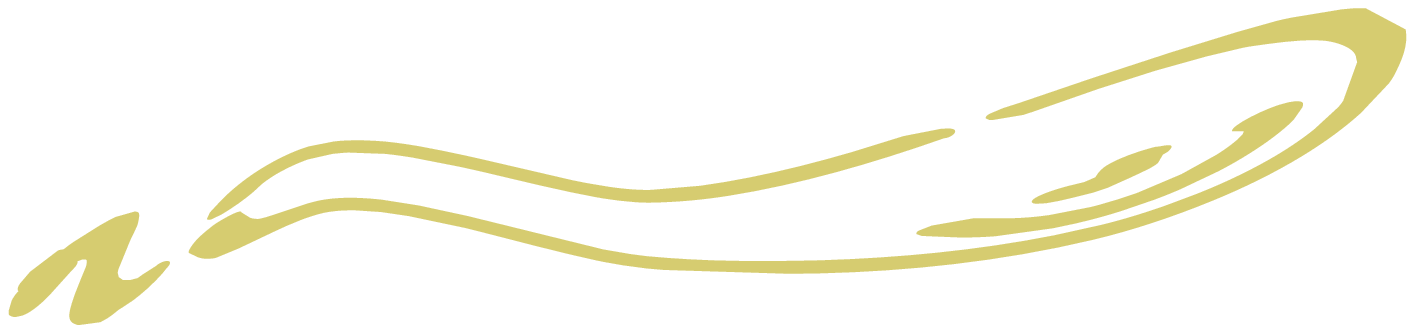}
\end{center}

\newpage
\rhead{}
\addcontentsline{toc}{chapter}{List of Publications}
\vspace{-3in}
\begin{center} \Large{List of Publications} \end{center}
\begin{enumerate}
\item \textsc{Strangelets in terrestrial atmosphere}, with S.K.Ghosh, S.Raha \& D.Syam, \textit{%
Journal of Physics}, 1999, \textbf{G25} L15

\item \textsc{The Chandrasekhar limit for quark stars}, With S.K.Ghosh \& S.Raha, \textit{%
Journal of Physics}, 2000 \textbf{G26} L1

\item \textsc{Can Cosmic Strangelets Reach the Earth ?}, With S.K.Ghosh  S.Raha \& D.Syam, \textit{%
Physical Review Letters}, 2000, \textbf{85} 1384

\item \textsc{Strange quark matter in cosmic rays and exotic events}, With S.K.Ghosh, A.Mazumdar, S.Raha \& D.Syam,
\textit{Astrophysics \& Space Science}, 2000, \textbf{274} 655

\item \textsc{Massive compact halo objects from the relics of the cosmic quark-hadron transition}, With 
A.Bhattacharyya, S.K. Ghosh, S.Raha, Bikash Sinha and H.Toki, \textit{Monthly Notices of the
Royal Astronomical Society}, 2003 \textbf{340} 284

\item \textsc{Relics of the cosmic quark-hadron phase transition and massive compact halo objects}, With
A.Bhattacharyya, S.K. Ghosh, S.Raha, Bikash Sinha and H.Toki, \textit{Nuclear Physics}, 2003, \textbf{A715} 827 

\item \textsc{Some aspects of strangeness in astrophysics and cosmology}, With A.Bhattacharyya, S.K. Ghosh,
S.Raha, Bikash Sinha and H.Toki, \textit{Nuclear Physics}, 2003, \textbf{A721}, 1028

\item \textsc{Quantum chromodynamics, phase transition in the early universe and quark nuggets}, 
With A.Bhattacharyya, S.K. Ghosh, S.Raha, Bikash Sinha and H.Toki, 2003, \textit{Pramana - Journal of Physics}
\textbf{60}, 909

\end{enumerate}

\end{document}